\documentclass[fleqn,usenatbib]{mnras}

\usepackage{graphics,graphicx}

\usepackage{amsmath,amsthm,amssymb}
\usepackage{xspace}
\usepackage{ulem}
\usepackage{color}
\usepackage{notoccite}
\usepackage{hyperref}

\def\gs{\mathrel{\raise0.35ex\hbox{$\scriptstyle >$}\kern-0.6em \lower0.40ex\hbox{{$\scriptstyle \sim$}}}}
\def\ls{\mathrel{\raise0.35ex\hbox{$\scriptstyle <$}\kern-0.6em \lower0.40ex\hbox{{$\scriptstyle \sim$}}}}

\def\Shihii{\mbox{$\Sigma_{\rm H{\sc I}+H{\sc II}}$}\xspace}

\def\Tk{\mbox{$T_{\rm{k}}$}\xspace}
\def\Td{\mbox{$T_{\rm{dust}}$}\xspace}

\def\Hpe{\mbox{$\Gamma_{\rm{PE}}$}\xspace}

\def\Hcrhi{\mbox{$\Gamma_{\rm{CR,H}\textsc{i}}$}\xspace}
\def\Hcrh2{\mbox{$\Gamma_{\rm{CR,H}_2}$}\xspace}

\def\Cions{\mbox{$\Lambda_{\rm{atoms+ions}}$}\xspace}
\def\Crec{\mbox{$\Lambda_{\rm{rec}}$}\xspace}
\def\Cff{\mbox{$\Lambda_{\rm{f-f}}$}\xspace}
\def\Ch2{\mbox{$\Lambda_{\rm{H}_2}$}\xspace}
\def\Cco{\mbox{$\Lambda_{\rm{CO}}$}\xspace}
\def\Cgd{\mbox{$\Lambda_{\rm{gas-dust}}$}\xspace}
\def\Ccii{\mbox{$\Lambda_{\rm{C}\rm{\textsc{ii}}}$}\xspace}

\def\Coi{\mbox{$\Lambda_{\rm{O}\rm{\textsc{i}}}$}\xspace}

\def\nH{\mbox{$n_{\rm{H}}$}\xspace}
\def\nHplus{\mbox{$n_{\rm{H}^+}$}\xspace}
\def\colH{\mbox{$N_{\rm{H}}$}\xspace}
\def\ne{\mbox{$n_{\rm{e}}$}\xspace}

\def\xe{\mbox{$x_{\rm{e}}$}\xspace}
\def\next{\mbox{$n_{\rm{H,ext}}$}\xspace}
\def\Pe{\mbox{$P_{\rm{ext}}$}\xspace}
\def\nh2{\mbox{$n_{\rm{H}_2}$}\xspace}
\def\Sh2{\mbox{$\Sigma_{\rm{mol}}$}\xspace}
\def\nhi{\mbox{$n_{\rm{H}\textsc{i}}$}\xspace}

\def\aCO{\mbox{$\alpha_{\rm CO}$}\xspace}
\def\fh2{\mbox{$f'_{\rm mol}$}\xspace}
\def\Mh2{\mbox{$M_{\rm{H}_2}$}\xspace}

\def\Mgmc{\mbox{$m_{\rm{GMC}}$}\xspace}
\def\g0{\mbox{$G_0'$}\xspace}
\def\ga{\mbox{$G_{0{\rm ,att}}'$}\xspace}
\def\Z{\mbox{$Z'$}\xspace}
\def\Zn{\mbox{$Z'$}}
\def\cri{\mbox{$\zeta_{\rm{CR}}$}\xspace}
\def\crihi{\mbox{$\zeta_{\rm{CR,H}\textsc{i}}$}\xspace}
\def\crimol{\mbox{$\zeta_{\rm{CR,H}_2}$}\xspace}
\def\Av{\mbox{$A_{\rm{v}}$}\xspace}
\def\Avtr{\mbox{$A_{\rm{v}}^{\rm{(tr)}}$}\xspace}
\def\sv{\mbox{$\sigma_{\rm{v}}$}\xspace}
\def\Re{\mbox{$R_{\rm{GMC}}$}\xspace}

\def\SFRsd{\mbox{$\Sigma_{\rm{SFR}}$}\xspace}
\def\gassd{\mbox{$\Sigma_{\rm{gas}}$}\xspace}
\def\starssd{\mbox{$\Sigma_{\rm{*}}$}\xspace}

\def\rh2{\mbox{$R_{\rm{H}_2}$}\xspace}
\def\rh2{\mbox{$R_{\rm{H}_2}$}\xspace}
\def\rrh2{\mbox{$r_{\rm{H}_2}$}\xspace}

\def\rcl{\mbox{$R_{\rm{GMC}}$}\xspace}
\def\rl10{\mbox{$R_{\rm{L10}}$}\xspace}
\def\rp06{\mbox{$R_{\rm{P06}}$}\xspace}

\def\oi{\mbox{[\rm{O}{\sc i}]}\xspace}

\def\lsun{\mbox{$L_\odot$}\xspace}
\def\cmps{\mbox{cm$^{-2}$}\xspace} 
\def\cmpc{\mbox{cm$^{-3}$}\xspace} 
\def\kms{\mbox{km\,s$^{-1}$}\xspace}
\def\ps{\mbox{s$^{-1}$}\xspace}

\def\msun{\mbox{$\rm{M}_\odot$}\xspace}
\def\sfru{\mbox{$\rm{M}_\odot$~yr$^{-1}$}\xspace}

\def\hu{\mbox{ergs\,cm$^{-3}$\,s$^{-1}$}\xspace}

\def\h2{\mbox{{\sc H}$_2$}\xspace}
\def\hi{\mbox{{\rm{H}\textsc{i}}}\xspace}
\def\hii{\mbox{\rm{H}\textsc{ii}}\xspace}
\def\cii{\mbox{[\rm{C}\textsc{ii}]}\xspace}

\def\sigame{{\sc s\'igame}\xspace}
\def\lime{{\sc lime}\xspace}
\newcommand{\e}  [1]{\ensuremath{\times10^{#1}}}

\newcommand{\ave}[1]{\ensuremath{\langle #1 \rangle}}

\def\apj{ApJ}
\def\aj{AJ}
\def\apjl{ApJL}
\def\aap{A\&A}
\def\nat{Nature}
\def\mnras{MNRAS}
\def\araa{ARA\&A}

\title[Simulating CO emission with S\'IGAME]{SImulator
of GAlaxy Millimetre/submillimetre Emission (\sigame\thanks{\sigame means
`follow me' in Spanish.}): CO emission from massive $z=2$ main-sequence
galaxies}

\author[K. P. Olsen]{Karen P. Olsen,$^{1}$
\thanks{\small{Email: \href{mailto:karen@dark-cosmology.dk}{karen@dark-cosmology.dk} (KPO)}; 
\href{mailto:t.greve@ucl.ac.uk}{t.greve@ucl.ac.uk} (TRG); 
\href{mailto:brinch@nbi.dk}{brinch@nbi.dk} (CB)}, 
Thomas R. Greve,$^{2}$$\dagger$
Christian Brinch,$^{3,4}$$\dagger$
Jesper Sommer-Larsen,$^{1,5,6}$ 
\newauthor
Jesper Rasmussen,$^{1,7}$
Sune Toft,$^{1}$
Andrew Zirm$^{1}$\\
$^{1}$Dark Cosmology Centre, Niels Bohr Institute, University of Copenhagen, Juliane Maries Vej 30, DK-2100 Copenhagen, Denmark\\
$^{2}$Department of Physics and Astronomy, University College London, Gower Street, London WC1E 6BT, UK\\
$^{3}$Centre for Star and Planet Formation (Starplan) and Niels Bohr Institute, University of Copenhagen,\\
Juliane Maries Vej 30, DK-2100 Copenhagen, Denmark\\
$^{4}$DeIC, Technical University of Denmark, Building 309, DK-2800 Kgs. Lyngby, Denmark\\
$^{5}$Excellence Cluster Universe, Boltzmannstr. 2, 85748 Garching, Germany\\
$^{6}$Marie Kruses Skole, Stavnsholtvej 29-31, DK-3520 Farum, Denmark\\
$^{7}$Department of Physics, Technical University of Denmark, Building 309, DK-2800 Kgs. Lyngby, Denmark}

\begin{document}

\date{}

\pagerange{\pageref{firstpage}--\pageref{lastpage}} \pubyear{2015}

\maketitle

\label{firstpage}

\begin{abstract}
We present \sigame (SImulator of GAlaxy Millimetre/submillimetre Emission), a
new numerical code designed to simulate the $^{12}$CO rotational line 
spectrum of galaxies.  Using sub-grid physics recipes to post-process the
outputs of smoothed particle hydrodynamics (SPH) simulations, a molecular gas
phase is condensed out of the hot and partly ionised SPH gas. The gas is  
subjected to far-UV
radiation fields and CR ionisation rates which are set to scale with the local
star formation rate volume density.  Level populations
and radiative transport of the CO lines are solved with the 3-D radiative
transfer code \lime. We have applied \sigame to cosmological SPH
simulations of three disk galaxies at $z=2$ with stellar masses in the range
$\sim0.5-2\times10^{11}$\,\msun and star formation rates $\sim40-140$\,\sfru. 
Global CO luminosities and line ratios are in agreement with observations of 
disk galaxies at $z\sim2$ up to and
including $J = 3-2$ but falling short of the few existing $J=5-4$ observations. 
The central $5\,{\rm kpc}$ regions of our galaxies have
CO $3-2/1-0$ and $7-6/1-0$ brightness temperature ratios
of $\sim0.55-0.65$ and $\sim0.02-0.08$, respectively, 
while further out in the disk the ratios drop
to more quiescent values of $\sim0.5$ and $<0.01$.
Global CO-to-H$_2$ conversion (\aCO) factors are $\simeq 1.5\,{\rm
\msun\,pc^{-2}\,(K\,km\,s^{-1})^{-1}}$, i.e., $\sim 2-3\times$ below the
typically adopted values for disk galaxies, and \aCO increases with radius, 
in agreement with observations of nearby galaxies.  
Adopting a top-heavy Giant Molecular Cloud (GMC) mass spectrum does not significantly change 
the results. 
Steepening the GMC density profiles leads to higher global line ratios for
$J_{\rm up} \ge 3$ and CO-to-H$_2$ conversion factors ($\simeq 3.6\,{\rm
\msun\,pc^{-2}\,(K\,km\,s^{-1})^{-1}}$).  
\end{abstract}


\section{Introduction}
Inferring the physical properties of the molecular gas in galaxies is an
important prerequisite for understanding their star formation. Carbon monoxide
(CO) has proven to be a valuable tracer of the molecular interstellar medium
(ISM), its rotational transitions being sensitive to a range of gas densities
and temperatures through collisional excitation by \h2.  Observations of CO
lines have therefore been used extensively to probe the conditions of the
molecular ISM in galaxies at high and low redshifts.  However, the number of $z
> 1$ galaxies detected in CO is still relatively modest -- little more than
about 200 at the time of writing -- with only a fraction of those having
multiple CO transitions observed \citep[see review
by][]{carilli-and-walter-2013}.  Furthermore, the vast majority of high-$z$ CO
measurements are of extreme objects such as submillimetre-selected galaxies
\citep[SMGs, see][]{frayer98,neri03,bothwell13,yun15} 
and powerful QSOs with infra-red luminosities in excess of
$10^{12}\,{\rm \lsun}$ \citep{barvainis94,weiss07,riechers11}. 
Very few normal star forming
galaxies at high redshifts, i.e., galaxies on the galaxy `main sequence' (MS),
have been observed in CO to date, and as a result little is known, however,
about the ISM conditions of normal star-forming galaxies at high redshifts.
Some of the best studied cases are the four BzK-selected galaxies BzK$-$4171,
BzK$-$16000, and BzK$-$21000 at $z\simeq 1.5$, where up to four CO transitions,
including CO(1$-$0), have been observed
\citep{dannerbauer09,aravena10,daddi15}.  While the initial analysis of the
low-$J$ ($J_{\rm up} \le 3$) transitions in these galaxies suggested a
quiescent ISM reminiscent of the Milky Way (MW) \citep{dannerbauer09}, the more
recent CO$(5-4)$ observations require a second much denser, and likely warmer,
ISM component, which is thought to be directly associated with star forming
clumps within the galaxies \citep{daddi15}. Studies such as these illustrate
the importance of sampling both the low- and high-$J$ parts of the CO 
Spectral Line Energy Distribution (SLED) in
order to obtain a comprehensive picture of the molecular gas in galaxies.

With the Atacama Large Millimetre/submillimetre Array (ALMA) now in operation,
the collection of CO data is gaining rapid pace as is our understanding of the
molecular gas in distant galaxies.  Galaxy evolution simulations that include
molecular line emission (such as CO) are an important tool with which to
interpret the observed molecular emission of galaxies caught in various
evolutionary stages.  The past decade saw the development of the first detailed
simulations of CO line emission from galaxies, which consisted of SPH galaxy
simulations \citep{narayanan06,narayanan11,narayanan14,greve08}, or
semi-analytical models
\citep{obreschkow09,obreschkow11,lagos12,munoz13,popping14}, combined with
sub-grid physics prescriptions to model the molecular gas phase and its CO
content.  The simulations were used to examine the CO emission properties of
high-$z$ mergers and starbursts \citep{narayanan08a,narayanan09}, active
galactic nuclei (AGN) and quasars \citep{narayanan06,narayanan08,narayanan08b},
as well as moderately star-forming galaxies
\citep{greve08,narayanan11,lagos12,popping14}. For example, simulating a large
set of galaxies, \cite{narayanan14} proposed a calibration scheme in which the
global CO SLED is parameterised by the star formation rate (SFR) surface
density, the latter being a more readily observed quantity. Simulations of CO
have also been used to examined how well CO can probe the relationship between
gas surface density and SFR, i.e., the so-called Kennicutt-Scmidt relations,
and the validity of the CO-to-H$_2$ conversion factor ($X_{\rm CO}$) used to
infer \h2 column densities from measured CO $J=1-0$ line intensities
\citep[e.g.,][]{narayanan11,narayanan13}.

Here we present a new numerical framework for simulating the molecular line
emission from star-forming galaxies. The code, SImulator of GAlaxy
Millimetre/submillimetre Emission (\sigame), combines SPH (or grid-based)
simulations of galaxies with sub-grid physics prescriptions for the \h2/\hi
fraction and the ISM thermal balance on scales of individual Giant Molecular Clouds (GMCs).
\sigame accounts for a FUV and cosmic ray (CR) intensity field that vary with local
SFR density within the galaxy. 
In this paper, we apply \sigame to cosmological SPH simulations of three
massive, normal star-forming galaxies at $z=2$ (i.e., so-called main-sequence
galaxies), and model their CO rotational line spectrum using a publicly
available 3D radiative transfer code.  We show that \sigame reproduces observed
low-$J$ CO line luminosities and provides new estimates of the \aCO factor for
main-sequence galaxies at $z\sim2$, while at the same time predicting their CO
line luminosities at high-$J$ ($J_{\rm up}>6$) transitions where observations
are yet to be made. 

The structure of the paper is as follows. Section $2$ describes the
cosmological SPH simulations used, along with the basic properties of the three
$z=2$ star-forming galaxies extracted from the simulations. A detailed
description of \sigame and its application to the three simulations is
presented in Section $3$, along with a bench test of the code on MW-like
galaxies at $z=0$.  The resulting CO emission maps and SLEDs 
of the $z=2$ simulations are presented in Section $4$, and
compared to CO observations of similar galaxies at $z\sim 1-2.5$.  In section
$5$, we explore how the results in Section $4$ change when adopting alternative
ISM models. Section $6$ discusses the strengths and weaknesses of \sigame in
the context of other molecular line (CO) simulations and recent observations.
Finally, in Section $7$ we summarise the main steps of \sigame, and list our
main findings regarding the CO line emission from massive, star-forming
galaxies at $z\simeq 2$.  We adopt a flat cold dark matter ($\Lambda$CDM)
scenario with $\Omega_m=0.3$, $\Omega_\Lambda=0.7$ and $h=0.65$ 
\citep{planck14}.

\section{Cosmological Simulations}\label{cosmological_simulations}
\subsection{SPH simulations}
We employ a cosmological TreeSPH code for simulating galaxy formation and
evolution, though in principle, grid-based hydrodynamic simulations could be
incorporated equally well.  The TreeSPH code used for the simulations is in most
respects similar to the one described in \cite{somm05} and \cite{romeo06}. A
cosmic baryon fraction of $f_{\rm{b}}=0.15$ is assumed, and simulations are
initiated at a redshift of $z=39$.  The implementation of star formation and
stellar feedback, however, has been manifestly changed. 

Star formation is assumed to take place in cold gas ($\Tk \la 10^4$\,K) at
densities $\nH > 1$\,\cmpc. The star formation efficiency (or probability that a
gas particle will form stars) is formally set to 0.1, but is, due to effects of
self-regulation, considerably lower.
Star formation takes place in a stochastic way, and in a star formation event,
$1$ SPH gas particle is converted completely into $1$ stellar SPH particle,
representing the instantaneous birth of a population of stars according to a
Chabrier (2003) stellar initial mass function (IMF; \citealt{chabrier03}) -- see
further below.

The implementation of stellar feedback is based on a sub-grid super-wind model,
somewhat similar to the `high-feedback' models by \cite{stinson06}. These
models, though, build on a supernova blast-wave approach rather than super-wind
models. Both types of models invoke a Chabrier (2003) IMF, which is somewhat
more top-heavy in terms of energy and heavy-element feedback than, e.g., the
standard Salpeter IMF. The present models result in galaxies characterised by
reasonable $z=0$ cold gas fractions, abundances and circum-galactic medium
abundance properties. They also improve considerably on the ``angular momentum
problem'' relative to the models presented in, e.g., \cite{somm03}. The models
will be described in detail in a forthcoming paper. 

\subsection{The model galaxies}\label{model_galaxies}
Three model galaxies, hereafter referred to as G1, G2 and G3 in order of
increasing SFR, were extracted from the above SPH simulation and re-simulated
using the `zoom-in' technique described in \citep[e.g.,][]{somm03}. The emphasis
in this paper is on massive ($M_* \gtrsim 5\times 10^{10}\,\msun$) galaxies, and the
three galaxies analysed are therefore larger, rescaled versions of galaxies
formed in the 10/$h$\,Mpc cosmological simulation described in \cite{somm03}.
The linear scale-factor is of the order 1.5, and since the CDM power spectrum is
fairly constant over this limited mass range the rescaling is a reasonable
approximation. 

Galaxy G1 was simulated at fairly high resolution, using a total of $1.2\times
10^6$ SPH and dark matter particles, while about $9\times 10^5$ and $1.1\times
10^6$ particles were used in the simulations of G2 and G3, respectively.  For
the G1 simulation, the masses of individual SPH gas, stellar and dark matter
particles are $m_{\rm{SPH}}=m_*\approx6.3\e{5}$ $h^{-1}$M$_{\odot}$ and
$m_{\rm{DM}}$=$3.5\e{6}$ $h^{-1}$M$_{\odot}$, respectively.  Gravitational
(cubic spline) softening lengths of 310, 310 and 560 $h^{-1}$pc, respectively,
were employed. Minimum gas smoothing lengths were about $50\,h^{-1}$pc.  For the
lower resolution simulations of galaxies G2 and G3, the corresponding particle
masses are $m_{\rm{SPH}}=m_*\approx4.7\e{6}$\,$h^{-1}$M$_{\odot}$ and
$m_{\rm{DM}}= 2.6\e{7}$ $h^{-1}$M$_{\odot}$, respectively, and the gravitational
softening lengths were 610, 610 and 1090 $h^{-1}$pc. Minimum gas smoothing
lengths were about 100 $h^{-1}$pc.  
\begin{figure}
\centering
\includegraphics[width=1\columnwidth]{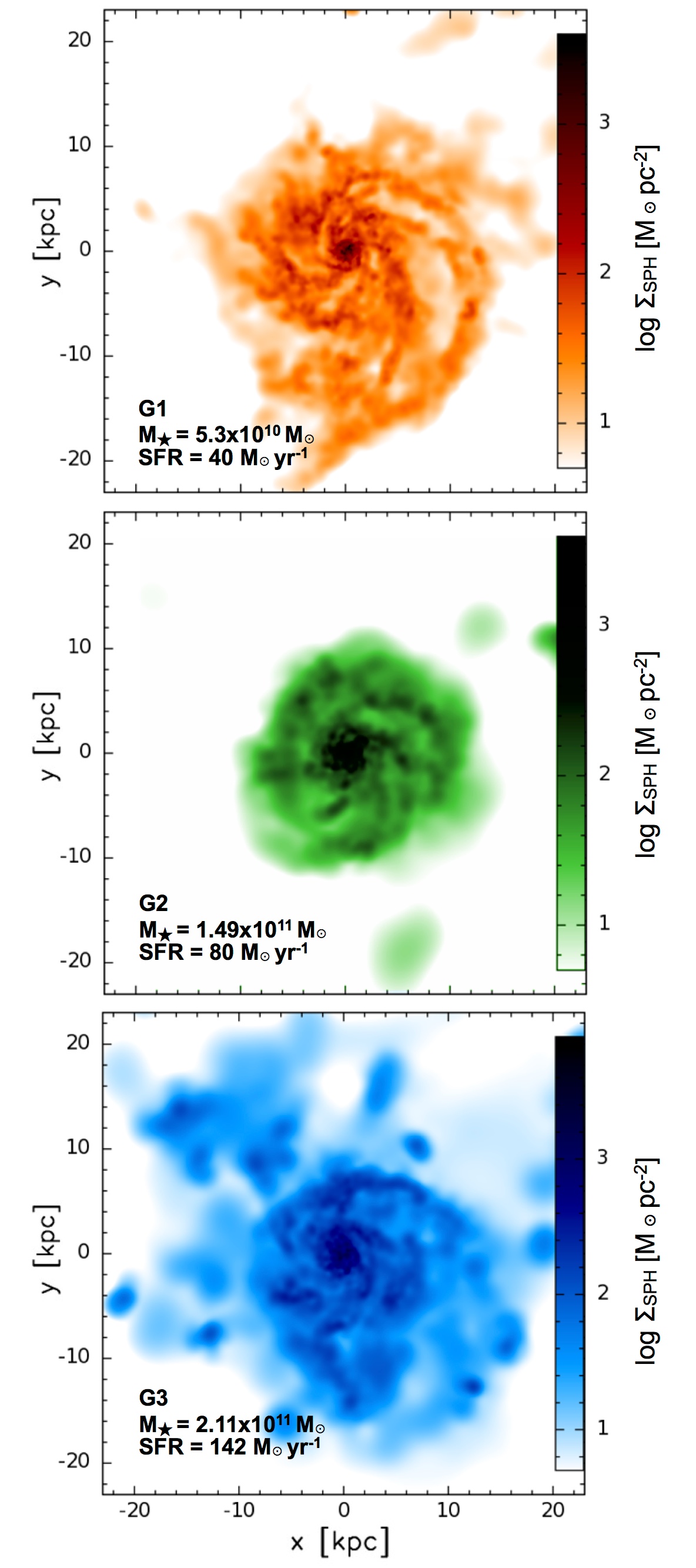}
\caption{SPH gas surface density maps of the three model galaxies G1 (top), G2
(middle), and G3 (bottom) viewed face-on.  The stellar masses and SFRs of each
galaxy are indicated  (see also Table\,\ref{model_galaxies_t1}).  The maps have
been rendered with the visualization tool SPLASH version 2.4.0 \citep{price07}
using the gas smoothing lengths provided by the simulations. 
}
\label{model_galaxies_f1}
\end{figure}

Due to effects of gravitational softening, typical velocities in the innermost
parts of the galaxies (typically at radii less than about
2$\epsilon_{\rm{SPH}}$, where $\epsilon_{\rm{SPH}}$ is the SPH and star particle
gravitational softening length) are somewhat below dynamical values \citep[see,
e.g.][]{somm98}. The dynamical velocities will be of the order
$v_{\rm{dyn}}=\sqrt{G M(R)/R}$, where $G$ is the gravitational constant, $R$ is
the radial distance from the centre of the galaxy and $M(R)$ is the total mass
located inside of $R$. Indeed, it turns out that for the simulated galaxies
considered in this paper SPH particle velocities inside of
2$\epsilon_{\rm{SPH}}$ are only about 60-70\% of what should be expected from
dynamics. To coarsely correct for this adverse numerical effect, for SPH
particles inside of 2$\epsilon_{\rm{SPH}}$ the velocities are corrected as
follows: For SPH particles of total velocity less than $v_{\rm{dyn}}$, the
tangential component of the velocity is increased such that the total velocity
becomes equal to $v_{\rm{dyn}}$. Only the tangential component is increased in
order not to create spurious signatures of merging. With this correction
implemented, the average ratio of total space velocity to dynamical velocity of
all SPH particles inside of 2$\epsilon_{\rm{SPH}}$ equals unity.

\bigskip

Figure \ref{model_galaxies_f1} shows surface density maps of the SPH gas in G1,
G2, and G3, i.e., prior to any post-processing by \sigame.  The gas is seen to
be strongly concentrated towards the centre of each galaxy and structured in
spiral arms containing clumps of denser gas.  The spiral arms reach out to a
radius of about $20$\,kpc in G1 and G3, with G2 showing a more compact
structure that does not exceed $R\sim15$\,kpc.  Table \ref{model_galaxies_t1}
lists key properties of the simulated galaxies, namely their SFR, stellar mass
($M_*$), SPH gas mass ($M_{\rm SPH}$), SPH gas mass fraction ($f_{\rm
SPH}=M_{\rm SPH}/(M_*+M_{\rm SPH})$), and metallicity \Z. These quantities were
measured within a radius ($R_{\rm cut}$, also given in
Table\,\ref{model_galaxies_t1}) corresponding to where the radial cumulative
stellar mass function has flattened out. The metallicity is in units of solar
metallicity and is calculated from the abundances of C, N, O, Mg, Si, S, Ca and
Fe in the SPH simulations, and adjusted for the fact that not all heavy metals
have been included according to the solar element abundancies measured by
\cite{asplund09}.  
\begin{center}
\begin{table}
\centering
\caption{Physical properties of the three simulated galaxies G1, G2, and G3} 
\begin{tabular}{@{}lllllll@{}} 
\hline
\hline
 	 	&	SFR 		& 	$M_{\ast}$	 		 & $M_{\rm SPH}$ 		&	$f_{\rm SPH}$	&	$\Z$	& 	$R_{\rm cut}$  \\ 
		&[$\rm{M}_\odot$~yr$^{-1}$]& 	[$10^{11}\,\rm{M}_\odot$]	 & [$10^{10}\,\rm{M}_\odot$]	&					&		& 	[kpc]  \\ 
\hline
G1				&	40	& 	$0.53$		&	$2.07$		 &	28\%		&	1.16	&	$20$		\\ 
G2				& 	80	&	$1.49$		&	$2.63$		 &	15\%		&	1.97	&	$15$		\\ 
G3 				& 	142	&	$2.11$		&	$4.66$		 &	18\%		&	1.36	&	$20$		\\ 
\hline
\end{tabular}
\\
	{\bf Notes.} All quantities are determined within a radius $R_{\rm cut}$,
	which is the radius where the cumulative radial stellar mass function of
	each galaxy becomes flat. The gas mass ($M_{\rm SPH}$) is the total SPH gas
	mass within $R_{\rm cut}$.  The metallicity ($\Z=Z/Z_{\odot}$) is the mean
	of all SPH gas particles within $R_{\rm cut}$.
\label{model_galaxies_t1}
\end{table}
\end{center}

\begin{figure*}
\centering
\includegraphics[width=2\columnwidth]{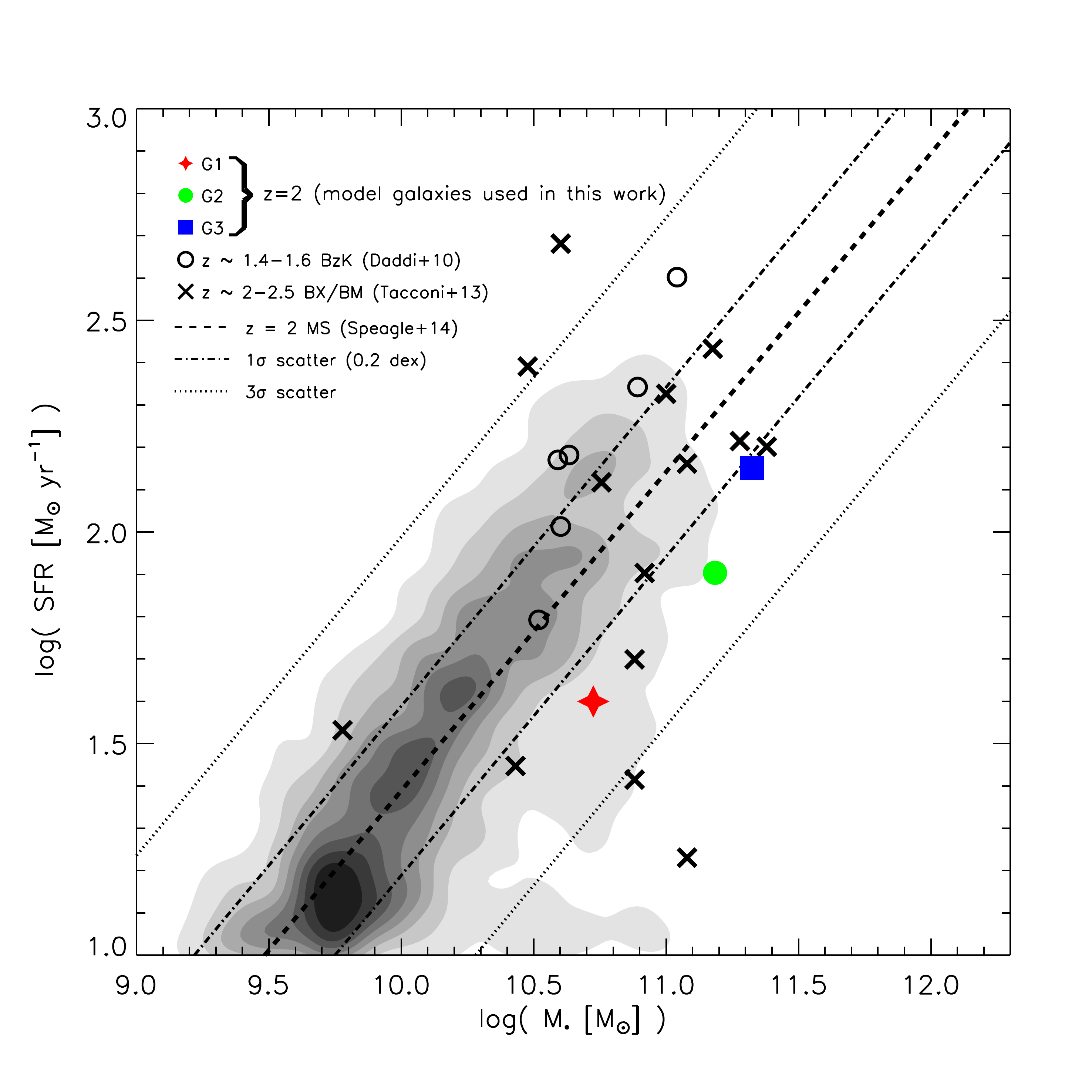}
\caption{Position of the three model galaxies studied here (G1, G2 and G3 with
filled a red star, a green circle and a blue square respectively), on a SFR--M$_{\ast}$ diagram. The grey
filled contours show the $z\sim 2$ number density of $3754$
$1.4<z<2.5$ galaxies from the NEWFIRM Medium-Band Survey \citep{whitaker11}.
The ${\rm SFR-M_{\ast}}$ relation at $z\simeq 2$ as determined by \citet{speagle14} is
indicated by the dashed line, with the $1\sigma$ and $3\sigma$ scatter of the
relation shown by the dot-dashed and dotted lines, respectively.  
Also shown are six $z\sim1.4-1.6$ BzK galaxies (black circles; \citealt{daddi10}) and 14
$z\sim2-2.5$ Bx/BM galaxies (black crosses; \citealt{tacconi13}). The BzK
galaxies are from top to bottom: BzK$-$12591, BzK$-$21000, BzK$-$16000,
BzK$-$17999, BzK$-$4171 and BzK$-$2553 (following the naming convention of
\citealt{daddi10}).
} 
\label{model_galaxies_f_M_SFR}
\end{figure*}

The location of our three model galaxies in the SFR-$M_*$ diagram is shown in
Figure \ref{model_galaxies_f_M_SFR} along with a sample of 3754 $1.4 < z < 2.5$
main-sequence galaxies selected in near-IR from the NEWFIRM Medium-Band Survey
\citep{whitaker11}.  The latter used a Kroupa IMF but given its similarity with
a Chabrier IMF no conversion in the stellar mass and SFR was made (cf.,
\citet{papovich11} and \citet{zahid12} who use conversion factors of 1.06 and
1.13, respectively).  Also shown is the determination of the main-sequence
relation at $z\simeq2$ by \cite{speagle14}, and the 1 and $3\sigma$ scatter
around it.  G1, G2, and G3 are seen to lie within the $3\sigma$ scatter around the $z\simeq 2$ main
sequence, albeit offset somewhat towards lower SFRs. This latter tendency is
also found among a subset of CO-detected BX/BM galaxies at $z\sim2-2.5$
\citep{tacconi13}, highlighted in Figure \ref{model_galaxies_f_M_SFR} along
with a handful of $z\sim1.5$ BzK galaxies also detected in CO \citep{daddi10}.
The BX/BM galaxies are selected by a UGR colour criteria \citep{adelberger04},
while the BzK galaxies are selected by the BzK colour criteria \citep{daddi04}.
Based on the above we conclude that, in terms of stellar mass and SFR, our
three model galaxies are representative of the star-forming galaxy population
detected in CO at $z\sim2$.

\section{Modeling the ISM with \sigame}\label{modeling}
\subsection{Methodology overview}\label{subsection:methodology}
Here we give an overview of the major steps that go into \sigame,
along with a brief description of each. The full details of each step are given
in subsequent sections and in appendices \ref{apB} through \ref{apD}. We stress,
that \sigame operates entirely in the post-processing stage of an SPH
simulation, and can in principle easily be adapted to any given SPH galaxy
simulation as long as certain basic quantities are known for each SPH particle
in the simulation, namely: position ($\boldsymbol{r}$), velocity ($\vec{v}$), atomic
hydrogen density (\nH), metallicity (\Z), kinetic temperature (\Tk), smoothing
length ($h$), and star formation rate (SFR). The key steps involved in
\sigame are:

\begin{enumerate}
\item Cooling of the SPH gas. The initially hot ($T_{\rm k}\sim 10^{3-7}\,{\rm
K}$) SPH gas particles are cooled to temperatures typical of the warm neutral
medium ($\lesssim 10^4\,{\rm K}$) by atomic and ionic cooling lines primarily. 

\item Inference of the molecular gas mass fraction ($\fh2 = m_{\rm H_2}/m_{\rm
SPH}$) of each SPH particle after the initial cooling in step 1.  \fh2 
for a given SPH particle is calculated by taking into account its temperature,
metallicity, and the local CR and FUV radiation field impinging on it.

\item Distribution of the molecular gas into GMCs. Cloud masses and sizes are
obtained from random sampling of the observed GMC mass-spectrum in nearby
quiescent galaxies and applying the local GMC mass-size relation.

\item GMC thermal structure. A radial density profile is adopted for each GMC
and used to calculate the temperature structure throughout individual clouds,
taking into account heating and cooling mechanisms relevant for neutral and
molecular gas, when exposed to the local CR and attenuated FUV fields. 

\item Radiative transport of CO lines. Finally, the CO line spectra are calculated 
separately for each GMC with a radiative transfer code and accumulated on a common 
velocity axis for the entire galaxy.
\end{enumerate}

Determining the temperature (i) and the molecular gas mass fraction
(ii) of a warm neutral gas SPH particle cannot be done independently of
each other, but must be solved for simultaneously in an iterative fashion (see
Sections \ref{WCNM} and \ref{HI_to_H2}).  As already mentioned, we shall apply
\sigame to the SPH simulations of galaxies G1, G2, and G3 described in Section
\ref{cosmological_simulations}, and in doing so we will use them to illustrate
the workings of the code.

\subsection{The Warm and Cold Neutral Medium}\label{WCNM}
\begin{figure} 
\includegraphics[width=1.1\columnwidth]{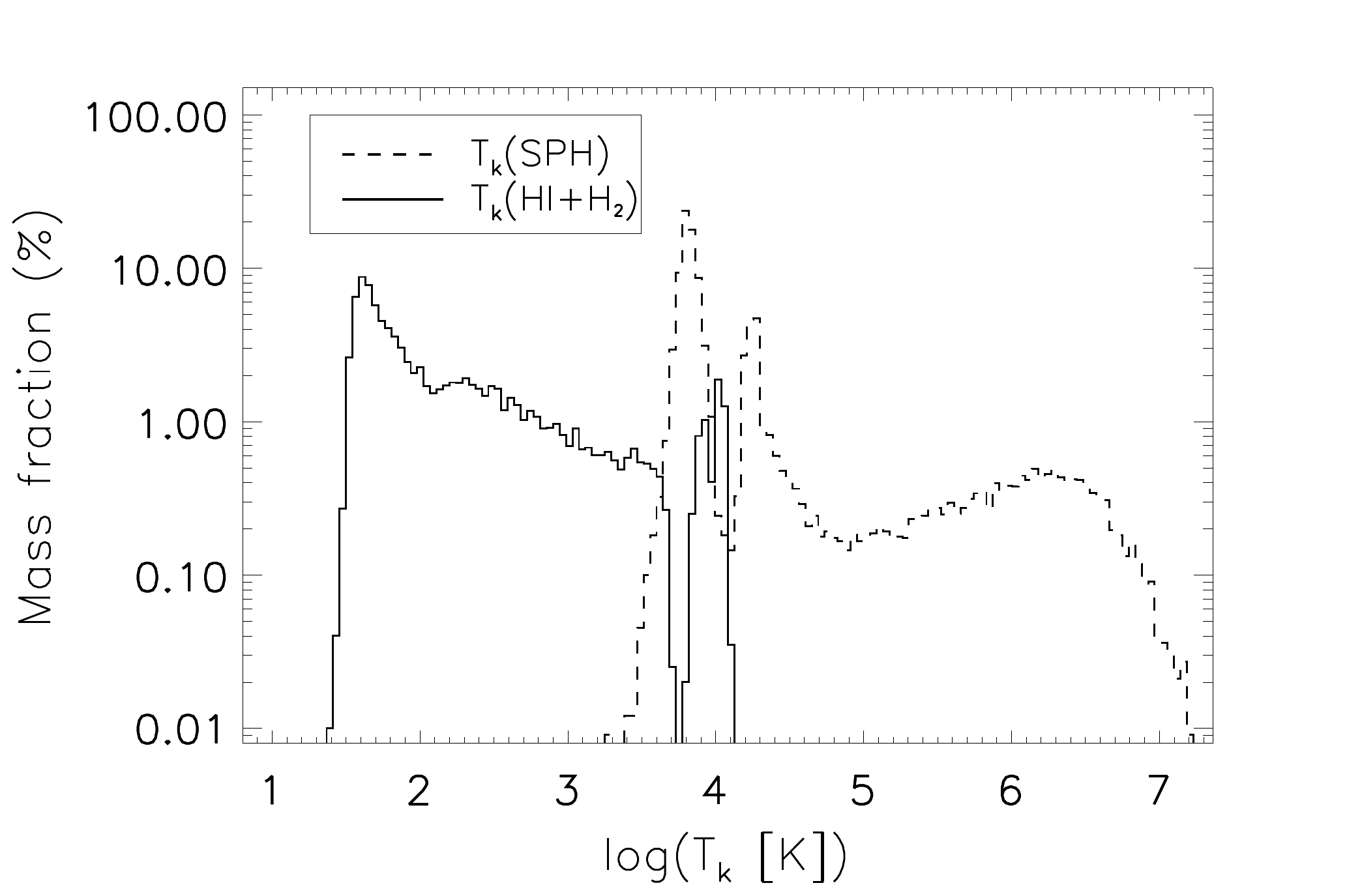}
\caption{The distributions of gas kinetic temperature before (dashed histogram)
and after (solid histogram) applying the heating and cooling mechanisms of
eq.\,\ref{WCNM_e1} to galaxy G1. The original hot SPH gas is seen to span a
temperature range from about $10^3$\,K up to $\sim10^7$\,K, while once the gas
has been cooled the temperature distribution only barely exceeds $\sim10^4$\,K.
}
\label{WCNM_f1}
\end{figure}

In SPH simulations of galaxies the gas is typically not cooled to temperatures
below a few thousand Kelvin \citep{springel03}. This is illustrated in Figure
\ref{WCNM_f1}, which shows the SPH gas temperature distribution (dashed
histogram) in G1 (for clarity we have not shown the corresponding temperature
distributions for G2 and G3, both of which are similar to that of G1). Minimum
SPH gas temperatures in G1, G2 and G3 are about 1200\,K, 3100\,K and 3200\,K,
respectively, and while temperatures span the range $\sim10^{3-7}\,{\rm K}$, the
bulk ($\sim 80-90\%$) of the gas mass in all three galaxies is at $\Tk \lesssim
10^5\,{\rm K}$.

At these temperatures the gas will be in atomic or ionised form, and H atoms that
attach to dust grain surfaces via chemical bonds (chemisorbed) will evaporate from
the grains before \h2 can be formed. \h2 can effectively only exist at
temperatures below $\sim 10^3$\,K, assuming a realistic desorption energy of
$3\times 10^4$\,K for chemisorbed H atoms \citep[see][]{cazaux04}.  The first
step of \sigame is therefore to cool some portion of the hot SPH gas
down to $\Tk \lesssim 10^3\,{\rm K}$, i.e., a temperature range characteristic of a warm
and cold neutral medium for which we can meaningfully employ a prescription for
the formation of H$_2$. 

\sigame employs the standard cooling and heating mechanisms pertaining to a
hot, partially ionised gas. Cooling occurs primarily via emission lines from H,
He, C, O, N, Ne, Mg, Si, S, Ca, and Fe in their atomic and ionised states, with
the relative importance of these radiative cooling lines depending on the
temperature \citep{wiersma09}. In addition to these emission lines, electron
recombination with ions can cool the gas, as recombining electrons take away
kinetic energy from the plasma, a process which is important at temperatures $ >
10^3$\,K \citep{wolfire03}. At similar high temperatures another important
cooling mechanism is the scattering of free electrons off other free ions,
whereby free-free emission removes energy from the gas \citep{draine11}.
Working against the cooling is heating caused by cosmic rays via the expulsion
of bound electrons from atoms or the direct kinetic transfer to free electrons
via Coulomb interactions \citep{Simnett69}.  
By ignoring heating of the gas by photo-ionization, we shall adopt an 
assumption typically used for galactic ISM \citep{gnat07,smith08}, 
though see \cite{wiersma09} who demonstrate how cooling rates are 
reduced when including photoionization in the intergalactic medium 
and proto-galaxies.

We arrive at a first estimate of the temperature of the neutral medium by
requiring energy rate equilibrium between the above mentioned heating and
cooling mechanisms:
\begin{equation}
\Hcrhi = \Cions + \Crec + \Cff,
\label{WCNM_e1} 
\end{equation}
where \Cions is the cooling rate due to atomic and ionic emission lines, \Crec
and \Cff are the cooling rates from recombination processes and free-free
emission as described above, and \Hcrhi is the CR heating rate in
atomic, partly ionised, gas.  The detailed analytical expressions employed by
\sigame for these heating and cooling rates are given in appendix \ref{apB}.

The abundances of the atoms and ions included in \Cions either have to be
calculated in a self-consistent manner as part of the SPH simulation or set by
hand. For our set of galaxies, the SPH simulations follow the abundances of H,
C, N, O, Mg, Si, S, Ca and Fe, while for the abundances of He and Ne, we adopt
solar mass fractions of $0.2806$ and $10^{-4}$, respectively, as used in
\cite{wiersma09}.

\Hcrhi depends on the {\it primary} CR ionization rate (\cri), a
quantity that is set by the number density of supernovae since they are thought
to be the main source of CRs \citep{ackermann13}. In \sigame, this is accounted
for by parameterizing \cri as a function of the {\it local} star formation rate
density (SFRD) as it varies across the simulated galaxy. The details of this
parameterization are deferred to Section \ref{Tk_GMC}.

In eq.\,\ref{WCNM_e1}, all terms but \Cions depend on the ionization fraction
($x_e = n_e/n_{\rm HI}$) of the gas, which in turn depends on the {\it
total} CR ionization rate (i.e., \cri corrected for secondary
ionizations of H and He), the gas temperature (\Tk), and the H\,{\sc i} density
(\nhi) (see appendix \ref{apB}).  The ionization fraction is calculated taking
into account the ionization of H and He (with a procedure kindly provided by
I.\ Pelupessy; see also \cite{pelupessy05}). Since, \nhi is set by the
molecular gas mass fraction (\fh2), which in turn also depends on \Tk (see
Section \ref{HI_to_H2} on how \fh2 is calculated), eq.\ \ref{WCNM_e1} has to be
solved in an iterative fashion until consistent values for \Tk, \ne, and \fh2
are reached. Example solutions are given in Figure \ref{apB1} in Appendix
\ref{apB}.  

\smallskip

The temperature distribution that results from solving eq.\ \ref{WCNM_e1} for
every SPH particle in the G1 simulation is shown in Figure \ref{WCNM_f1} (very
similar distributions are obtained for G2 and G3). The gas has been cooled to
$\Tk\lesssim 10^4$\, K, with temperatures extending down to $\sim 25\,{\rm K}$.
This new gas phase represents both the warm neutral medium (WNM) and the cold
neutral medium (CNM), and from it we derive the molecular gas phase.

\subsection{\hi to \h2 conversion}\label{HI_to_H2}
For the determination of the molecular gas mass fraction associated with each
SPH gas particle we use the prescription of \cite{pelupessy06}, inferred by
equating the formation rate of \h2 on dust grains with the photodissociation
rate of \h2 by Lyman-Werner band photons, and taking into account the
self-shielding capacity of \h2 and dust extinction. We ignore \h2 production in
the gas phase \citep[cf.,][]{christensen12} since only in diffuse low
metallicity ($\lesssim0.1Z_{\odot}$) gas is this thought to be the dominant
formation route \citep{norman97}, and so should not be relevant in our model
galaxies that have mean metallicities $\Z > 1$ (see
Table\,\ref{model_galaxies_t1}) and very little gas with $\Z<0.1$ (see
Figure \ref{apD1} in Appendix \ref{apD}). We adopt a steady-state for the
H\,{\sc i}$\rightarrow$H$_2$ transition, meaning that we ignore any time
dependence owing to temporal changes in the UV field strength and/or disruptions
of GMCs, both of which can occur on similar time scales as the \h2 formation.
This has been shown to be a reasonable assumption for environments with
metallicities $\gtrsim 0.01\,Z_{\odot}$ \citep{narayanan11,krumholz11}.

\smallskip

The first step is to derive the FUV field strength, \g0, which sets the H\,{\sc
i}$\rightarrow$H$_2$ equilibrium. In \sigame, \g0 consists of a
spatially varying component that scales with the local SFRD (${\rm
SFRD_{local}}$) in different parts of the galaxy on top of a constant component
set by the total stellar mass of the galaxy.  This is motivated by
\citet{seon11}  who measured the average FUV field strength in the MW
($G'_{\rm 0,MW}$) and found that about half comes from star light directly with
the remainder coming from diffuse background light.  We shall assume that in the
MW the direct stellar contribution to $G'_{\rm 0,MW}$ is determined by the
average SFRD ($\rm SFRD_{MW}$), while the diffuse component is fixed by the
stellar mass (${\rm M}_{*,\rm{MW}}$). From this assumption, i.e., by calibrating
to MW values, we derive the desired scaling relation for \g0 in our simulations: 
\begin{equation}
\g0 = G'_{\rm 0,MW} \left( 0.5\frac{{\rm SFRD}_{\rm{local}}}{{\rm SFRD}_{\rm{MW}}} + 0.5\frac {{\rm M}_*}{{\rm M}_{\rm *,MW}}\right), 
\label{equation:g0} 
\end{equation}
where $G'_{\rm 0,MW} = 0.6$\,Habing \citep{seon11}, and ${\rm
M}_{*,\rm{MW}}=6\e{10}$ \,\msun \citep{mcmillan11}.  For ${\rm SFRD}_{\rm{MW}}$
we adopt $0.0024$\,\sfru\,kpc$^{-3}$, inferred from the average SFR within the
central 10\,kpc of the MW \citep[$0.3$\,\sfru; ][]{heiderman10} and within a
column of height equal to the scale height of the young stellar disk
\citep[$0.2\,{\rm kpc}$; ][]{bovy12} of the MW disk. ${\rm SFRD_{local}}$ is the
SFRD ascribed to a given SPH particle, and is calculated as the volume-averaged
SFR of all SPH particles within a $5\,{\rm kpc}$ radius. Note, that the stellar mass
sets a lower limit on \g0, which for G1, G2, and G3 are $0.22$, $0.62$, and
$0.88$\,Habing, respectively. 

\smallskip

Next, the gas upon which the FUV field impinges is assumed to reside in
logotropic clouds, i.e., clouds with radial density profiles given by
$n(r)=\next\left( r/R \right)^{-1}$, where \next is the density at the cloud
radius $R$. 
For a logotropic
density profile, the external density is given by $n_{\rm{H,ext}} = 2/3
\ave{\nH}$, where we approximate $\ave{\nH}$ with the original SPH gas density.
From \cite{pelupessy06} we then have that the molecular gas mass
fraction, including heavier elements than hydrogen, of each cloud is given
by\footnote{We will use lower case $m$ when dealing with individual SPH
particles. Furthermore, \fh2 is not to be confused with $f_{\rm mol}$
describing the total molecular gas mass fraction of a galaxy and to be
introduced in Section \ref{results}.}:
\begin{equation}
\fh2\equiv \frac{m_{\rm mol}}{m_{\rm SPH}}=\exp{\left[ -4 \frac{\Avtr}{\ave{\Av}} \right]}.
\label{HI_to_H2_e1}
\end{equation}
Here \ave{\Av} is the area-averaged visual extinction of the cloud and \Avtr is
the extinction through the outer layer of neutral hydrogen.  The area-averaged
extinction, \ave{\Av}, is calculated from the metallicity and average cloud
density, $\ave{\nH}$: 
\begin{equation} 
	\ave{\Av} = 7.21\e{-22}\Z \langle n_{\rm H}\rangle R,
\end{equation}
where $\ave{\nH} R$ is set by the well known density-size scaling
relation for virialised clouds, normalised by the external boundary pressure, \Pe, i.e.:
\begin{equation}
	\ave{\nH} = n_{\rm ds}\left( \frac{\Pe/k_{\rm B}}{10^4\,\cmpc\,{\rm K}}\right)^{1/2} \left ( \frac{R}{\rm pc}\right )^{-1}.
	\label{nR_Pe}
\end{equation}
For the normalization constant we adopt $n_{\rm ds}=10^3$\,\cmpc, as inferred
from studies of molecular clouds in the MW \citep{larson81,wolfire03,heyer04},
although we note that \cite{pelupessy06} uses 1520\,\cmps.  The external
hydrostatic pressure for a rotating disk of gas and stars is calculated at
mid-plane following \cite{swinbank11}:
\begin{equation}
	P_{\rm tot}\approx\frac{\pi}{2}\rm{G}\gassd \left[ \gassd+\left( \frac{\sigma_{\rm gas\perp}}{\sigma_{\rm *\perp}} \right) \starssd\right], 
	\label{eq:Pe}
\end{equation}
where $\sigma_{\rm gas\perp}$ and $\sigma_{\rm *\perp}$ are the local vertical
velocity dispersions of gas and stars respectively, and $\Sigma$ denotes surface
densities of the same.  These quantities are all calculated directly from the
simulation output, using the neighbouring SPH particles within 5\,kpc, 
weighted by mass, density
and the cubic spline kernel \citep[see also][]{monaghan05}. 
The external cloud pressure that enters in eq.\,\ref{nR_Pe}, is assumed to be
equal to $P_{\rm tot}/(1+\alpha_0+\beta_0)$ for relative cosmic and magnetic
pressure contributions of $\alpha_0=0.4$ and $\beta_0=0.25$
\citep{elmegreen89,swinbank11}.  For the MW, $\Pe/k_{\rm B}\sim10^4\,{\rm \cmpc\,K}$
\citep{elmegreen89}, but as shown in Figure\,\ref{apD1} in Appendix \ref{apD}, our
model galaxies span a wide range in $\Pe/k_{\rm B}$ of
$\sim10^2-10^7\,{\rm \cmpc\,K}$.  

For \Avtr, the following expression is provided by \cite{pelupessy06}:
\begin{equation} 
\Avtr = 1.086 \nu \xi_{\rm{FUV}}^{-1} \times \ln \left[ 1+\frac{\frac{G_0'}{\nu\mu S_\text{H} (T_{\rm k})} \sqrt{ \frac{\xi_{\text{FUV}}}{Z' T_{\rm k}} }}{\next/135~\cmpc } \right], 
\label{HI_H2_e2}
\end{equation}
where $\xi_{\text{FUV}}$ is the ratio between dust FUV absorption cross section
($\sigma$) and the effective grain surface area ($\sigma_d$), and is set to
$\xi_{\text{FUV}} = \sigma/\sigma_{\rm d} = 3$. $S_\text{H} = (1+0.01 T_{\rm
k})^{-2}$ is the probability that \hi atoms stick to grain surfaces, where
$T_k$ is the kinetic gas temperature (determined in an iterative way as
explained in Section \ref{WCNM}). Furthermore, $\nu=\next R \sigma (1+\next R
\sigma)^{-1}$, and $\mu$ \citep[set to $3.5$ as suggested by][]{pelupessy06} is a
parameter which incorporates the uncertainties associated with primarily
$S_\text{H}$ and $\sigma_{\rm d}$.

\smallskip

\begin{figure}
\includegraphics[width=1.1\columnwidth]{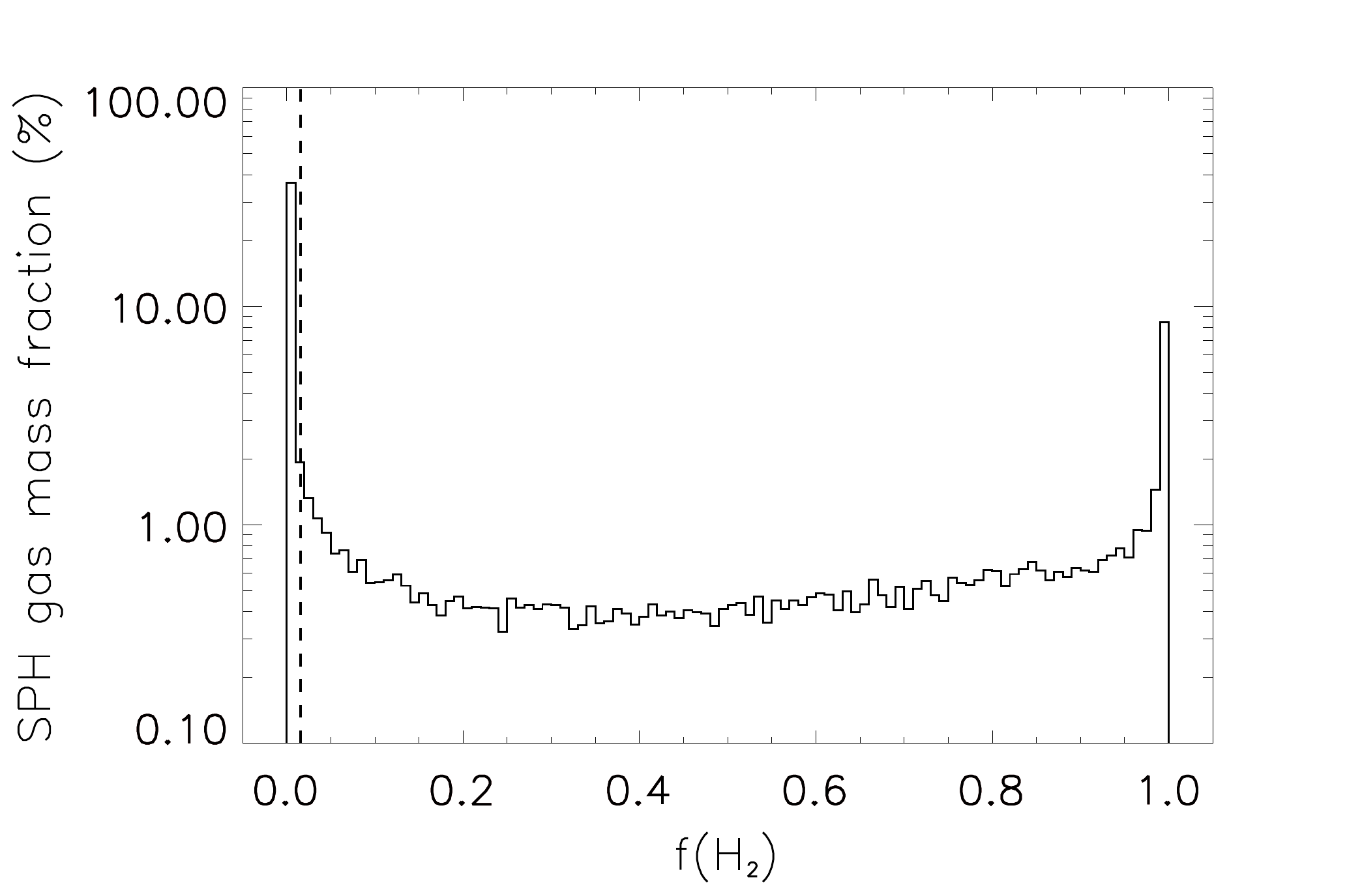}
\caption{The distribution of the \h2 gas mass fraction of the SPH particles in
G1 calculated according to eq.\ \ref{HI_to_H2_e1} (solid line histogram).
Similar distributions are found for G2 and G3. The lower limit on \fh2, defined
as described in Section \ref{split}, is indicated by the dashed vertical line.} 
\label{figure:f_H2}
\end{figure}

Following the above prescription \sigame determines the molecular gas mass
fraction and thus the molecular gas mass ($m_{\rm mol}=\fh2 m_{\rm SPH}$)
associated with each SPH particle. Depending on the environment (e.g., \g0,
\Z, \Tk, \Pe,...), the fraction of the SPH gas converted into \h2 can take
on any value between 0 and 1, as seen from the \fh2 distribution of G1 in Figure
\ref{figure:f_H2}.  Overall, the total mass fraction of the SPH gas in G1, G2,
and G3 (i.e., within $R_{\rm cut}$) that is converted to molecular gas is
$29$\,\%, $52$\,\% and $34$\,\%, respectively. 

\subsection{Structure of the molecular gas}\label{structure}
Having determined the molecular gas mass fractions, \sigame proceeds by
distributing the molecular gas into GMCs, and calculates their masses and sizes,
along with internal density and temperature structures, as
described in the following. 

\subsubsection{GMC masses and sizes}\label{split}
\begin{figure}
\hspace{-0.3cm}
\includegraphics[width=1.1\columnwidth]{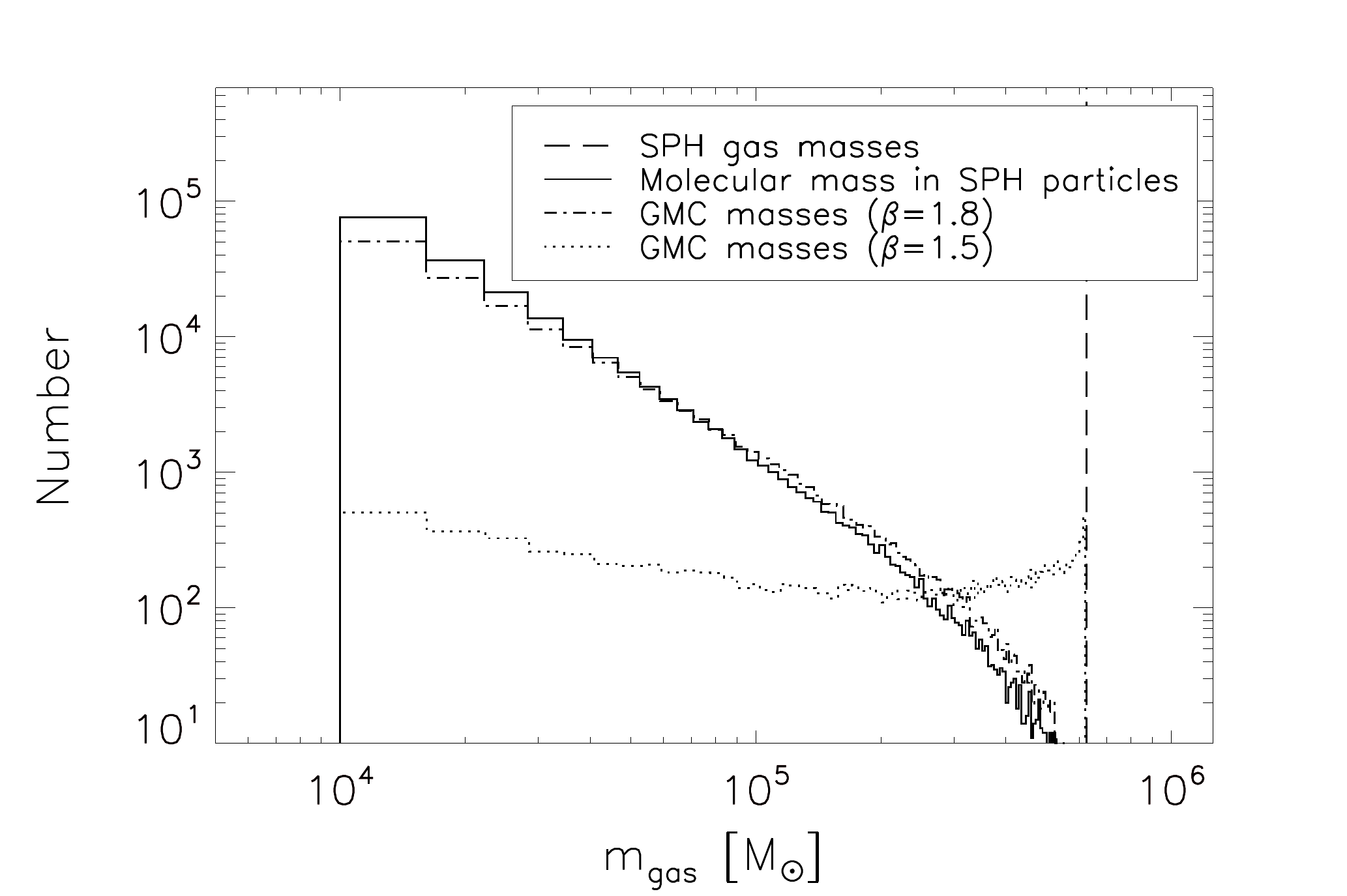}
\caption{The distribution of GMC masses in G1 obtained by applying eq.\
\ref{equation:mass-spectrum}, with $\beta = 1.8$ (solid histogram) and $\beta =
1.5$ (dash-dotted histogram), compared to the total molecular gas masses associated with
SPH particles (distribution shown as dotted histogram). The dashed vertical line
indicates the SPH gas mass resolution of the simulation ($=6.3\times
10^5\,h^{-1}\,{\rm \msun}$ in the case of G1).}
\label{structure_f1}
\end{figure}

The molecular gas associated with a given SPH particle is divided into 
GMCs by randomly sampling a power-law mass spectrum of the form:
\begin{equation}
	\frac{dN}{dm_{\rm GMC}}\propto \Mgmc^{-\beta}.
	\label{equation:mass-spectrum}
\end{equation}
For GMCs in the MW disk and Local Group galaxies $\beta\simeq1.8$ \citep{blitz07}, and
is the value adopted by \sigame in this paper unless otherwise stated. Lower and upper mass
cut-offs at $10^4\,\msun$ and $10^6\,\msun$, respectively, are enforced in order
to span the mass range observed by \cite{blitz07}. A similar approach was
adopted by \citet{narayanan08,narayanan08a}. Note that, in the case of G1 the upper
cut-off on the GMC masses is in fact set by the mass resolution of the SPH
simulation ($6.3\times 10^5\,h^{-1}\,{\rm \msun}$). For G1, typically $\lesssim
30$ GMCs are created in this way per SPH particle, while for G2 and G3, which
were run with SPH gas particle masses almost an order of magnitude higher, as
much as $\sim 100$ GMCs could be extracted from a given SPH particle. 
Figure \ref{structure_f1} shows the resulting mass distribution of
all the GMCs in G1, along with the distribution of molecular gas mass associated
with the SPH gas particles prior to it being divided into GMCs. The net effect
of re-distributing the \h2 mass into GMCs is a mass distribution dominated by
relatively low cloud masses, which is in contrast to the relatively flat SPH \h2
mass distribution. Note, the lower cut-off at $\Mgmc=10^4\,\msun$ implies that
if the molecular gas mass associated with an SPH particle (i.e., $m_{\rm
mol}=\fh2 m_{\rm SPH}$) is less than this lower limit it will not be
re-distributed into GMCs. Since $m_{\rm SPH}$ is constant in our simulations
($6.3\times 10^5\,h^{-1}\,\msun$ for G1 and $4.7\times 10^6\,h^{-1}\,\msun$ for
G2 and G3) the lower limit imposed on $m_{\rm GMC}$ translates directly into a
lower limit on \fh2 ($0.016$ for G1 and 0.002 for G2 and G3, shown as a dashed
vertical line for G1 in Figure \ref{figure:f_H2}).  As a consequence, 0.2, 0.005
and 0.01\,\% of the molecular gas in G1, G2, and G3, respectively, does not end
up in GMCs. These are negligible fractions and the molecular gas they represent
can therefore be safely ignored.

\begin{figure}
\hspace{-0.3cm}
\includegraphics[width=1.1\columnwidth]{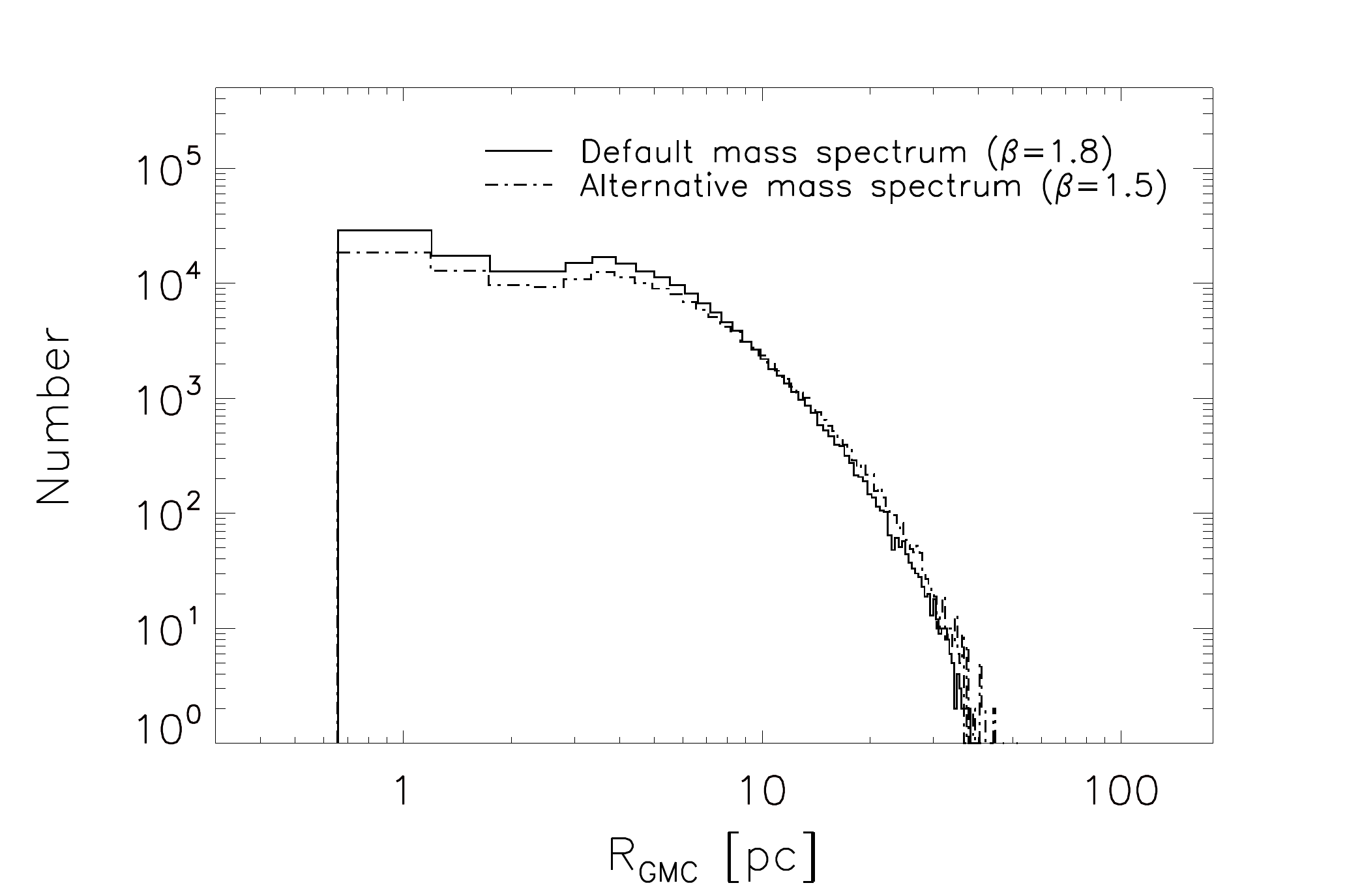}
\caption{The distribution of GMC radii in G1 for the adopted cloud mass
spectrum with $\beta = 1.8$ (solid histogram) and for a slightly modified
spectrum with $\beta = 1.5$ (dash-dotted histogram).  Similar distributions are
found for G2 and G3.} 
\label{structure_f2}
\end{figure}

The GMC sizes are derived from the virial theorem, which relates the radius of a GMC
(\Re) to its mass and external pressure (\Pe) according to:
\begin{equation} 
	\frac{\Re}{\text{pc}} = \left( \frac{\Pe/k_{\rm B}}{10^4\,\cmpc\,{\rm K}}\right)^{-1/4} \left (\frac{\Mgmc}{290\,\msun}\right )^{1/2}.
	\label{structure_e1}
\end{equation}
Similarly, the internal velocity dispersion ($\sigma_v$) of the clouds is given by:  
\begin{equation} 
	\frac{\sigma_v}{{\rm km\,s^{-1}}} = 1.2 \left( \frac{\Pe/k_{\rm B}}{10^4\,\cmpc\,{\rm K}}\right)^{1/4}\left ( \frac{\Re}{\rm{pc}} \right )^{1/2}
	\label{structure_e2}
\end{equation}
\citep[e.g.,][]{elmegreen89,swinbank11}.
Figure \ref{structure_f2} shows the resulting distribution of GMC radii in G1
(solid histogram). The minimum and maximal cloud radii found in G1 are $\sim 0.07\,{\rm
pc}$ and $\sim 100\,{\rm pc}$, respectively, and are set by the pressure and the 
imposed limits on $m_{\rm GMC}$.

Observations have indicated that the shape of the GMC mass spectrum might be
different in gas-rich systems where a high-pressure ISM leads to the
characteristic mass of star-forming clumps of molecular gas being much higher
than what is observed in normal spirals \citep[e.g.][]{swinbank11,leroy15}.  In
Section \ref{dif_ISM} we therefore examine the effects of adopting a more
top-heavy GMC mass distribution, corresponding to $\beta=1.5$ (shown as
dot-dashed histogram in Figure \ref{structure_f1}). 
The total amount of molecular gas in our galaxies does not change significantly
between $\beta = 1.8$ and $\beta = 1.5$, and the CO simulation results are
robust against (reasonable) changes in the mass-spectrum.

The GMCs are placed randomly around the position of their `parent' SPH
particle, albeit with an inverse proportionality between radial displacement
and mass of the GMC. The latter is done in order to retain the mass
distribution of the original galaxy simulation as best as possible. The GMCs
are assigned the same bulk velocity, $\bar{v}$, \Z, \g0 and \cri as their
`parent' SPH particle.

\subsubsection{GMC density structure}\label{plummer}
In order to ensure a finite central density in our GMCs, \sigame adopts a
Plummer radial density profile \citep[e.g.,][]{plummer11,whitworth01}: 
\begin{equation} 
	\frac{n_{\text{H}_2} (R)}{\cmpc} = 3.55 \left (\frac{m_{\rm GMC}}{\msun}\right ) \left (\frac{R_p}{{\rm pc}}\right )^{-3}
        \left( 1+\frac{R^2}{R_p^2} \right)^{-5/2}, 
	\label{equation:plummer-profile}
\end{equation}
where $R_p$ is the so-called Plummer radius, which we set to $R_p=0.1 \Re$.
The latter allows for a broad range in gas densities throughout the clouds,
from $\sim 10^{5}$\,\cmpc in the central parts to a few $10\,\cmpc$ further
out. Note, eq.\ \ref{equation:plummer-profile} accounts for the presence of
helium in the GMC.  

\begin{figure}
\hspace{-0.6cm}
\includegraphics[width=1.1\columnwidth]{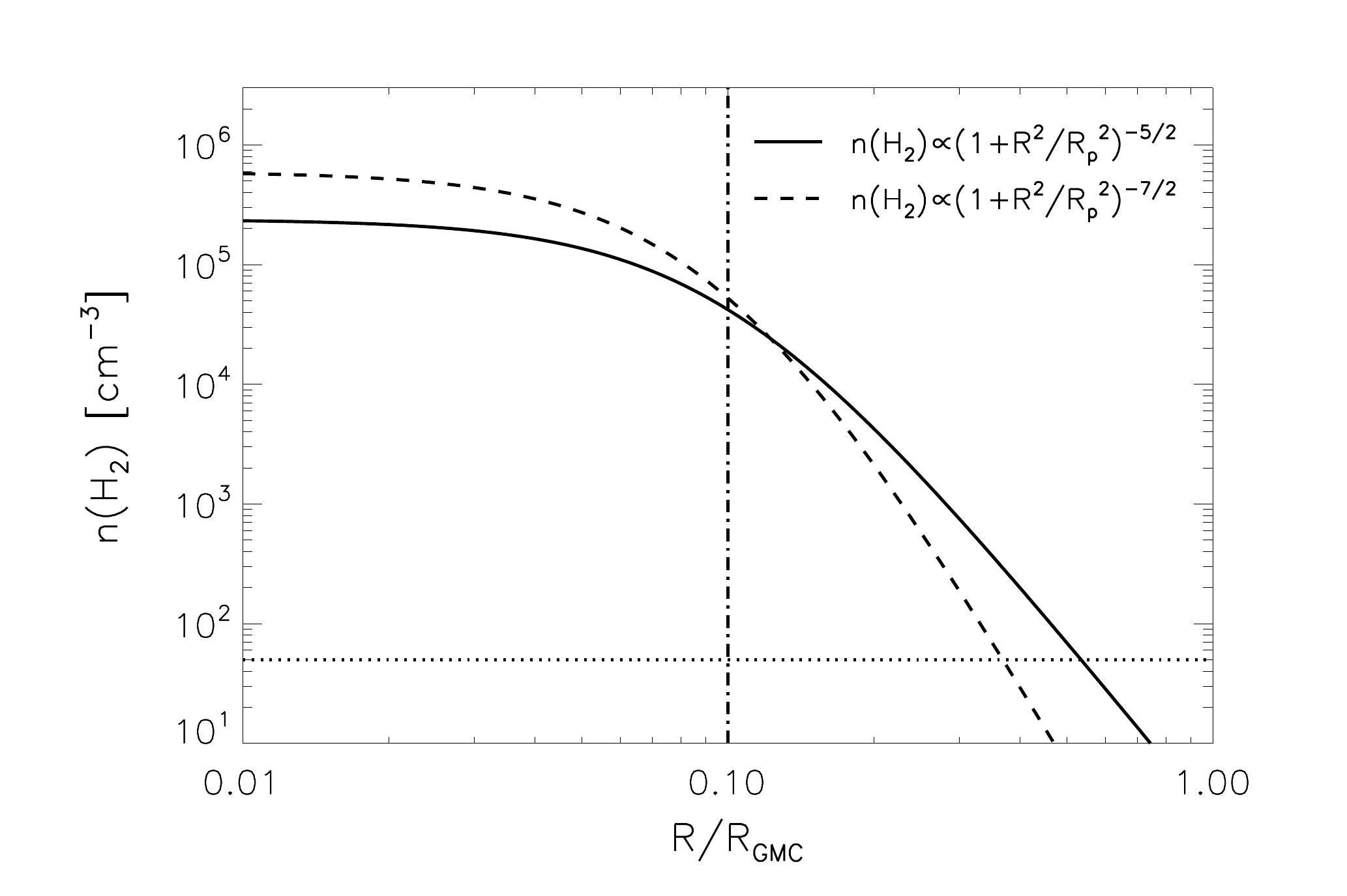}
\caption{Plummer density profiles (eq.\ \ref{equation:plummer-profile}) with
exponents $-5/2$ (solid line) and $-7/2$ (dashed line) for a GMC mass of
$10^4$\,\msun and an external pressure of $\Pe/k_{\rm B}=10^4\,$K\,cm$^{-3}$.
The corresponding cloud radius is $\rcl=5.9$\,pc (eq.\ \ref{structure_e1}).
The vertical dash-dotted line marks the Plummer radius ($=0.1R_{\rm GMC}$), and
the horizontal dotted line the \nh2 value ($50\,{\rm cm^{-3}}$) below
which CO is assumed to photo-dissociate in our simulations.}
\label{fig:plummer}
\end{figure}

In Section \ref{dif_ISM} we examine the impact on our simulation results
if a steeper GMC density profile with an exponent of $-7/2$ is adopted instead
of the default $-5/2$ in eq.\ \ref{equation:plummer-profile}.
This modified Plummer profile, as well as the default profile, are shown in
Figure \ref{fig:plummer} for a GMC with mass of $10^4$\,\msun and external
pressure $10^4\,{\rm K\,cm^{-3}}$.

\subsubsection{GMC thermal structure}\label{Tk_GMC}
Having established the masses, sizes, and density structure of the GMCs,
\sigame solves for the kinetic temperature throughout the clouds by balancing
the relevant heating and cooling mechanisms as a function of cloud radius. 

GMCs are predominantly heated by FUV photons (via the photo-electric effect)
and cosmic rays. For the strength of the FUV field impinging on the GMCs, we
take the previously calculated values at the SPH gas particle position (eq.\
\ref{equation:g0}).  We shall assume that the CR ionization rate scales in a
similar manner, since CRs have their origin in supernovae and are therefore
related to the star formation and the total stellar mass of the galaxy: 
\begin{align}
\cri &= \zeta_{\rm CR,MW} \left( 0.5\frac{{\rm SFRD}_{\rm{local}}}{{\rm SFRD}_{\rm{MW}}} + 0.5\frac {{\rm M}_*}{{\rm M}_{\rm *,MW}}\right) 
\label{equation:CR}
\end{align}
where ${\rm SFRD}_{\rm{local}}$, ${\rm SFRD}_{\rm{MW}}$ and ${\rm M}_{\rm
*,MW}$ are as described in Section \ref{HI_to_H2}, and \cri is scaled to the
`canonical' MW value of \cri$_{\rm ,MW}=3\e{-17}$\,\ps
\citep[e.g.][]{webber98}.  While the FUV radiation is attenuated by dust and
therefore does not heat the GMC centres significantly, cosmic rays can
penetrate these dense regions and heat the gas there \citep{papa11}.  For this
reason \sigame attenuates the FUV field throughout the GMCs, while the CR
ionization rate remains constant for a given cloud. The extinction of the FUV
field at a certain point within a GMC is derived by integrating the Plummer
density profile from that point and out to \Re. This H column density is
converted into a visual extinction ($A_{\rm V} = \colH/2.2\e{21}\,\cmps$) so
that the attenuated FUV field becomes:
\begin{align}
\ga=\g0 e^{-1.8A_{\rm V}},
\label{ga}
\end{align}
where the factor of $1.8$ is the adopted conversion from visual to FUV
extinction \citep{black77}. 

\sigame calculates the gas kinetic temperature throughout each GMC via the
following heating and cooling rate balance:
\begin{equation}
\Hpe+\Hcrh2 = \Ch2+\Cco+\Coi+\Ccii+\Cgd.
\label{Tk_GMC_e1}
\end{equation}
\Hpe is the photo-electric heating by FUV photons, and \Hcrh2 is the cosmic ray
heating in molecular gas.  \Ch2 is the cooling rate of the two lowest \h2
rotational lines (S(0) and S(1)), and \Cco is the cooling rate of the combined
CO rotational ladder. \Cgd is the cooling rate due to interactions between gas
molecules and dust particles, which only becomes important at densities above
$10^4$\,\cmpc \citep[e.g.,][]{goldsmith01,glover12}.  \Ccii and \Coi are the
cooling rates due to \cii and \oi line emission, respectively.  The abundances
of carbon and oxygen used in the prescriptions for their cooling rates, scale
with the cloud metallicity, while the CO cooling scales with the relative CO to
neutral carbon abundance ratio, set by the molecular density (see Appendix
\ref{apC}).

\Cgd depends on the temperature difference between the gas and the dust (see
eq.\ \ref{equation:dust-gas-cooling}). The dust temperature, \Td, is set by the
equilibrium between the absorption of FUV and the emission of IR radiation by
the dust grains. We adopt the approximation given by \citet{tielens05}:
\begin{equation}
\frac{\Td}{{\rm K}} \simeq 33.5 \left( \frac{a}{1\mu {\rm m}}\right)^{-0.2} \left( \frac{\ga}{10^4\,{\rm Habing}} \right)^{0.2},
\label{Tk_GMC_e2}
\end{equation}
where \ga is the dust-attenuated FUV field (eq.\ \ref{ga}) and $a$ is the grain
size, which we set to 1\,$\mu$m for simplicity.  Values for \Td using eq.\
\ref{Tk_GMC_e2} range from 0 to 8.9\,K, but we enforce a lower limit on \Td
equal to the $z=2$ CMB temperature of $8.175$\,K. \Td is therefore essentially
constant ($\sim8-9$\,K) throughout the inner region of the GMC models, similar
to the value of $\Td=8$\,K adopted by \cite{papa13} for CR-dominated cores.
Analytical expressions for all of the above heating and cooling rates are given
in Appendix \ref{apC}, which also shows their relative strengths as a function
of density for two example GMCs (Figure \ref{apC1}).

\bigskip

Figure \ref{apD3} in Appendix \ref{apD} shows the resulting \Tk versus \nh2
behaviour for 80 GMCs spanning a broad range of GMC masses, metallicities, and
star formation rate densities.  As seen in the displayed GMC models, some
general trends can be inferred from the $\Tk-\nh2$ diagrams. At fixed
metallicity and mass, an increase in \g0 (and therefore also in \cri), leads to
higher temperatures throughout the models.  In the outer regions this is due
primarily to the increased photoelectric heating, while in the inner regions,
heating by the unattenuated cosmic rays takes over as the dominating heating
mechanism (Figure\,\ref{apC1}). Keeping \g0 (and \cri) fixed, lower \Tk-levels
and shallower $\Tk-\nh2$ gradients are found in GMCs with higher metallicities.
Both these trends are explained by the fact that the \cii and \oi cooling rates
scale linearly with \Z (see Appendix\,\ref{apC}). 

Moving from the outskirts and inward towards the GMC centres, \Tk drops as the
attenuation of \g0 reduces the photoelectric heating.  However, the transition
from cooling via \cii to the less efficient cooling mechanism by CO lines,
causes a local increase in \Tk at $\nh2\sim10^3-10^{4.5}$\,\cmpc, as also seen
in the detailed GMC simulations by \cite{glover12}. The exact density at which
this `bump' in \Tk occurs depends strongly on the mass of the GMC.  Our choice
of the Plummer model for the radial density profile means that the extinction,
$A_{\rm V}$, at a certain density increases with GMC mass. This in turn
decreases the FUV heating, and as a result the $\Tk-\nh2$ curve moves to lower
densities with increasing GMC mass.

As the density increases towards the cloud centres (i.e., $n_{\rm H_2} \gtrsim
10^{4-5}$\,\cmpc) molecular line cooling and also gas-dust interactions become
increasingly efficient and start to dominate the cooling budget.  The $\Tk-\nh2$
curves are seen to be insensitive to changes in \Z, which is expected since the
dominant heating and cooling mechanisms in these regions do not depend on \Z.
Eventually, in the very central regions of the clouds, the gas reaches
temperatures close to that of the ambient CMB radiation field, irrespective of
the overall GMC properties and the conditions at the surface (Figure
\ref{apC1}). 

\subsubsection{GMC grid models} \label{gmcgrid}
The $\Tk-\nh2$ curve for a given GMC is determined by the following quantities: 
\begin{itemize}
\item \g0 and \cri, which govern the gas heating and are set by the local star
formation rate density (SFRD$_{\rm local}$) and the total stellar mass ($M_*$)
according to eqs.  \ref{equation:g0} and \ref{equation:CR}.
\item \Mgmc and \Pe which determine the effective radius of a cloud (eq.\
\ref{structure_e1}) and thus its density profile (eq.\
\ref{equation:plummer-profile}). 
\item The local metallicity (\Z), which influences the fraction of \h2 gas and
plays an important role in cooling the gas. 
\end{itemize}

These local parameters together with the most important global parameters used by \sigame 
are listed in Table \ref{sigame_par}. 
The GMC ensemble distributions of \g0, \cri, \Mgmc, $\Pe/k_{\rm B}$, and \Z for
each of the galaxies G1, G2 and G3 are shown in Figure \ref{apD1}. 

There are more than $100,000$ GMCs in a single model galaxy and, as Figure
\ref{apD1} shows, they span a wide range in \g0, \cri, \Mgmc, $\Pe/k_{\rm B}$,
and \Z.  Thus, in order to shorten the computing time, we calculated $\Tk-\nh2$
curves for a set of 630 GMCs, chosen to appropriately sample the distributions
at certain grid values (listed in Table \ref{grid_par}, and marked by vertical
black lines in Figure \ref{apD1}). 
Every GMC in our simulations was
subsequently assigned the $\Tk-\nh2$ curve of the GMC grid model closest to it
in the (\g0, $m_{\rm GMC}$, \Z) parameter space.
In the default GMC grid, we keep $\Pe/k_{\rm B}$ fixed 
to a MW-like value of $10^4\,\rm{K}\,\cmpc$, thereby also anchoring the scaling relations 
in eq. \ref{structure_e1} and \ref{structure_e2} for size and velocity dispersion. 
This is done to minimise the number of radiative transfer calculations, 
effectively keeping the amount of molecular gas mass dependent on local pressure, 
but removing local pressure as a free parameter in, and hence simplifying, 
the calculation of CO emission.

In addition to the default grid described above, 
we made two separate tests to explore the GMC 
parameter space more fully. 
These tests are described in the following and results of their 
application are presented in Section \ref{dif_ISM}. 
The external cloud pressure, $\Pe/k_{\rm B}$, spans a wide range from $\sim10^3$ to 
$10^9\,\rm{K}\,\cmpc$ as shown in Figure \ref{apD1}. 
In a separate test, we therefore constructed the same GMC grid, but for
fixed pressures of $\Pe/k_{\rm B}=10^{5.5}\,$ and $10^{6.5}\,$K\,\cmpc 
and interpolated among the resulting three values of $\Pe/k_{\rm B}$ 
($[10^{4},10^{5.5},10^{6.5}]\,$K\,\cmpc) as 
a demonstration of how to incorporate local pressure in \sigame. 
We also test the effect of adopting the alternative radial GMC 
density profile defined in Section \ref{plummer}.
$\Tk-\nh2$ curves were calculated for each possible combination of the
parameter grid values listed in Table\,\ref{grid_par} for both density profiles, 
giving us a total of
$3\times2\times630=3780$ GMC grid models.

\begin{center}
\begin{table*}
\centering
\caption{Parameters, global and local, used by \sigame, together with relevant equations in this paper.}
\begin{tabular}{l |   l l} 
\hline
\hline
Global parameters 				&	[CO/\h2], $\beta$ (GMC mass spectrum), GMC density profile & Eqs.\,\ref{equation:mass-spectrum} and \ref{equation:plummer-profile} \\
Local parameters 				&	\Mgmc, \g0, \cri, \Z, \Pe, SFRD	&	Eqs.\,\ref{equation:mass-spectrum}, \ref{equation:g0}, \ref{equation:g0} and \ref{eq:Pe} \\
Derived internal GMC parameters	&	\nH, \Tk, \xe, \sv				&	Eqs.\,\ref{equation:plummer-profile} and \ref{Tk_GMC_e1}\\
\hline
\end{tabular}
\label{sigame_par}
\end{table*}
\end{center}

\begin{center}
\begin{table}
\centering
\caption{Grid parameter values} 
\begin{tabular}{l |   l} 
\hline
\hline
\g0	[Habing]  			&	0.5, 1, 4, 7, 10, 13, 16, 19, 23, 27	 \\ 
$\log(\Mgmc\,[\msun])$		&	4.0, 4.25, 4.5, 4.75, 5.0, 5.25, 5.5, 5.75, 6.0	\\ 
$\log(Z/Z_{\odot})$		& 	-1, -0.5, 0, 0.5, 1, 1.4, 1.8	\\ 
$\log(\Pe/k_{\rm B}\,{\rm [K\,\cmpc]})$ 	&	4.0, 5.5, 6.5	 \\ 
\hline
\end{tabular}
\label{grid_par}
\end{table}
\end{center}

\subsection{Radiative transfer of CO lines}\label{radiative_transfer}

\sigame assumes a fixed CO abundance equal to the Galactic value of
[CO/\h2]$=2\times 10^{-4}$ \citep[][and see Section \ref{dis_models} for a
justification of this value]{lee96,sofia04} everywhere in the GMCs except for
$\nh2<50$\,\cmpc, where CO is not expected to survive photo-dissociation
processes \citep[e.g.][]{narayanan08a}. \sigame calculates the CO line
radiative transfer for each GMC individually and derives the CO line
emission from the entire galaxy.

\subsubsection{Individual GMCs}
For the CO radiative transfer calculations we use a slightly modified version
of the LIne Modeling Engine \cite[\lime ver.\ 1.4;][]{brinch10} - a 3D
molecular excitation and radiative transfer code.  \lime has been
modified in order to take into account the redshift dependence of the CMB
temperature, which is used as boundary condition for the radiation field during
photon transport, and we have also introduced a redshift correction in the
calculation of physical sizes as a function of distance. We use collision rates
for CO (assuming \h2 is the main collision partner) from \cite{yang2010}.  In
\lime, photons are propagated along grid lines defined by Delaunay
triangulation around a set of appropriately chosen sample points in the gas,
each of which contain information on \nh2, $T_k$, $\sigma_v$, $[{\rm CO}/\h2]$
and $\boldsymbol{v}$. \sigame constructs such a set of sample points throughout each
GMC: about 5000 points distributed randomly out to a radius of $50$\,pc, i.e.,
beyond the effective radius of typical GMCs in G1, G2, and G3 (Figure
\ref{structure_f2}) and in a density regime below the threshold density of
$50$\,\cmpc adopted for CO survival (see previous paragraph). The concentration
of sample points is set to increase towards the centre of each GMC where the
density and temperature vary more drastically.  

For each GMC, \lime generates a CO line data cube, i.e., a series of CO
intensity maps as a function of velocity. The velocity-axis consists of 50
channels, each with a spectral resolution of $1.0$\,\kms, thus covering the
velocity range $v=[-25,+25]\,{\rm km\,s^{-1}}$. The maps are $100\,{\rm pc}$ on
a side and split into 200 pixels, corresponding to a linear resolution of
$0.5\text{\,pc/pixel}$ (or an angular resolution of
$5.9\times10^{-5}\,\arcsec\text{/pixel}$ at $z=2$).  Intensities are corrected
for arrival time delay and redshifting of photons.

Figure \ref{apD4} shows the area- and velocity-integrated CO Spectral Line
Energy Distributions (SLEDs) for the same 80 GMCs used in Section \ref{Tk_GMC}
to highlight the $\Tk-\nh2$ profiles (Figure \ref{apD3}). The first thing to
note is that the CO line fluxes increase with \Mgmc, which is due to the
increase in size, i.e., surface area of the emitting gas, with cloud mass
(Section \ref{split}). Turning to the shape of the CO SLEDs, a stronger \g0
(and \cri) increases the gas temperature and thus drives the SLEDs to peak at
higher $J$-transitions.  Only the higher, $J_{\rm up}>4$, transitions are also
affected by metallicity, displaying increased flux with increased \Z.  
For GMCs with high \g0, the high metallicity
levels thus cause the CO SLED to peak at $J_{\rm up}>8$.

\subsubsection{The effects of dust}\label{dust_properties}
Dust absorbs the UV light from young O and B stars and re-emits in the far-IR,
leading to possible `IR pumping' of molecular infrared sources.  However, due
to the large vibrational level spacing of CO, the molecular gas has to be at a
temperature of at least $159$\,K, for significant IR pumping of the CO
rotational lines to take place, when assuming a maximum filling factor of 1, as
shown by \cite{carroll81}.  Most of the gas in our GMC models is at
temperatures below $100$\,K, with only a small fraction of the gas, in the very
outskirts, of the GMCs reaching $\Tk>159$\,K, as seen in Figure \ref{apD3} in
Appendix \ref{apD}. This happens only if the metallicity is low ($\Z\leq0.1$)
or in case of a combination between high FUV field ($\g0\geq4$) and moderate
metallicity. 

In principle, the high-$J$ CO lines could be subject to extinction by dust, but
this effect is significant only in extremely dust-enshrouded sources
\citep{papa10a}, and therefore unlikely to be relevant for our
simulations. 

While \lime is
capable of including dust in the radiative transfer calculations, provided that a table of dust
opacities as function of wavelength be supplied together with the input model, 
we have chosen not to include dust in the simulations presented here.

\begin{figure*}
\centering
\includegraphics[width=2.1\columnwidth]{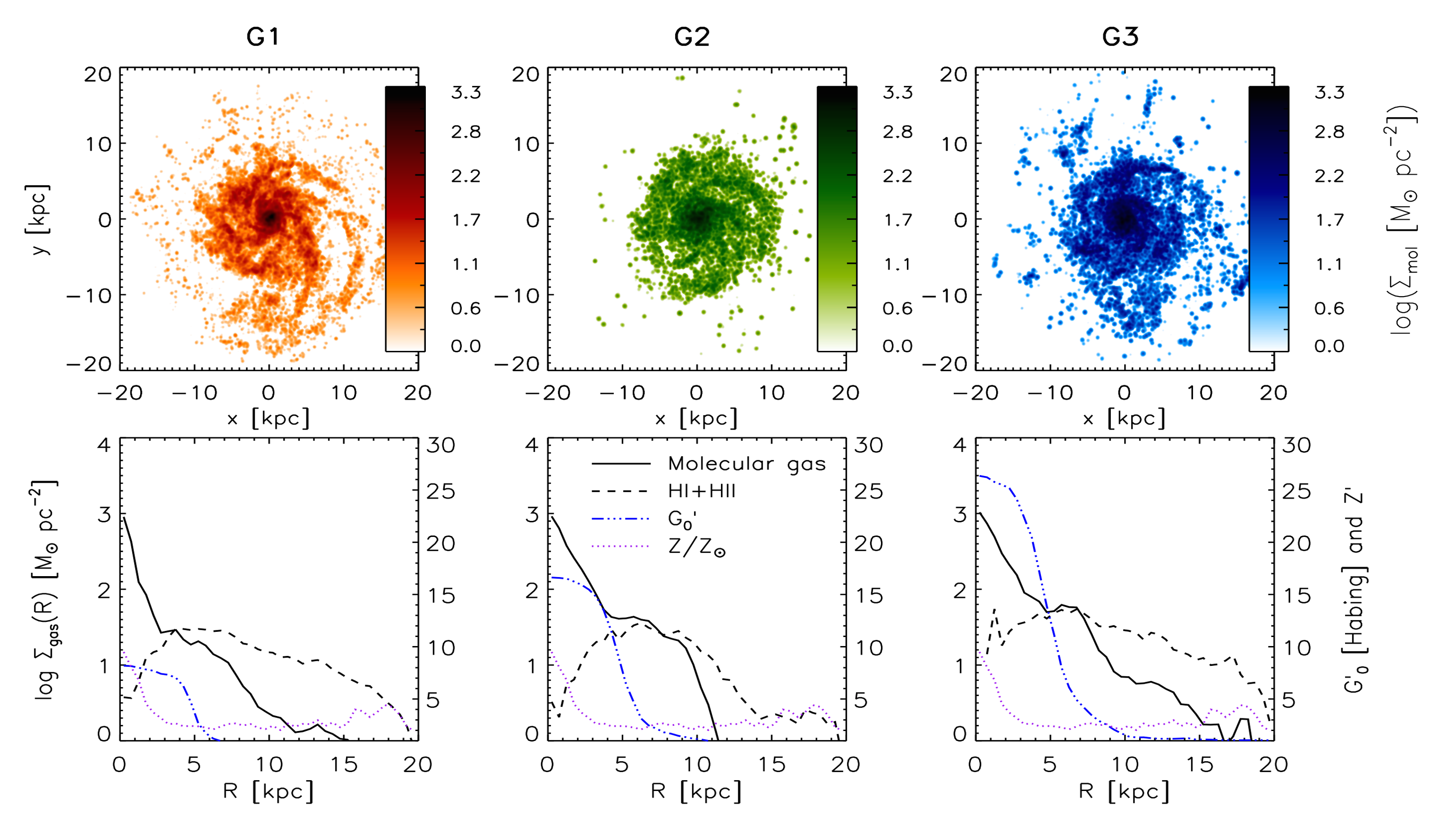}
\caption{Top row: Molecular surface density maps of our model galaxies seen
face-on.  The maps have been smoothed using a circular Gaussian with full width
at half maximum (FWHM) of $3$ pixels corresponding to $0.24$\,kpc. The
molecular gas surface density maps are seen to trace the spiral arms and inner
disk as well as a few dense clumps further out. Bottom row: azimuthally
averaged radial profiles of the molecular (solid curve) and \hi+\hii (dashed
curve) gas surface densities, of the mean metallicity (dotted curve) and of \g0
(dot-dashed curve) -- determined from 50 radial bins stretching from 0 to
15\,kpc from the centre of each galaxy. The molecular gas is the result of
applying the recipes in Section \ref{HI_to_H2} to the SPH simulations presented
in Section \ref{model_galaxies}, while the $\hi+\hii$ gas is the initial SPH
gas mass minus the derived molecular gas mass (both include the contribution
from helium).  We estimate \g0 by averaging over the FUV fields impinging on
all GMCs in each radial bin.}
\label{h2_map}
\end{figure*}

\subsubsection{CO emission maps}\label{combining}
A given GMC is assigned the CO emission line profile of the GMC model that
corresponds to the nearest grid point in the (\g0, \Mgmc, \Z) parameter space.
A spatial grid of $400\times400$\,pixels is overlayed on each galaxy viewed
face-on, and the line profiles of all GMCs within each pixel are added to a
common velocity axis, thus resulting in a position-velocity datacube.  CO
moment 0 maps are then constructed by integrating the flux within each pixel in
velocity. These maps are $40\,{\rm kpc}$ on each side and thus have a resolution of
$\sim 100\,{\rm pc/pixel}$.  By expanding the pixel size to encompass the
entire galaxy, the global CO SLED is derived. 

The approach described above assumes that the molecular line emission
from each GMC is radiatively decoupled from all other GMCs, due to the
significant velocity gradients across the galaxies ($\sim200-500$\,\kms) and
the relatively small internal line widths of the individual GMCs 
($\sim2-34$\,\kms for the GMCs in G1, G2 and G3).

\begin{figure*}
\centering
\includegraphics[width=1.85\columnwidth]{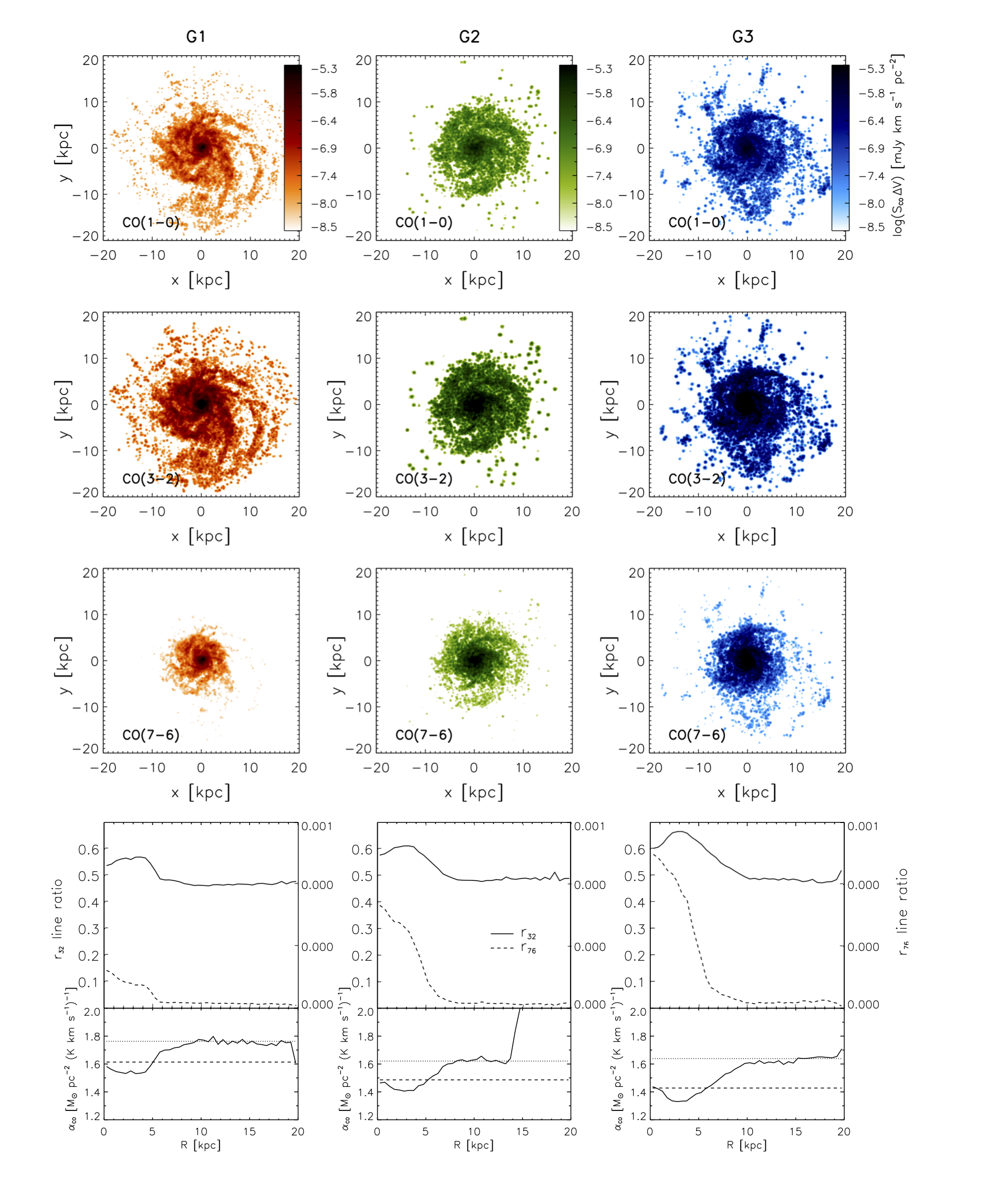}
\caption{The top three rows show the moment 0 maps of the CO(1$-$0),
CO(3$-$2) and CO(7$-$6) emission from G1, G2 and G3. The CO maps have been
smoothed using a circular Gaussian with full width at half maximum (FWHM) of 3
pixels corresponding to $0.24$\,kpc, and as with the SPH and molecular gas
surface density maps, a logarithmic scale has been applied in order to better
display extended emission.  The bottom row shows the azimuthally averaged CO
3$-$2/1$-$0 and 7$-$6/1$-$0 brightness temperature line ratios (denoted
$r_{32}$ and $r_{76}$, respectively) as functions of projected radius for each
of the three galaxies.  A radial bin-size of $0.5\,{\rm kpc}$ was used. Also
shown are the azimuthally averaged radial profiles of CO-to-\h2 conversion
factor $\alpha_{\rm CO}(R)=\Sigma_{\h2}/I_{{\rm CO}(1-0)}$ in units of
\msun\,pc$^{-2}$\,(K\,km\,\ps)$^{-1}$ with a dashed line indicating the global
\aCO factor and a dotted line for the disk-averaged \aCO factor (see Section \ref{CO_mom0}).
}
\label{CO_mom0_f_mom0}
\end{figure*}

\section{Simulating massive $z=2$ main sequence galaxies}\label{results}
In this section we examine the \h2 surface density and CO emission maps
resulting from applying \sigame to the three SPH galaxy simulations G1, G2, and
G3 at $z=2$ (Section \ref{cosmological_simulations}), and we compare with existing CO
observations of main sequence galaxies at $z\sim 1-2.5$.  
We also tested \sigame on three MW-like galaxies (see Appendix \ref{apMW}), 
finding a general agreement with the CO line observations of MW and other similar local galaxies.
As mentioned in
previous sections, our default grid will be that corresponding to a Plummer
density profile and a pressure of $\Pe/k_{\rm B}=10^4\,{\rm \cmpc\,K}$, combined
with a GMC mass spectrum of slope $\beta=1.8$, unless otherwise stated. 

\subsection{Total molecular gas content and \h2 surface density maps}\label{H2_cont}
The total molecular gas masses of G1, G2 and G3 -- obtained by summing up the
GMC masses associated with all SPH particles within each galaxy (i.e., $M_{\rm
mol} = \sum \Mgmc = \sum \fh2 m_{\rm SPH}$) -- are $7.1\e{9}$, $1.7\e{10}$, and
$2.1\e{10}$\,\msun, respectively, corresponding to about $34$, $59$ and
$45$\,\% of the original total SPH gas masses of the galaxies within $R_{\rm
cut}=20\,$kpc.\footnote{These global molecular-to-SPH gas mass fractions are
calculated as $M_{\rm mol}/M_{\rm SPH} = \sum \Mgmc / \sum m_{\rm SPH}$, where
the sums are over all SPH particles.} The global molecular gas mass fractions
(i.e., $f_{\rm mol} =M_{\rm mol}/(M_*+M_{\rm mol}$) are 11.8, 10.2 and 9.1\,\%
for G1, G2, and G3, respectively.

This is $\sim4-5\times$ below the typical molecular gas mass fraction ($\sim
40-60\,\%$) inferred from CO observations of main sequence galaxies at $z\sim
1-3$ with similar stellar masses and star formation rates as our simulations
\citep[e.g.,][]{daddi10,magnelli12,tacconi13}.

The reason for this discrepancy is the low SPH gas mass fractions of our
simulated galaxies to begin with ($f_{\rm SPH} =M_{\rm SPH}/(M_*+M_{\rm
SPH})=18-28$\,\%), which obviously restricts $f_{\rm mol}$ to lower values.
Such low gas fractions have been observed in local star forming galaxies.  For
example, assuming a MW-like \aCO factor, \cite{saintonge11} derived $f_{\rm
mol}\sim6.7\pm4.5$\,\% for a sample of 119 CO$(1-0)$ detected normal star
forming galaxies and a stack of 103 non-detections.

\smallskip

Figure \ref{h2_map} (top panels) shows the molecular gas surface density (\Sh2)
maps of G1, G2, and G3, where \Sh2 is calculated over pixels $80\,{\rm
pc}\times 80\,{\rm pc}$ in size. The pixel size was chosen in order to avoid
resolving the GMCs, which typically have sizes $\ls 40\,{\rm pc}$ (see
Figure \ref{structure_f2}).
The molecular gas is seen to extend out to radii of $\sim 10\,{\rm kpc}$ and
beyond, but generally the molecular gas concentrates within the inner regions
of each galaxy. The distribution of molecular gas broadly follows the central
disk and spiral arms where the SPH surface density ($\Sigma_{\rm SPH}$) is also
the highest (Figure \ref{model_galaxies_f1}). The correspondence is far from
one-to-one, however, as seen by the much larger extent of the SPH gas, i.e.,
regions where \h2 has not formed despite the presence of atomic and ionised
gas.  This point is further corroborated in the bottom panels of Figure
\ref{h2_map}, which show azimuthally-averaged radial surface density profiles
of the molecular gas and of the $\hi+\hii$ gas. The latter is simply the
initial SPH gas mass with the molecular gas mass subtracted, and has been
corrected (just like the molecular gas phase) for the mass contribution from
helium.

In order to compare \Shihii and \Sh2 with observations, we have set the
radial bin width to $0.5$\,kpc -- the typical radial bin size used for nearby
spirals in the work of \cite{leroy08}\footnote{The \h2 surface density maps in
Figure \ref{h2_map} are averaged over areas $80\times 80\,{\rm pc}$ in size, and
therefore give higher peak surface densities than the radial profiles which are
averaged over $\sim 0.5\,{\rm kpc}$ wide annuli.}. The radially binned \Sh2
reaches $\sim 800-1000$\,\msun\,pc$^{-2}$ in the central regions of our
simulated galaxies, which is comparable to observational estimates of \Sh2 (of
several 100\,\msun\,pc$^{-2}$) towards the centres of nearby spirals
\citep{leroy08}.  In all three galaxies, the H{\sc i}+H{\sc ii} surface density dips
within the central $\sim 1-2\,{\rm kpc}$, coinciding with a strong peak in \Sh2.
Thus, despite the marked increase in the FUV radiation field towards the centre,
the formation of \h2 driven by the increase in gas pressure is able to overcome
photodestruction of \h2 through absoption of Lyman or Werner band radiation. The central \h2 surface
densities are similar for all galaxies and is a direct consequence of very
similar SPH gas surface densities in the centre combined with molecular gas mass
fractions approaching 1.  From $R\sim 2\,{\rm kpc}$ and out to $\sim
10\,{\rm kpc}$, the $\hi+\hii$ surface density remains roughly constant with
values of $\sim 40$, $\sim 70$ and $\sim 100$\,\msun\,pc$^{-2}$ for G1, G2 and
G3, respectively. 

Radial profiles of the \hi and molecular gas surface density that are
qualitatively very similar to our simulations have been observed in several
nearby star-forming disk galaxies \citep[e.g.,][]{leroy08,bigiel08}.  In local
galaxies, however, the \hi surface density, including helium, rarely exceeds
$\sim 10$\,\msun\,pc$^{-2}$, while in our simulations we find \hi+\hii surface
densities that are $4-10\times$higher, which is due to the substantial fraction
of ionised gas in our simulated galaxies.

\subsection{CO line emission maps and resolved excitation
conditions}\label{CO_mom0} 
Moment 0 maps of the CO(1$-$0), CO(3$-$2) and
CO(7$-$6) emission from G1, G2 and G3 are shown in Figure \ref{CO_mom0_f_mom0}.
Both the CO(1$-$0) and CO(3$-$2) emission are seen to trace the \h2 gas
distribution well (Figure \ref{h2_map}), while the CO(7$-$6) emission only
trace gas in the central $\sim 7\,{\rm kpc}$ of the galaxies. 

Also shown in Figure \ref{CO_mom0_f_mom0} (bottom row) are the azimuthally
averaged CO 3$-$2/1$-$0 and 7$-$6/1$-$0 brightness temperature line ratios
(denoted $r_{\rm 32}$ and $r_{\rm 76}$, respectively) as a function of radius
for G1, G2 and G3.  The profiles show that the gas is more excited in the
central $\sim 5\,{\rm kpc}$, where typical values of $r_{\rm 32}$ and $r_{\rm
76}$ are $\sim 0.55-0.65$ and $\sim 0.02-0.08$, respectively, compared to
$r_{\rm 32}\sim 0.5$ and $r_{76}< 0.01$ further out in the disk. This radial
behaviour of the line ratios does not reflect the \h2 gas surface density,
which peaks towards the centre rather than flattens, and gradually trails off
out to $R\sim 12\,{\rm kpc}$ instead of dropping sharply at $R\sim 4-6\,{\rm
kpc}$ (Figure \ref{CO_mom0_f_mom0}). Rather, $r_{\rm 32}$ and $r_{\rm 76}$ seem
to mimick the radial behaviour of \g0 (and thus \cri), which makes sense since
\g0 and \cri are the most important factors for the internal GMC temperature
distribution (Figure \ref{apD3}). The central values for $r_{32}$ and $r_{76}$
increase when going from G1 to G3, as expected from the elevated levels of star
formation density (and of \g0 and \cri, accordingly) in their central regions.
Beyond $\sim 6\,{\rm kpc}$ the line ratios are constant ($\sim 0.5$ and $<
0.01$) and the same for all three galaxies, due to relatively similar \g0 (and
\cri) there.
 
The decrease in $r_{\rm 32}$ towards the centre has also 
been observed in nearby galaxies. 
For M\,51, a spiral galaxy with a smaller SFR than our model galaxies 
(see Appendix\,\ref{apMW}), 
\cite{vlahakis13} measured a median ratio of $r_{\rm 32}=0.54$ for pixels covering 
the central kpc region, but typical ratios of $0.2-0.4$ in the arm and inter-arm 
regions. 
In comparison, our model galaxies display larger $r_{32}$ ratios and a 
slightly less pronounced drop of $0.1$ or less when going from 
the central kpc region to the outskirts of the disk.
\cite{mao10} observed $125$ nearby galaxies of different types 
and found global $r_{32}$ values to be $0.61\pm0.16$
in normal galaxies, and $>0.89$ in starbursts and (U)LIRGs. \cite{iono09} and
\cite{papa12} found slightly lower $r_{32}$ in their samples of (U)LIRGs, with
mean values of $0.48\pm0.26$ and $0.67\pm0.62$ respectively. 
The $r_{32}$ profiles of our model galaxies reveal more highly excited gas 
in their centres than in the disks, with central values similar to the  
observed values towards (U)LIRGs by \cite{papa12} but below those reported by \cite{mao10}.
Finally, \cite{geach12} employed Large Velocity Gradient (LVG) radiative transfer models 
to examine $r_{32}$ in different environments. For quiescent clouds similar to 
very low-excitation gas clouds found in M31, 
they showed that $r_{32}$ can be as low as $\sim0.13$, 
while dense ($n(\h2)>10^4\,\cmpc$) star-forming clouds typically have $r_{32}\sim0.88$.
Most likely a result of their moderate star formation rates, the $r_{32}$ 
radial profiles of our model galaxies lie in between these more extreme cases.

\subsection{The CO-to-\h2 conversion factor}\label{section:alpha-CO}
The CO-to-\h2 conversion factor (\aCO) connects CO(1$-$0) line luminosity (in
surface brightness temperature units) with the molecular gas mass ($M_{\rm
mol}$) as follows:
\begin{equation}
\alpha_{\rm CO}=\frac{M_{\rm mol}}{L'_{\rm CO(1-0)}},
\label{alpha_eq}
\end{equation}
From the CO$(1-0)$ surface brightness and \h2 surface density maps of our
simulated galaxies, we calculate the average \aCO within radial bins from the
galaxy centres (Figure \ref{CO_mom0_f_mom0}). 
The resulting radial profiles show that \aCO is essentially constant as a function
of radius, taking on values in the range
$\sim1.3-1.8$\,\msun\,pc$^{-2}$\,(K\,km\,s$^{-1}$)$^{-1}$ across all three
galaxies, except in the outskirts of G2 where \aCO shoots up to higher values. 
The high values of \aCO in the outskirts of G2 
reflect the very low molecular gas surface densities 
(see Figure\,\ref{h2_map}) and resulting lack of CO line emission, 
rendering \aCO a meaningless quantity in this region.
In order to compare with the stellar properties in 
Table\,\ref{model_galaxies_t1}, we will use only the region within $R=15\,$kpc 
for the further analysis of CO line emission in G02. 
Small systematic changes in \aCO with radius are seen
in all three simulated galaxies as \aCO is systematically below (above) the
global \aCO value, marked with dashed lines, at $\ls 5\,{\rm kpc}$ ($\gs 5\,{\rm kpc}$).
\cite{blanc13} measured a drop in \aCO by a factor of two when going
from $R\sim7$\,kpc to the $R<2$\,kpc central region of the Sc galaxy NGC\,628, assuming a constant
gas depletion timescale when converting SFR surface densities into gas masses.
For comparison, the radial \aCO profiles of our model galaxies 
only drop by about $10\%$ from $R\gtrsim7$\,kpc 
to the $R<2$\,kpc central region.

\cite{sandstrom13} measured and examined the resolved \aCO values in a sample of 26 
nearby spiral galaxies. 
A disk-averaged \aCO value was calculated for each galaxy, by tiling them with 
pixels of spacing $37.5\arcsec$ (corresponding to pixel sizes of $0.7-3.9\,{\rm kpc}^2$)
and calculating the mean of \aCO values for pixels with 
a sufficient signal-to-noise ratio. 
It was found that, on average, the galaxies 
have a central ($R\leq1$\,kpc) \aCO value a factor of two 
below the disk-averaged \aCO value. 
We calculate disk-averaged \aCO for our model galaxies in a similar way by 
tiling the galaxies seen face-on with $1\,{\rm kpc^2}$-sized pixels 
and taking the average over pixels within the 
cut-out radii $R_{\rm cut}$ given in Table\,\ref{model_galaxies_t1}. 
The resulting values are shown with dotted lines in Figure 9 and 
are no more than $\sim1.2\times$ the central ($R\leq1\,$kpc) \aCO. 
Our simulated galaxies, therefore do not quite reproduce the drop in 
\aCO typically observed when going from the disk to the central regions 
of local, spiral galaxies. 
Although, we note that for three galaxies from the \cite{sandstrom13} sample 
(NGC\,3938, NGC\,3077 and NGC\,4536) \aCO changes by only $10-16\,\%$  
from the disk to the centre, which is in line with our simulations. 

The \aCO factor is expected to depend on \Z, as higher metallicity means higher C and O
abundances as well as more dust that helps shield CO from photodestruction by
FUV light, thereby leading to a possibly lowering of \aCO. A comparison of the
\aCO radial profiles with those of \g0 and \Z in the bottom panel of Figure
\ref{h2_map}, suggests that the transition in \aCO is caused by a change in \g0
rather than a change in \Z, since \aCO and \g0 generally start to drop at
around $R\sim6$\,kpc while \Z already drops drastically at 1\,kpc from the
centre.  Our modeling therefore implies that \aCO is controlled by \g0 rather
than \Z in normal star-forming galaxies at $z\sim2$, in agreement with the
observations by \cite{sandstrom13} who do not find a strong correlation with
\Z. 

From the total molecular gas masses and CO$(1-0)$ luminosities of G1, G2, and
G3, we derive global \aCO factors of $\aCO=1.6$, $1.5$ and
$1.4$\,\msun\,pc$^{-2}$\,(K\,km\,s$^{-1}$)$^{-1}$, respectively.  These values
are lower (by a factor $\sim 3$) than the inner disk MW value ($\alpha_{\rm CO,
MW} \simeq 4.3\pm 0.1\,{\rm \msun\,pc^{-2}\,(K\,km\,s^{-1})^{-1}}$), 
and closer to the typical
mergers/starburst \aCO-values ($\sim 0.2\times \alpha_{\rm CO, MW}$) inferred
from CO dynamical studies of local ULIRGs
\citep[e.g.,][]{solomon97,downes98,bryant99} and $z\sim 2$ SMGs
\citep[e.g.,][]{tacconi08}. Our \aCO-values are also below those inferred from
dynamical modeling of $z\sim1.5$ BzKs \citep[$\alpha_{\rm CO}=3.6\pm 0.8\,{\rm
\msun\,pc^{-2}\,(K\,km\,s^{-1})^{-1}}$; ][]{daddi10}.

The \aCO factors of the $z\sim1-1.3$ star-forming galaxies
studied by \cite{magnelli12} occupy the same region of the $M_*$--\,SFR plan as
our model galaxies, but have \aCO factors at least a factor $\sim6$ higher.
Their \aCO are $\sim5-20\,{\rm \msun\,pc^{-2}\,(K\,km\,s^{-1})^{-1}}$ when
converting dust masses to gas masses using a metallicity-dependent gas-to-dust
ratio, and \cite{magnelli12} further find that the same galaxies have relatively
low dust temperatures of $\lesssim28$\,K compared to the galaxies further above
the main sequence of \cite{rodighiero10}.  The fact that the main-sequence
galaxies of \cite{magnelli12} are all of near-solar metallicity as ours, cf.
Table \ref{model_galaxies_t1}, suggests that other factors, such as dust
temperature caused by strong FUV radition, can be more important than
metallicity in regulating \aCO, in line with our study of the \aCO radial
profiles.
 
\begin{figure*}
\centering
\includegraphics[width=2.0\columnwidth]{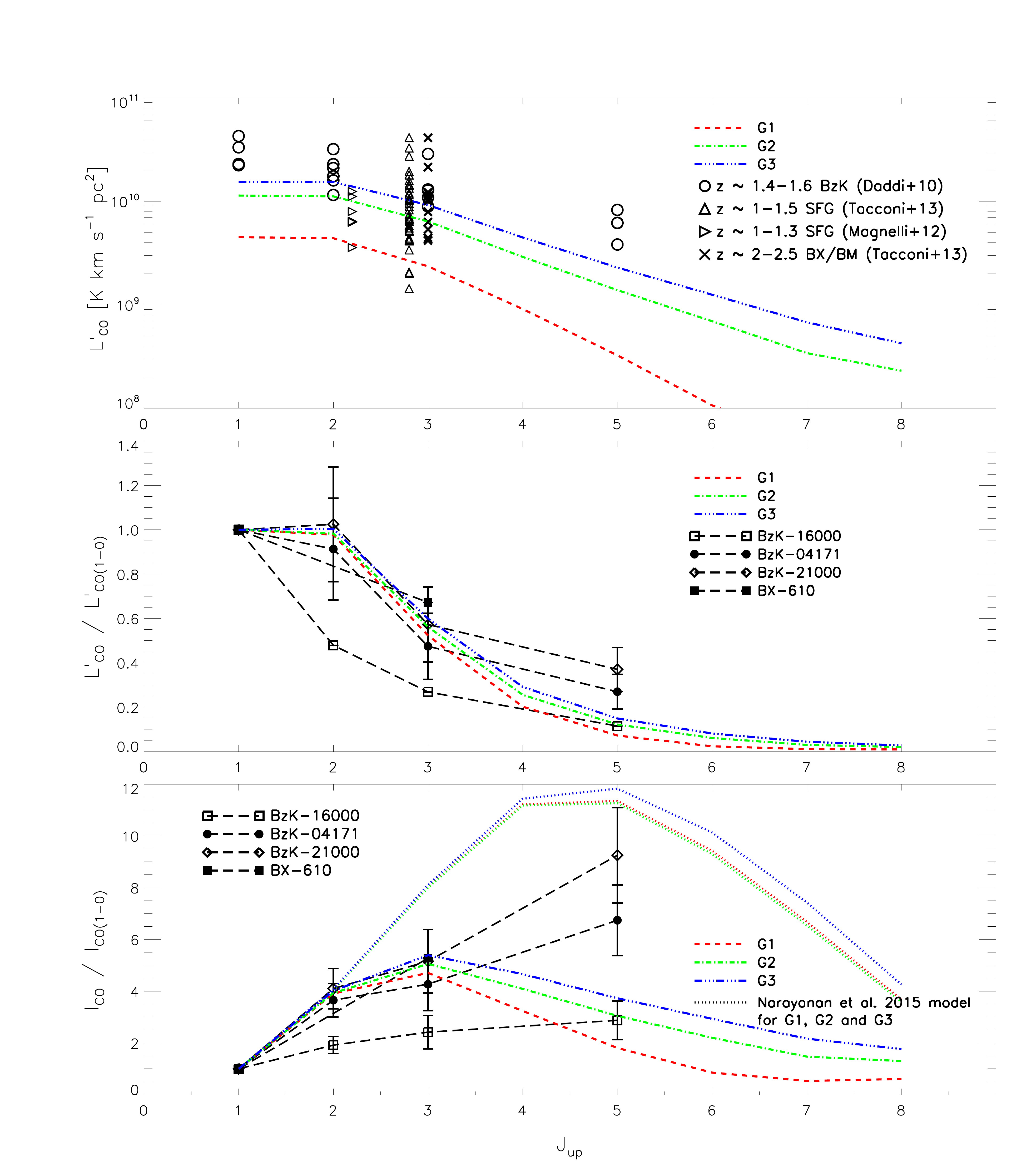}
\caption{Global CO SLEDs of our three model galaxies G1, G2 and G3 shown as 
red (dashed), green (dash-dot) and blue (dash-dot-dot-dot) curves, respectively.  
The SLEDs are given as absolute line
luminosities in units of K\,\kms\,pc$^{-2}$ (top panel), as brightness
temperature ratios normalised to the CO(1$-$0) transition (middle panel), and as
velocity-integrated intensity ratios normalised to CO(1$-$0) (bottom panel).
The model CO SLEDs are compared with observations of $z\sim1.4-1.6$ BzK galaxies
\citep[open circles;][]{dannerbauer09,daddi10,daddi15,aravena10,aravena14}, $z\sim1-1.5$
star-forming galaxies \citep[CO(3$-$2); empty triangles;][]{tacconi13},
$z\sim1-1.3$ star-forming galaxies \citep[CO(1$-$0) and CO(2$-$1); right-facing
triangles;][]{magnelli12}, and $z\sim2-2.5$ BX/BM galaxies \citep[CO(3$-$2);
crosses;][]{tacconi13}.  Four BzK galaxies (BzK$-$16000, BzK$-$4171, BzK$-$21000 
and BzK$-$610) have been observed in CO$(1-0)$ and at least one additional
transition to date, and are highlighted in the bottom two panels by connecting 
dashed lines and individual symbols. Also shown in the bottom panel, 
with dotted lines, are the line ratio predictions of
\protect\cite{narayanan14} (D14), calculated for the \SFRsd of our galaxies, 
G1 to G3 from bottom to top (see Section \ref{dis_models}).} 
\label{CO_sled_f_sled}
\end{figure*}

\subsection{Global CO line luminosities and spectral line energy distributions} \label{CO_sled}
The global CO SLEDs of G1, G2 and G3 are shown in Figure \ref{CO_sled_f_sled}
in three different incarnations: 1) total CO line luminosities ($L'_{\rm
CO_{J,J-1}}$), 2) brightness temperature ratios ($T_{\rm B,CO_{J,J-1}}/T_{\rm
B,CO_{1,0}}=L'_{\rm CO_{J,J-1}}/L'_{\rm CO_{1,0}}$) and 3) line intensity
ratios ($I_{\rm CO_{J,J-1}}/I_{\rm CO_{1,0}} = L'_{\rm CO_{J,J-1}}/L'_{\rm
CO_{1,0}} \times \left ( \nu_{\rm CO_{J,J-1}}/\nu_{\rm CO_{1,0}}\right )^2$).
We have also compiled relevant CO observations of normal star-forming galaxies
at $z \sim 1-3$ in order to facilitate a comparison with our simulated CO
SLEDs. In addition to the $z\sim 1-1.5$ BzK and $z\sim 2-2.5$ BX/BM samples by
\cite{daddi10} and \cite{tacconi13}, respectively, this includes 7 star-forming
galaxies (SFGs) at $z\sim1.2$ \citep{magnelli12} and 39 SFGs at $z\sim1-1.5$
\citep{tacconi13}. The galaxies in the sample from \citet{magnelli12} are all
detected in CO(2$-$1) as well as in {\it Herschel}/PACS bands, and have
$\log({M_*/\msun})=10.36-11.31$ and SFR\,$\sim29-74$\,\sfru. The $z\sim 1-1.5$
SFG sample from \citet{tacconi13} comes from the Extended Growth Strip
International Survey (EGS), is covered by the CANDELS and 3D-HST programs
($J-H$ bands and H$\alpha$ respectively), and has
$\log({M_*/\msun})=10.40-11.23$ and SFR\,$\sim28-630$\,\sfru.

The $z\sim2-2.5$ BX/BM galaxies of \cite{tacconi13} are not only closest in
redshift to our simulated galaxies but also occupy the same region of the
SFR$-M_*$ plane (see Figure\,\ref{model_galaxies_f_M_SFR}). We find that G2 and
G3, the two most massive ($M_*>10^{11}$\,\msun) and star-forming (SFR
$\gtrsim80$\,\sfru) galaxies of our simulations, have CO(3$-$2) luminosities of
$6.4\e{9}$ and $9.2\e{9}\,{\rm K\,km\,\ps\,pc^2}$, respectively, in the range of
CO(3$-$2) luminosities of the BX/BM galaxies.  G1 is just below this range and
about a factor of 4 below the average BX/BM luminosity
($\sim1.2\pm0.9\times10^{10}\,{\rm K\,km\,\ps\,pc^2}$).  All three simulated
galaxies have CO(2$-$1) and CO(3$-$2) luminosities consistent with those of the
$z\simeq1-1.5$ SFGs observed by \citet{magnelli12} and \citet{tacconi13}, which
span a range in CO(2$-$1) and CO(3$-$2) luminosity of $(3.6-12.5)\times10^9$
and $(1.4-41.3)\times10^9\,{\rm K\,km\,s^{-1}\,pc^2}$, respectively.  Finally,
we see that the $z\sim 1.5$ BzK galaxies in general have higher CO line
luminosities across all observed CO transitions (up to $J_{\rm up} = 5$),
although at the $J=2-1$ and $J=3-2$ transitions there is overlap with G3.
As mentioned in Section \ref{H2_cont}, the molecular gas mass fractions of our
galaxies is a factor $\sim 4-5$ below the mean of the observed galaxies at
$z\sim1-2.5$ with which we compare, and we propose this as the main reason for
the comparatively low CO luminosities. 

Differences in the global CO excitation conditions between G1, G2 and G3 are
best seen in the CO(1$-$0) normalised luminosity (and intensity) ratios, i.e.,
middle (and bottom) panel in Figure \ref{CO_sled_f_sled}. The CO SLEDs of all
three galaxies follow each other quite closely up to the $J=3-2$ transition
where the SLEDs all peak; at $3<J_{\rm up}<7$ the SLEDs gradually diverge.
While the metallicity distributions in G1, G2 and G3 are similar (see Figure \ref{apD1}), 
the rise in \g0 presents a likely cause to the increasing high-$J$ flux 
when going from G1 to G3, as higher \g0 leads to more flux primarily in the $J>4$ 
transitions (see Figure \ref{apD4}).  

Our simulated galaxies are seen to have CO 2$-$1/1$-$0, 3$-$2/1$-$0, and
5$-$4/1$-$0 brightness temperature ratios of $r_{\rm 21}\simeq 1$, $r_{\rm
32}\simeq 0.6$, and $r_{\rm 54}\simeq 0.15$, respectively. The first two ratios
compare extremely well with the line ratios measured for BzK$-$4171 and
BzK$-$21000, i.e., $r_{\rm 21}\simeq 0.9-1$, and $r_{32}\simeq 0.5-0.6$
\citep{dannerbauer09,aravena10,daddi10,daddi15}, and suggest that our
simulations are able to emulate the typical gas excitation conditions
responsible for the excitation of the low-$J$ lines in normal $z\sim 1-3$ SFGs.
In contrast, $r_{\rm 54}=0.3-0.4$ observed in BzK$-$4171 and BzK$-$21000
\citep{daddi15}, is nearly $2\times$\,higher than our model predictions.
\cite{daddi15} argue that this is evidence for a component of denser and
possibly warmer molecular gas, not probed by the low-$J$ lines.  In this
picture, we would expect CO(4$-3$) to probe both the cold, low-excitation gas as
well as the dense and possibly warm star-forming gas traced by the CO(5$-$4)
line, and we would expect CO(6$-$5) to be arising purely from this more highly
excited phase and thus departing even further from our models. 

However, significant scatter in the CO line ratios of main-sequence galaxies is
to be expected, as demonstrated by the significantly lower line ratios observed
towards BzK$-$16000: $r_{\rm 21}\simeq 0.4$, $r_{\rm 32}\simeq 0.3$, and $r_{\rm
54}\simeq 0.1$ \citep{aravena10,daddi10,daddi15} and, in fact, the average
$r_{\rm 54}$ for all three BzK galaxies above \citep[$\simeq 0.2$;][]{daddi15} is
consistent with our models.  It may be that G2 and G3, and perhaps even G1, are
more consistent with the CO SLEDs of the bulk $z\sim 1-3$ main-sequence
galaxies.  We stress, that to date, no $z\sim 1-3$ main sequence galaxies have
been observed in the $J=4-3$ nor the $6-5$ transitions, and observations of
these lines, along with low- and high-$J$ lines in many more BzK and
main-sequence $z\simeq 1-3$ galaxies are needed in order to fully delineate the
global CO SLEDs in a statistically robust way. 

\section{Testing different ISM models} \label{dif_ISM}
In this section we investigate the effects on our simulation results when
adopting i) a more top-heavy GMC mass spectrum with a slope of $\beta=1.5$, ii)
a steeper GMC density profile, i.e., a Plummer profile exponent of $-7/2$, and
iii) a GMC model grid that includes \Pe as a fourth parameter. In order to
carry out option iii), the GMC model grid was produced for $\Pe/k_{\rm B} =
10^{4}\,{\rm cm^{-3}\,K}$ (default), $10^{5.5}\,{\rm cm^{-3}\,K}$ and
$10^{6.5}\,{\rm cm^{-3}\,K}$, allowing for a look-up table of CO SLEDs in (\g0,
$m_{\rm GMC}$, \Z, \Pe) parameter space.  

\begin{figure}
\centering
\includegraphics[width=1\columnwidth]{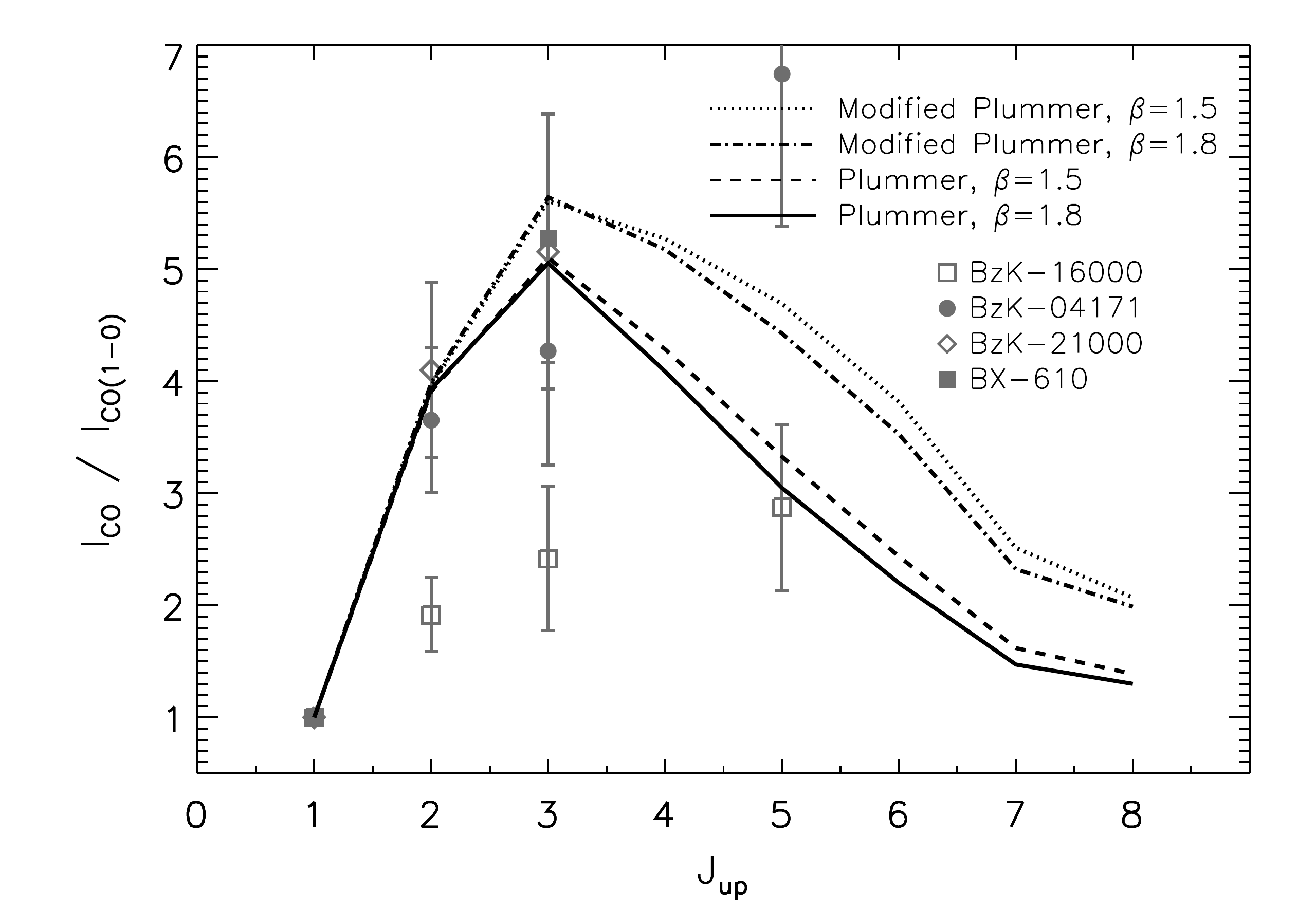}\\
\includegraphics[width=1\columnwidth]{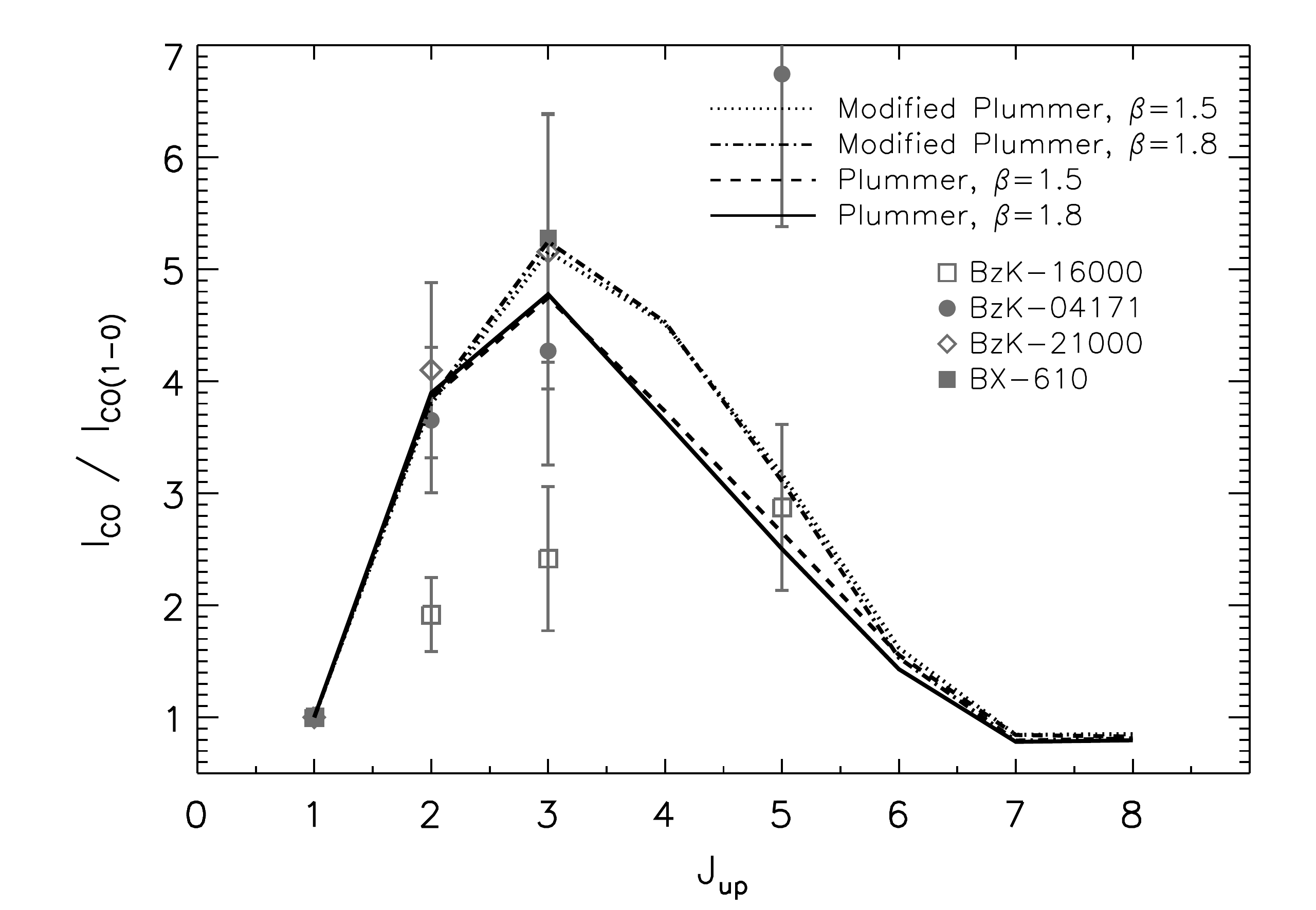}
\caption{Global CO SLEDs of our model galaxy G2 for different choices of ISM
prescriptions. {\it Top}: For a pressure fixed of $10^4\,{\rm \cmpc\,K}$.  {\it
Bottom}: Using the \Pe as a fourth parameter in the GMC model grid (see Section
\ref{gmcgrid}).}
\label{CO_sled_ism}
\end{figure}

We examine the effects that options i)-iii) have on the global CO SLED of
galaxy G2 (Figure \ref{CO_sled_ism}) and how combinations of option i) and ii)
change the values of the global CO-to-H$_2$ conversion factor for all three
simulated galaxies (Table \ref{aCOt}).  Also, we show the impact that changes
ii) and iii) have on the CO SLEDs of the individual GMC grid models in Figures
\ref{apD5} and \ref{apD6}. 

Changing the GMC mass spectrum from $\beta=1.8$ to $1.5$ leaves the CO SLED
virtually unchanged with only a marginal increase in the line flux ratios for
$J_{\rm up}\geq4$. This holds true regardless of what has been assumed for ii)
and iii) (compare solid vs.\ dashed and dot-dashed vs.\ dotted curves in Figure
\ref{CO_sled_ism}). Also, changing $\beta$ from 1.8 to 1.5 does not lead to any
significant change in \aCO (Table \ref{aCOt}).

Adopting the modified Plummer density profile with an exponent of $-7/2$
instead of $-5/2$ results in significantly higher global line ratios for
$J_{\rm up} \ge 3$ (see dotted vs.\ dashed curve and dot-dashed vs.\ solid
curves in Figure \ref{CO_sled_ism}). This is due to the higher central
densities achieved for the modified Plummer profile which, as shown in Figure
\ref{apD5}, leads to an increase in the excitation of the higher $J$ lines
relative to that of the low-$J$ lines. Significantly higher \aCO values
($\simeq 3.6\,{\rm \msun\,pc^{-2}\,(K\,km\,s^{-1})^{-1}}$) are obtained when
adopting a modified Plummer profile (Table \ref{aCOt}). These are in excellent
agreement with the \aCO values inferred for $z\simeq 1.5-2$ BzK galaxies by
\citet{daddi10}. In our simulations, the higher \aCO values are due the
steeper density profile which leads to smaller GMC sizes and therefore smaller
CO$(1-0)$ fluxes (Figure\,\ref{apD5}) and, ultimately, a lowering of the total
CO$(1-0)$ luminosity of the galaxy.

Including external pressures of $\Pe/k_{\rm B}=10^{5.5}\,{\rm cm^{-3}\,K}$ and
$10^{6.5}\,{\rm cm^{-3}\,K}$ in the GMC model grid (see Section \ref{gmcgrid}),
means that these pressure values are assigned to GMCs with $\Pe/k_{\rm B}$ in
the ranges $10^{4.75}-10^6$ and $> 10^{6}\,{\rm cm^{-3}\,K}$, respectively.
The higher pressures now experienced by these subsets of GMCs results in
smaller radii since $R_{\rm GMC} \propto \Pe^{-1/4}$ for a given mass (see eq.\
\ref{structure_e1}). Their velocity dispersions and densities increase since
$\sigma_v \propto P_{\rm ext}^{1/8}$ and $n_{\rm H_2}(R=0) \propto P_{\rm
ext}^{3/4}$ for a given mass (see eqs.\ \ref{structure_e2} and
\ref{equation:plummer-profile}). The net effect this has on the global CO SLED
is depicted in the bottom panel of Figure \ref{CO_sled_ism}.  Line ratios with
$J_{\rm up}=3$ are lowered, and increasingly so for higher $J$, resulting in a
steeper decline at high ($J_{\rm up}\geq4$) transitions. 

\begin{center}
\begin{table}
\centering
\caption{The global \h2-to-CO conversion factors (in units of ${\rm
\msun\,pc^{-2}\,(K\,km\,s^{-1})^{-1}}$), averaged over G1, G2, and G3, for combinations
of assumptions i) and ii). The pressure has been kept fixed at $\Pe/k_{\rm
B}=10^4\,{\rm\cmpc\,K}$.}
\begin{tabular}{l|ll} 
\hline
\hline
 	 				&	$\beta=1.8$     	& 	$\beta=1.5$      	\\\hline 
Plummer profile				&	$1.5\pm0.1$		& 	$1.5\pm0.1$ 		\\ 
Modified Plummer profile		&	$3.6\pm0.4$		& 	$3.5\pm0.5$     	\\ 
\hline
\end{tabular}
\\
\label{aCOt}
\end{table}
\end{center}

\section{Comparison with other models}
\label{dis_models}
The sub-grid physics implemented by \sigame assumes that the scaling-laws and
thermal balance equations that have been established for GMCs in
our own Galaxy can be applied to the ISM conditions in high-$z$ galaxies.
In this respect, \sigame does not differ from most other numerical simulations
of the molecular line emission from galaxies
\citep[e.g.,][]{narayanan06,pelupessy06,greve08,christensen12,lagos12,munoz13,narayanan14,lagos12,popping14},
although there are of course differences in the range and level of detail of
the various sub-grid physics implementations.  Here we highlight and discuss
some of these differences between \sigame and other simulations.

First, however, we compare the \sigame CO SLEDs of G1, G2, and G3 with those
predicted by other CO emission simulations of similar main-sequence galaxies.
A direct comparison can be made with the models presented by
\cite{narayanan14}, where the global CO SLED is parametrised as a function of
the luminosity weighted SFR surface density ($\Sigma_{\rm SFR}$). 
We approximate a luminosity weighted $\Sigma_{\rm SFR}$ by calculating a mean of 
$\Sigma_{\rm SFR}$ for $1\,{\rm kpc}^2$ pixels across the galaxies seen face-on, 
weighted by the total SFR within the pixels. 
With this procedure, G1, G2 and G3 have $\Sigma_{\rm SFR}$ of 
$3.0$, $2.9$ and $3.8\,\sfru\,{\rm kpc}^2$, respectively, and the resulting 
CO SLEDs inferred from the \cite{narayanan14} parametrisation
are shown as dotted lines in the bottom panel of Figure \ref{CO_sled_f_sled}.
These are seen to peak at $J_{\rm up}=5$ and not $3$ as the \sigame CO SLEDs do.
Also, the line flux ratios are significantly higher for all transitions above 
$J_{\rm up}=2$. 
Both set of models agree on the $r_{21}$ ratio, but follow distinct CO SLED 
shapes for the remaining transitions. 
The models also seem to suggest one dominating ISM 
phase, whereas the observed CO SLED of the BzK galaxies 
indicate a more complex multi-phased ISM.

Combining a semi-analytical galaxy formation model with a photon-dominated
region code, \cite{lagos12} made predictions of the CO SLEDs of $z=2$ galaxies
as a function of their IR luminosities. They found that for galaxies with
infrared luminosities $L_{\rm IR}\sim10^{11.6}-10^{12.2}\,\lsun$ -- which is
the range found for G1, G2, and G3 if one adopts $L_{\rm
IR}/L_{\odot}=\rm{SFR}/[\msun\,\rm{yr}^{-1}]\times10^{10}$
(\citealt{kennicutt98} for a Chabrier IMF) -- the CO SLEDs (when
converted to velocity integrated flux ratios in order to compare with the
bottom panel of Figure \ref{CO_sled_f_sled}) peak at $J=3-2$,  
in agreement with our model galaxies. 
However, the line ratios of the CO SLEDs of \cite{lagos12} 
are consistently above ours up to $J_{\rm up}=7$, 
but cross over our models at $J_{\rm up}=7$ and go 
below at $J_{\rm up}=8$. 
In terms of CO line luminosity, our model G1 agrees best with the models of 
\cite{lagos12}, lying within the 10-90 percentile ranges ($80\%$ of their galaxy distribution) 
for the CO luminosities of their sample 
for $2<J_{\rm up}<8$, but slightly below for the first two transitions 
and for $J_{\rm up}>7$. G2 and G3, lying roughly a factor 2-4 above G1, 
are outside the 10-90 percentile ranges of the models from \cite{lagos12}.

\cite{popping14}, in their semi-analytical study of
MS galaxies at $z=0$, 1.2 and 2, found that 
for galaxies with far-infrared (FIR)
luminosities of $L_{\rm FIR}=10^{11}-10^{12}\,\lsun$, the CO SLED
peaks at CO(3$-$2) or CO(4$-$3) at $z=0$, but at CO(6$-$5) at $z=2$, 
as a result of denser and warmer gas in their model galaxies. 
The CO luminosities of our galaxies G2 and G3 are in broad agreement with 
the corresponding models by \cite{popping14} at CO(4$-$3) and (5$-$4) for similar SFRs. 
But at $J_{\rm up}<4$ our galaxies are above 
while at $J_{\rm up}>5$, our galaxies are below those of \cite{popping14} up until 
$J_{\rm up}=9$ where the SLEDs cross again.

\bigskip

\noindent$\bullet$ {\it Implementation of \g0 and \cri}\\
\sigame stands out from most other simulations to date in the way the FUV
radiation field and the CR flux that impinge on the molecular clouds are
modelled and implemented.  Most simulations adopt a fixed, galaxy-wide value of
\g0 and \cri, scaled by the total SFR or average gas surface density across the
galaxy \citep[e.g.,][]{lagos12,narayanan14}.  \sigame refines this scheme by
determining a spatially varying \g0 (and \cri) set by the local SFRD, as
described in Section \ref{HI_to_H2}. By doing so, we ensure that the molecular
gas in our simulations is calorimetrically coupled to the star formation in
their vicinity.  If we adopt the method of \cite{narayanan14} and calculate a
global value for \g0 by calibrating to the MW value, using SFR$_{\rm
MW}=2$\,\sfru and $G_{\rm 0,MW}=0.6$\,Habing, our galaxies would have \g0
ranging from about $12$ to $42$, whereas, with the SFR surface density scaling
of \cite{lagos12}, global \g0 values would lie between about $6$ and $9$.  For
comparison, the locally determined \g0 in our model galaxies spans a larger
range from $0.3$ to $27$ (see Figure \ref{apD1}).  \cite{narayanan14} determine
the global value of \cri as $2\e{-17}\Z\ps$, corresponding to values of
$3.7-6.2\e{-17}\ps$ in our model galaxies, when using the mass-weighted mean of
\Z in each galaxy.  Typical values adopted in studies of ISM conditions are around
$(1-2)\e{-17}$\ps \citep{wolfire10,glover12}, but again, the local values of
\cri in our galaxies span a larger range, from $1.3\e{-17}$ to
$1.3\e{-15}$\,\ps.
Adopting the method of \cite{narayanan14} for \g0 and \cri in our galaxies, leads to higher 
CO luminosities at $J_{\rm up}>2$, but very similar CO$(1-0)$ luminosities which together 
with slightly reduced molecular gas masses result in smaller \aCO factors by about $7-8$\,\%.
\bigskip

\begin{figure} 
\hspace{-0.5cm}
\includegraphics[width=1.1\columnwidth]{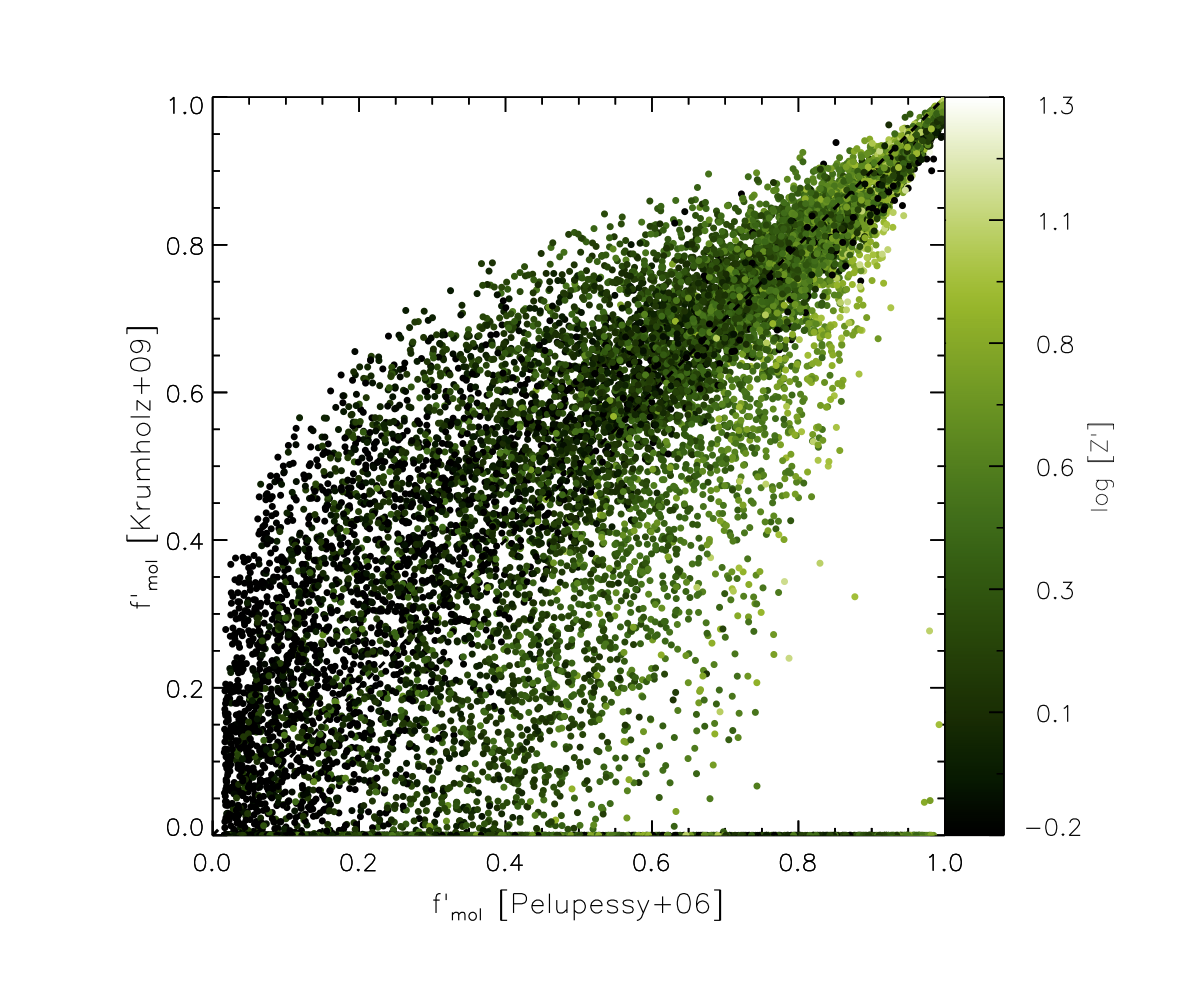}
\caption{Comparison of two methods for calculating the molecular gas mass
fraction, \fh2, of SPH particles in G1: That of \protect\cite{pelupessy06} with
external cloud pressure derived from hydrostatic mid-plane equilibrium as done
in this work (abscissa) and that of \protect\cite{krumholz09b} (ordinate) used
by e.g., \protect\cite{narayanan14}, color-coded by metallicity.}
\label{h2_models}
\end{figure}

\noindent$\bullet$ {\it The molecular gas mass fraction}\\ In \sigame the
molecular gas mass fraction (\fh2) is calculated following the work by
\cite{pelupessy06} (P06), in which \fh2 depends on temperature, metallicity,
local FUV field, and boundary pressure on the gas cloud in question.  Other
methods exist, such as those of \cite{blitz06} and \cite{krumholz09b} (K09).
The K09 method, which was adopted by e.g. \cite{narayanan14}, 
establishes \fh2 from local cloud metallicity, dissociating radiation field and 
column density.
In Figure \ref{h2_models} we
compare the K09 method to that of
P06.  The gas surface density that enters in the K09 method was estimated
within 1 kpc of each SPH particle, as used in our calculation of \Pe (see
Section \ref{HI_to_H2}). 

The two methods agree for molecular gas fractions $\gs 80\%$, which are seen to
be dominated by SPH particles with high metallicities (i.e., $\log \Z \gs
0.3$). At lower molecular gas fractions the agreement between the two methods
exhibit an increasing scatter. Also, systematic, metallicity dependent
differences are seen between the two methods.  For high metallicity gas, the
P06 method gives systematically higher molecular gas fractions than the K09
method. At low metallicities (i.e., $\log \Z \ls 0.2$) the K09 method tends
to result in higher molecular gas fractions than P06. The above is the result
of the K09 method having a much weaker dependance on \Z \citet{krumholz09b}
than the P06 method (see Section \ref{HI_to_H2}).  Despite these differences,
we find that using the K09 instead of the P06 method does not change the total
molecular gas mass significantly: $f_{\rm mol}$ for G1, G2 and G3 goes from
12.2, 10.0, 9.2\,\% to 11.4, 10.7 and 10.3\,\%, respectively. 

A common feature of the above \hi $\longrightarrow \h2$ prescriptions is that
they assume an instantaneous \hi to \h2 conversion.  However, as pointed out by
\cite{pelupessy06}, the typical timescales for \h2 cloud formation are of order
$\sim10^7$\,yr, which are comparable to the time-scales of a number of
processes, such as cloud-cloud collisions and star formation, that can
potentially alter or even fully disrupt typical molecular clouds and drive \h2
formation/destruction in both directions.  Ideally, therefore, a comprehensive
modeling of the \hi $\longrightarrow \h2$ transition should be time-dependent.
Relying on the findings of \cite{narayanan11} and \cite{krumholz11}, the static
solution for \h2 formation is a valid approximation for $\Z>0.01$, which is the
case for the three model galaxies studied here.

\bigskip

\noindent$\bullet$ {\it CO abundance}\\
\sigame assumes a constant CO abundance relative to H$_2$, which for the
simulations presented in this paper was set to the Galactic CO abundance, i.e.,
$[{\rm CO}/\h2]=2\times10^{-4}$.  In reality, \h2 gas can self-shield better
than the CO gas, creating an outer region or envelope of `CO-dark' gas
\citep{bolatto08}.  Observations in the MW by \cite{pineda13} indicate that the
amount of dark gas grows with decreasing metallicity and the resulting absence
of CO molecules.  By modeling a dynamically evolving ISM with cooling physics
and chemistry incorporated on small scales, \cite{smith14} showed that in a
typical MW-like disk of gas, the CO-dark gas mass fraction, $f_{\rm DG}=$, defined
as having integrated CO intensity $W_{\rm CO}<0.1$\,K\,km\,\ps, 
is about $42$\,\%, but up to $62$\,\% in a radiation field ten times that of
the solar neighbourhood.  While \cite{smith14} kept metallicity fixed to 
solar, \cite{wolfire10} found $f_{\rm DG}$ values of about
$0.5-0.7$ for the low metallicity cases with $\Z=0.5$ in $10^6$\,\msun clouds 
immersed in the local interstellar radiation field. 
Recently, Bisbas et al.\ (2015) argued that CR
ionization rates of $10-50\times\zeta_{\rm{CR,MW}}$ can effectively destroy
most CO in GMCs.  

Thus, it is expected that $[{\rm CO}/\h2]$ is lower in regions of low metallicity
and/or intense FUV and CR fields, effectively leading to underestimates of the \aCO
conversion factor when not accounted for in models. Since in this paper we have
restricted our simulations to main-sequence galaxies with solar or higher than
solar metallicities, adopting a constant Galactic $[{\rm CO}/\h2]$ seems a
reasonable choice, but see \cite{lagos12}, \cite{narayanan12} and
\cite{narayanan14} for alternative approaches.

\bigskip

\noindent$\bullet$ {\it GMC density and thermal structure}\\
When solving for the temperature structure of each GMC, \sigame includes
only the most dominant atomic and molecular species in terms of heating and
cooling efficiencies. 
\sigame also ignores heating via X-ray irradiation and turbulence. 
In particular, it has been shown with models that 
turbulent heating can completely control the mean temperature of 
molecular clouds, exceeding the heating from cosmic rays \citep{pan09}. 
Our comparison of observed CO SLEDs at $z=2$ with those of \sigame, 
suggest that our simulations are missing a warm gas component, 
which could be explained, at least in part, by the exclusion of turbulent heating.
We have, however, checked our \Tk$-$\nh2 curves in Figure \ref{apD3} against those of
\cite{glover12}, who performed time-resolved, high-resolution ($\delta m \simeq
0.05-0.5\,\msun$) SPH simulations of individual FUV irradiated molecular clouds
using a chemical network of 32 species (see their Figure 2). Overall, there is
great similarity in the \Tk vs.\ \nh2 behaviour of the two sets of
simulations. This includes the same main trend of decreasing temperature with
increasing hydrogen density, as well as the local increase in \Tk at $\sim
10^3-10^4$\,\cmpc and the subsequent decrease to $\Tk<10$\,K at
$\nh2>10^{5.5}$\,\cmpc.  Thus, our GMCs models, despite their simplified density
profiles and chemistry, seem to agree well with much more detailed simulations.

\bigskip

\section{Summary}
In this paper we have presented \sigame, a code that simulates the
molecular line emission of galaxies via a detailed post-processing of the
outputs from cosmological SPH simulations. 
A sequence of sub-grid prescriptions are applied to a simulation snapshot 
in order to derive the molecular gas density and temperature from the 
SPH particle information, which includes SFR, gas density, 
temperature and metallicity.

A key aspect of \sigame is the {\it localised} coupling between the star
formation and the energetics of the ISM, where the strength of the local  FUV
radiation field and CR ionization rate that impinge and heat a cloud scales with
the local star formation rate density.
The radial temperature profile of each GMC is calculated by balancing the
heating rate with cooling from \h2, CO, \oi, and \cii lines in addition to
gas-dust interactions. In these thermal balance calculations, \sigame takes
into account the local enrichment of the gas, which is a critical parameter
for the cooling rates.  The CO emission line spectrum from a grid of GMC models
is calculated using the 3D radiative transfer code \lime. 

We used \sigame to create line
emission velocity-cubes of the full CO rotational ladder for three cosmological
N-body/SPH simulations of massive ($M_* \gtrsim 10^{10.5}\,\msun$)
main-sequence galaxies at $z=2$. 

Molecular gas is produced more efficiently towards the centre of each galaxy, 
and while \hi surface gas densities (including helium) do not exceed $\sim100$\,\msun\,pc$^{-2}$ anywhere in the disk, 
central molecular gas surface densities reach $\sim1000$\,\msun\,pc$^{-2}$ 
on spatial scales of $100\,{\rm pc}\times 100\,{\rm pc}$, in good agreement with observations made at 
similar spatial resolution. This strong increase in molecular surface density 
is brought on by a similar increase in total gas surface density, 
overcoming the increase in photo-dissociating FUV field towards the centre of each galaxy.

Turning to the CO emission, the velocity-integrated moment 0 maps reveal distinct 
differences in the various transitions as molecular gas tracers. 
The morphology of molecular gas in our model galaxies is well reproduced in CO(1-0), 
but going to higher transitions, the region of CO emitting gas shrinks towards the galaxy centres.
The global CO SLEDs of our simulated galaxies all peak at $J=3-2$. 
Recent CO$(5-4)$ observations of $z\sim1.5$ BzK galaxies seem to suggest that 
these galaxies actually peak at higher $J$. 
The CO$(3-2)$ line luminosities of our model galaxies 
are within the range of 
corresponding observed samples at redshifts $z\sim1-2.5$, 
however on the low side. 
In particular, the model galaxies are below or at the CO luminosities 
of BzK-selected galaxies of comparable mass and SFR but at $z\sim1.5$. 
The low luminosities are most likely a consequence of molecular 
gas mass fractions in our galaxies being 
about $\sim4-5$ times below the observed values in the star-forming galaxies 
at $z=1-2.5$ used to compare with. 

Combining the derived \h2 gas masses with the CO$(1-0)$ line emission found, 
we investigate local variations in the CO-\h2 conversion factor \aCO. 
The radial \aCO profiles all show a decrease towards the galaxy centres, 
dropping by a factor of $\sim1.2$ in the central $R\leq2$\,kpc region compared to the disk average, 
the main driver being the FUV field rather than a gradient in density or metallicity. 
Global \aCO factors range from $1.4$ to $1.6$\,\msun\,pc$^{-2}$\,(K\,km\,s$^{-1}$)$^{-1}$ 
or about $0.3$ times the MW value, but closer to values for $z\sim1.5$ normal star-forming galaxies 
identified with the BzK colour criteria. 
Changing the GMC properties from what is observed in the MW and local galaxies to a 
steeper GMC density profiles and/or a shallower GMC mass spectra, 
resulted in elevated \aCO values of up to $0.9$ times the MW value.

The CO luminosity ratios of CO 3$-$2/1$-$0 and 7$-$6/1$-$0 ($r_{32}$ and $r_{76}$ respectively) 
drop off in radius about where the FUV radiation drops in intensity, and are thus likely controlled by 
the FUV field as is \aCO. The global ratios of $r_{21}\simeq 1$ and 
$r_{32}\simeq 0.6$ agree very well with observations of BzK galaxies, 
while the $r_{54}$ of about $0.15$ 
is low compared to recent observations in BzK$-$4171 and BzK$-$21000. 
The low flux from high CO transitions in our models compared to observations could be 
explained, at least in part, by our omission of turbulent heating in \sigame.
However, more observations of $J_{\rm up}>3$ lines towards high-$z$ main-sequence galaxies,
such as the BzKs, are still needed in order to determine the
turn-over in their CO SLEDs and better constrain the gas excitation. 

Finally, we note that \sigame in principle is able to simulate the
emission from a broad range of molecular and atomic lines in the far-IR/mm
wavelength regime provided measured collision rates exist, such as those found in the
LAMDA database\footnote{\url{http://www.strw.leidenuniv.nl/~moldata/}, \cite{schoier2005}}. 
In the future, we aim to apply
\sigame to a larger and more diverse sample of galaxies, possibly at higher redshift, and explore 
the emission from other important ISM diagnostic lines. 

\section*{Acknowledgments}
We are grateful to Inti Pelupessy for elaborations on his papers and help with
the ionization fractions, and to Desika Narayanan for stimulating discussions.
We thank Padelis Papadopoulos for useful discussions and suggestions along
the way. 
We also thank Julia Kamenetzky for providing us with CO line data from her paper.
The draft was much improved by thoughful reports from the anonymous referee. 
KPO gratefully acknowledge the support from the Lundbeck foundation and
TRG acknowledges support from a STFC Advanced Fellowship. 
S. T. acknowledges support from the ERC Consolidator Grant funding scheme 
(project ConTExt, grant number 648179). 
The Dark Cosmology
Centre and Starplan are funded by the Danish National Research Foundation.

\bibliographystyle{mnras}
\bibliography{bibs}

\appendix

\section{Thermal balance of the atomic gas phase}
\label{apB}
As explained in \S\,\ref{WCNM}, we cool the initial hot SPH gas by 
requiring:
$\Hcrhi = \Cions + \Crec + \Cff$.

\Hcrhi is the heating rate of the atomic gas due to cosmic ray ionizations
\citep{draine11}:
\begin{equation}
\begin{aligned}
	\Hcrhi = &1.03\e{-27}\nhi \left( \frac{\crihi}{10^{-16}} \right) \\
	&\times\left[ 1+4.06\left( \frac{\xe}{\xe+0.07}\right)^{1/2}\right]~~{\rm erg\,cm^{-3}\,s^{-1}}, 
	\label{test}
\end{aligned}
\end{equation}
where \crihi is the primary CR ionization rate of \hi atoms (determined locally
in our simulations according to eq.\ \ref{equation:CR}), and \xe is the
hydrogen ionization fraction calculated with a procedure kindly provided by I.
Pelupessy; see also Pelupessy (2005).  The term containing \xe in
eq.\,\ref{test} accounts for the fact that in highly ionised gas, electrons
created by primary CR ionization have a high probability of transferring their
kinetic energy into heat via long-range Coulomb scattering off free electrons.
For low ionization gas, this term becomes insignificant as a higher fraction of
the energy of the primary electrons goes to secondary ionizations or excitation
of bound states instead of heating.  

\Cions is the total cooling rate due to line emission from H, He, C, N, O, Ne, 
Mg, Si, S, Ca, and Fe, calculated using the publically available code of 
\cite{wiersma09} which takes \Tk, \nH 
and the abundances of the above elements as input. 
\cite{wiersma09} compute the cooling rates with the
photoionization package \texttt{CLOUDY} assuming CIE. They also adopt a value 
for the meta-galactic UV and X-ray field equal to that expected at $z\sim 2$ \citep{haardt01}.
At $z\sim2$, the emission rate of \hi ionizing radiation is higher by a factor of 
about $\sim 30$ than at $z=0$ \citep{puchwein15}, and thus plays an important role
in metal line cooling calculations.

\Crec is the cooling rate due to hydrogen recombination emission \citep{draine11}:
\begin{equation}
	\Crec = \alpha_B\ne\nHplus\langle E_{rr} \rangle~~{\rm ergs\,cm^{-3}\,s^{-1}},
\end{equation}
where $\alpha_B$ is the radiative recombination rate for hydrogen in the case of
optically thick gas in which ionizing photons emitted during recombination are
immediately re-absorbed. We adopt the approximation for $\alpha_B$ given by
\cite{draine11}: 
\begin{equation}
	\alpha_{\rm B} = 2.54\e{-13}T_4^{\left(-0.8163-0.0208\ln T_4 \right)}~~{\rm cm}^3\,\ps,
\end{equation}
where $T_4$ is defined as $\Tk/10^4$\,K. The density of ionised hydrogen,
\nHplus, is set equal to the electron density, \ne, and $E_{rr}$ is the
corresponding mean kinetic energy of the recombining electrons:
\begin{equation}
	\langle E_{rr}\rangle = \left[0.684-0.0416\,\ln T_4 \right]\,k_{\rm B}\Tk~~{\rm ergs}.	
\end{equation}

\Cff is the cooling rate due to free-free emission from electrons in a pure H
plasma (i.e., free electrons scattering off H$^+$), and is given by
\citep{draine11}:
\begin{equation}
	\Cff = 0.54 T_4^{0.37} k_B\Tk\ne\nHplus\alpha_{\rm B}~~\rm{ergs\,cm^{-3}\,s^{-1}},
\end{equation}
where the recombination rate, $\alpha_{\rm B}$, is calculated in the same way as for \Crec.

\begin{figure*}
\centering
\includegraphics[width=0.5\textwidth]{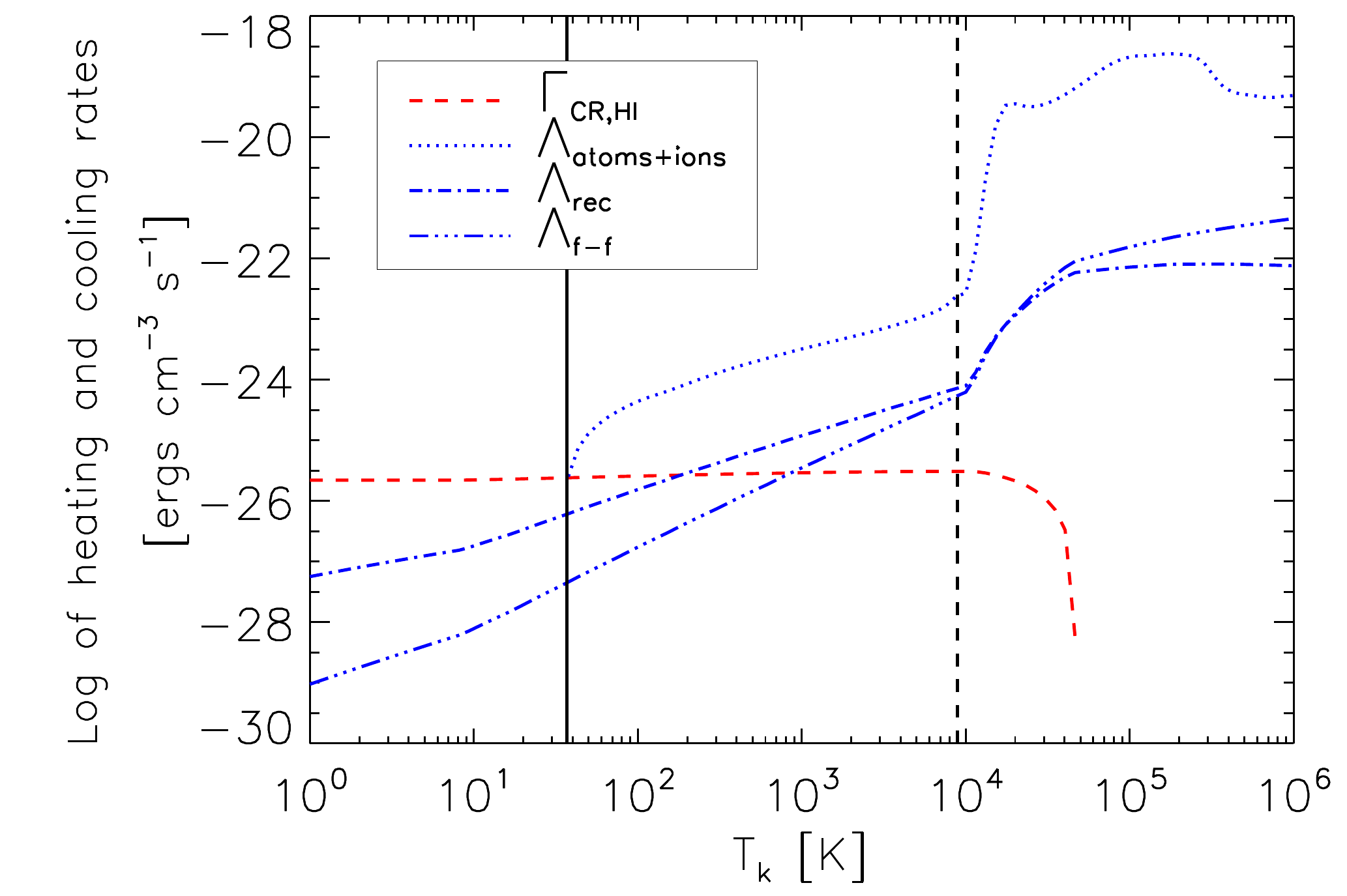}\includegraphics[width=0.5\textwidth]{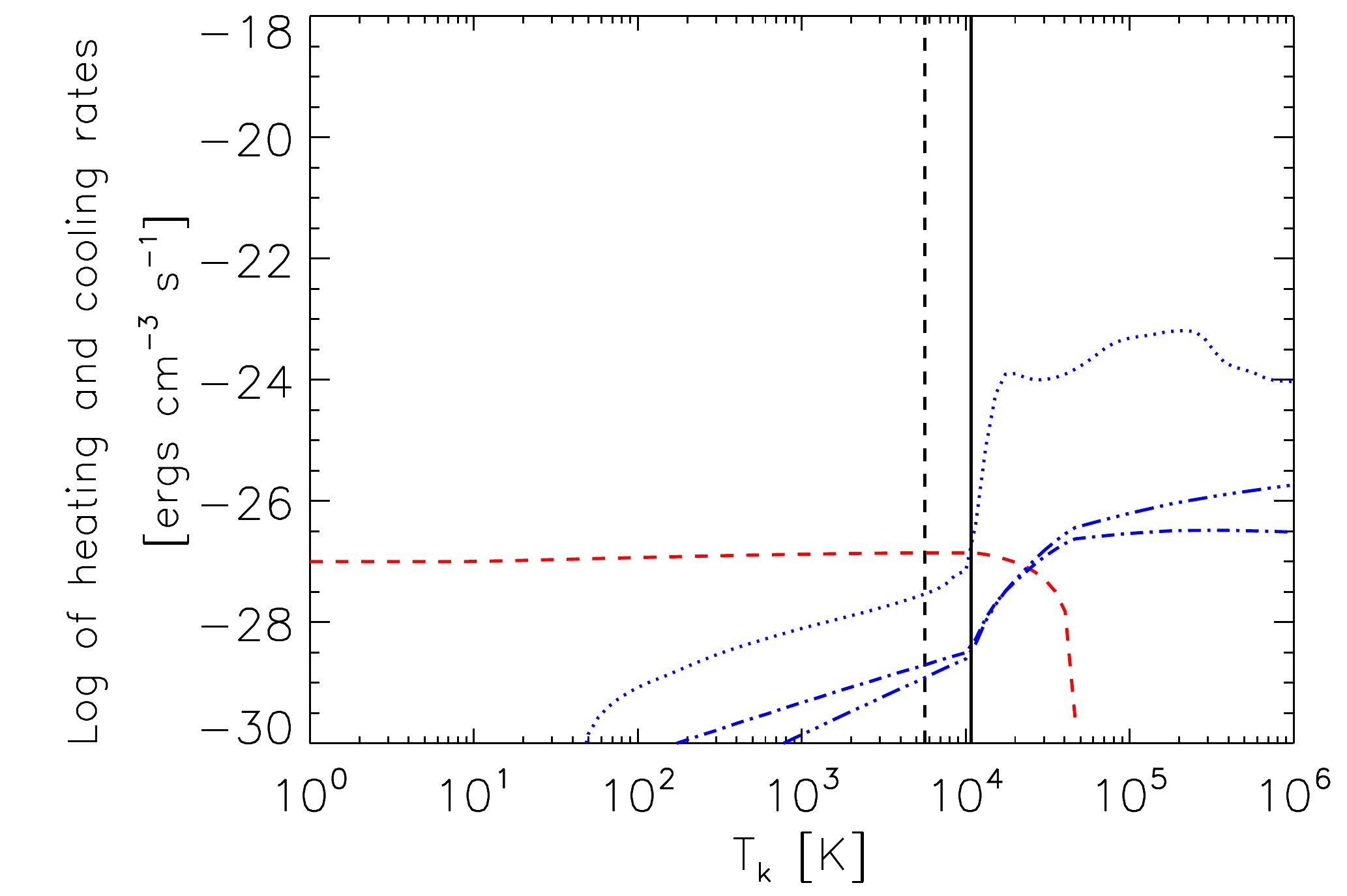}
\caption{Heating and cooling rates as functions of temperature for two SPH gas 
particles with [$\nH=14.62$\,\cmpc, $\cri=4.02\e{-16}$\,\ps, \xe($T_{\rm 
k,SPH}=8892\,K)=0.006$] (left) and [$\nH=0.09$\,\cmpc, $\cri=3.92\e{-16}$\,\ps, 
\xe($T_{\rm k,SPH}=5649\,K)=0.001$] (right). In both cases the metal line emission
is the main cooling agent. The left-hand side plot 
illustrates the case of an SPH particle of high gas density and metallicity, 
leading to relatively efficient metal line cooling and in a low equilibrium temperature 
of $\Tk=37$\,K.  The right-hand side plot shows a case of lower density and 
metallicity causing less cooling by metal lines, hence a higher equilibrium 
temperature despite a slightly lower CR heating rate. The dashed vertical lines in the two
panels indicate the original SPH gas temperatures of the SPH particles, and the
solid vertical lines mark their final equilibrium temperatures.
}
\label{apB1}
\end{figure*}

Figure \ref{apB1} shows the above heating and cooling rates pertaining to two
example SPH particles with similar initial temperatures ($\sim 10^4\,{\rm K}$).
Because of different ambient conditions (i.e., \nhi, \xe, \Z, and \cri) the
equilibrium temperature solutions for the two gas particles end up being
significantly different.

\section{Thermal balance of the molecular gas phase}
\label{apC}
As described in Section \ref{structure} \texttt{S\'IGAME} assumes that the molecular
gas resides exclusively in giant molecular clouds that have Plummer radial
density profiles (i.e., given by eq.\ \ref{equation:plummer-profile}).
Throughout the clouds the gas temperature is solved for according to the heating
and cooling equilibrium requirement $\Hpe+\Hcrh2 = \Ch2+\Cco+\Ccii+\Coi$+\Cgd
(eq.\ \ref{Tk_GMC_e1}).

\Hpe is the heating rate of the gas due to photo-electric ejection of electrons 
from dust grains by FUV photons, and is given by \citep{bakes94}:
\begin{equation}
	\Hpe = 10^{-24}\epsilon \ga \nH~~{\rm \hu},
\end{equation}
where \ga is the local attenuated FUV field in Habing units, derived following eq.\,\ref{ga}, 
and $\epsilon$ is the heating efficiency:
\begin{align}
	\epsilon = &\frac{4.87\times10^{-2}}{1+4\times10^{-3}(\ga T^{0.5}/n_e)^{0.73}}\\
		&+\frac{3.65\times10^{-2}}{1+4\times10^{-3}(\ga T^{0.5}/n_e)^{0.73}}, \nonumber
\end{align}
where \ne is the electron density, calculated as $\xe \nH$, with \xe again calculated using the procedure of 
I. Pelupessy.

\Hcrh2 is the heating rate by cosmic rays traveling through molecular gas \citep{stahler05}:
\begin{equation}
	\Hcrh2 = 1.068\e{-24}\left (\frac{\crimol}{10^{-16}}\right )\left( \frac{\nh2}{10^3\,\cmpc}\right)~~{\rm ergs\,cm^{-3}\,s^{-1}}
\end{equation}
where \crimol is the local CR primary ionization rate of \h2 molecules, which
is approximately $1.6\times$ higher that of \hi atoms \citep{stahler05}.

\Ch2 is the \h2 line cooling rate, and we use the parameterization made by
\cite{papa14} that includes the two lowest \h2 rotational lines (S(0) and S(1),
the only lines excited for  $\Tk\lesssim1000\,{\rm K}$):
\begin{align}
	\Ch2 = &2.06\times10^{-24} \frac{n_{\rm H_2}}{1+{\rm r}_{\rm op}} \left[ 1+\frac{1}{5}e^{510{\rm K}/T_{\rm k}}\left( 1+\frac{n_{0}}{n_{\rm H_2}} \right) \right]^{-1} \\
	&\times(1+R_{10})\rm{\,ergs\,cm^{-3}\,s^{-1}}, \nonumber
\end{align}
where ${\rm R}_{10}$ is defined as:
\begin{equation}
 	{\rm R}_{10} = 26.8\,{\rm r}_{\rm op}\left[ \frac{1+(1/5)e^{510{\rm K}/T_{\rm k}}\left( 1+\frac{n_0}{n_{\rm H_2}}\right)}{1+(3/7)e^{845\,{\rm K}/T_{\rm k}}\left( 1+\frac{n_1}{n_{\rm H_2}}\right)} \right],
\end{equation}
and $n_{\rm 0}\sim54$\,\cmpc and $n_{\rm 1}\sim10^3$\,\cmpc are the critical
densities of the S(0):2-0 and S(1):3-1 rotational lines. ${\rm r}_{\rm op}$ is
the ortho-\h2/para-\h2 ratio (set to 3 which is the equilibrium value).

For the cooling rates due to the [C\,{\sc ii}]$158\,{\rm \mu m}$ and [O\,{\sc
i}]$63\,{\rm \mu m}$+$146\,{\rm \mu m}$ fine-structure lines we adopt the
parameterizations by \citet{rollig06}. The C\,{\sc ii} cooling rate (\Ccii) is:
\begin{align}
	\Ccii = &2.02\e{-24}n\Z \\ 
	&\times \left[ 1+\frac{1}{2}e^{92{\rm K}/T_{\rm k}}\left(1+1300/\nH \right) \right]^{-1}~~{\rm ergs\,cm^{-3}\,s^{-1}}, \nonumber
\end{align}
where a carbon to hydrogen abundance ratio that scales with metallicity
according to $\chi_{\rm [C]}=1.4\e{-4}$\,\Z is assumed. For the parameterization
of the O\,{\sc i} cooling rate ($\Coi = \Lambda_{63\mu \rm{m}}+\Lambda_{146\mu
\rm{m}}$) we refer to eqs.\ A.5 and A.6 in \citet{rollig06} and simply note that
we adopt \citep[in accordance with][]{rollig06} an oxygen to hydrogen abundance
ratio of $\chi_{\rm [O]}=3\e{-4}\,\Z$.

\Cco is the cooling rate due to CO rotational transitions. We use the parameterization provided by \cite{papa13}:
\begin{equation}
	\Cco = 4.4\times10^{-24} \left(\frac{n_{\rm H_2}}{10^4} \right)^{3/2} \left( \frac{T_{\rm k}}{10\,{\rm K}} \right)^2 \left( \frac{\chi_{\rm CO}}{\chi_{\rm [C]}} \right)\,{\rm ergs\,cm^{-3}\,s^{-1}}, 
\end{equation} 
where $ \chi_{\rm CO}/\chi_{\rm [C]}$ is the relative CO to neutral carbon
abundance ratio, the value of which we determine by interpolation, assuming that
$\chi_{\rm CO}/\chi_{\rm [C]}=(0.97,0.98,0.99,1.0)$ for
$\nh2=5\times10^3,10^4,10^5,10^6$\,\cmpc, respectively \citep{papa13}.  

\Cgd is the cooling rate due to gas-dust interactions and is given by \citep{papa11}:
\begin{equation}
	\Cgd = 3.47\times10^{-33}\nH^2\sqrt{\Tk}(\Tk-T_{\rm dust})\,{\rm ergs\,cm^{-3}\,s^{-1}}, 
        \label{equation:dust-gas-cooling}
\end{equation}
where the dust temperature ($T_{\rm dust}$) is calculated using
eq.\,\ref{Tk_GMC_e2} (Section \ref{Tk_GMC}). 

\begin{figure*} 
\centering
\includegraphics[width=0.5\textwidth]{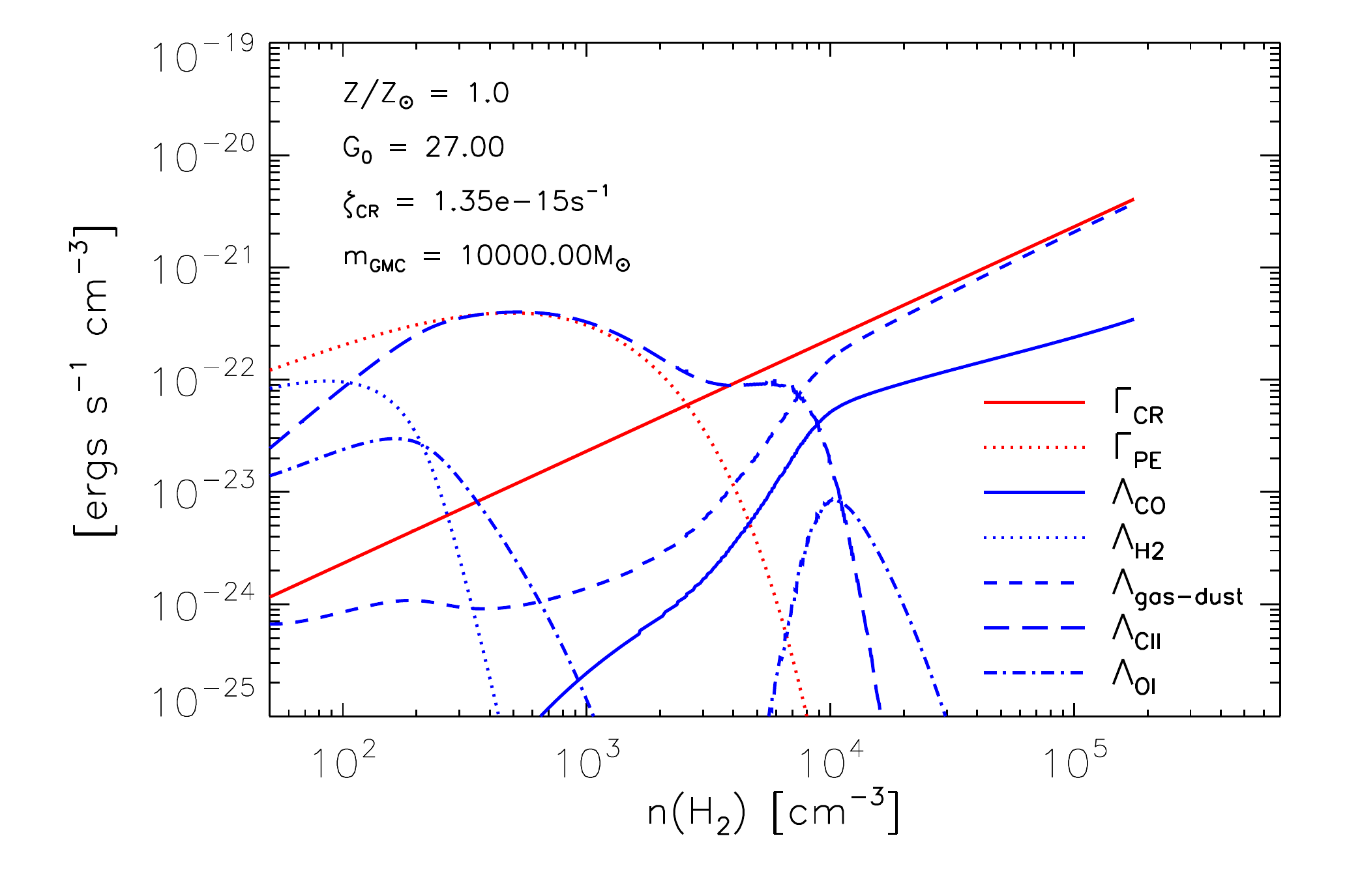}\includegraphics[width=0.5\textwidth]{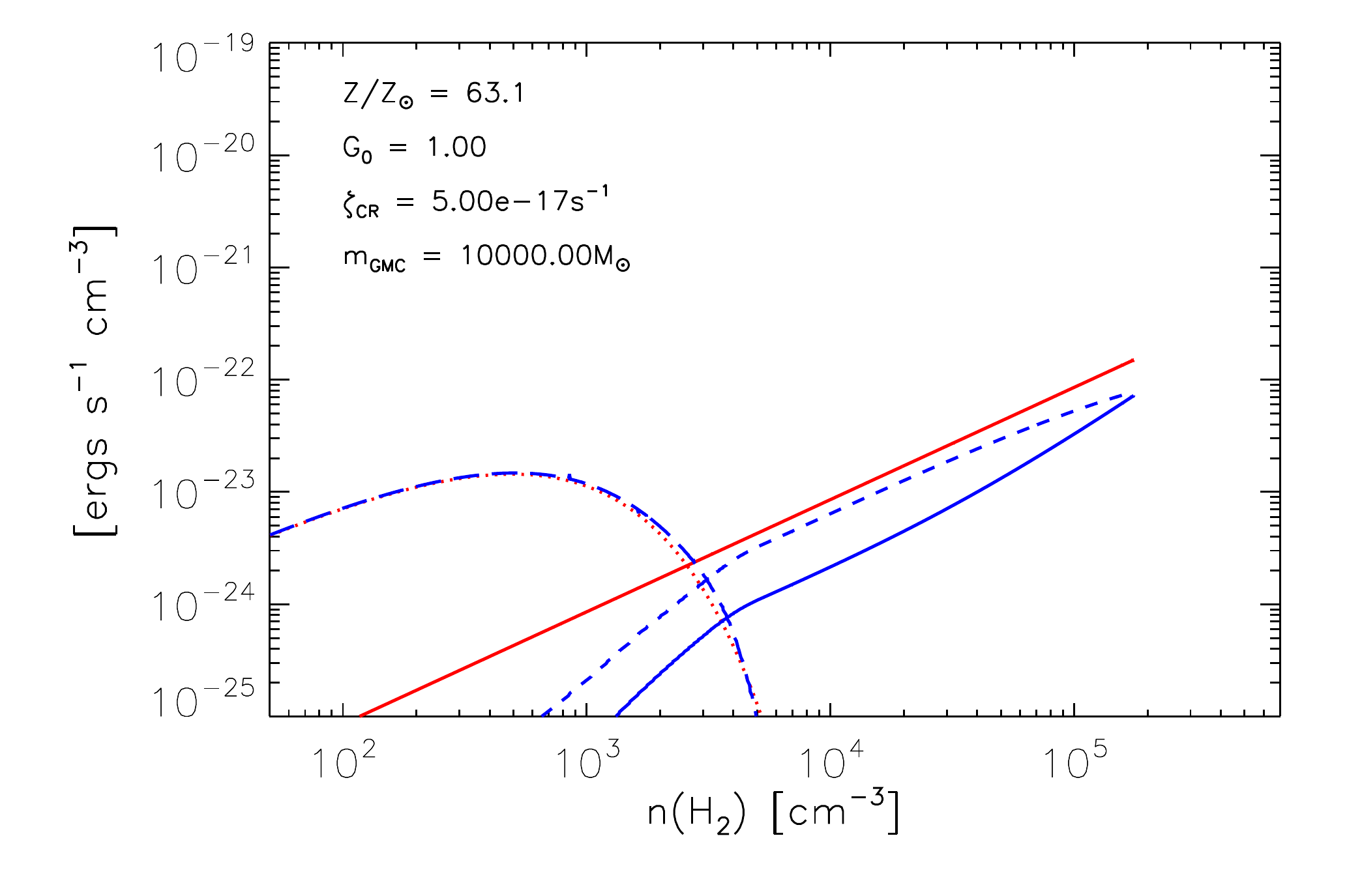}
\caption{Equilibrium heating (red curves) and cooling (blue curves) rates of the gas as
functions of \h2 density for two different GMC models with 
[$\Mgmc=10^4$\,\msun, $\Z=1$, $\g0=27$, $\cri=1.35\e{-15}\,\ps$, $\Pe=10^4\,$K\,\cmpc] (left) and 
[$\Mgmc=10^4$\,\msun, $\Z=63$, $\g0=1$, $\cri=5\e{-17}\,\ps$, $\Pe=10^4\,$K\,\cmpc] (right). 
In the case of high FUV and CR radiation fields but low metallicity (left), the heating 
is dominated by cosmic ray heating (red solid) in the inner region ($\nh2\gtrsim1500$\,\cmpc) and 
photoelectric heating (red dotted) in the outer region.
Cooling is dominated by gas-dust interactions (blue dashed) 
in the inner region ($\nh2\gtrsim10000$\,\cmpc) and by \cii as well as \h2 line cooling 
(blue long-dashed and dotted) in the outer region.
In the opposite case of low FUV and CR radiation fields but high metallicity (right), 
the same heating and cooling mechanisms are dominating the energy balance throughout the cloud, except 
in the inner region, where cooling by CO line emission (blue solid) is more important than it is in the 
case of low metallicity.}
\label{apC1}
\end{figure*}

\section{GMC models}
\label{apD}
\begin{figure*}
\hspace{-1cm}
\includegraphics[width=2.2\columnwidth]{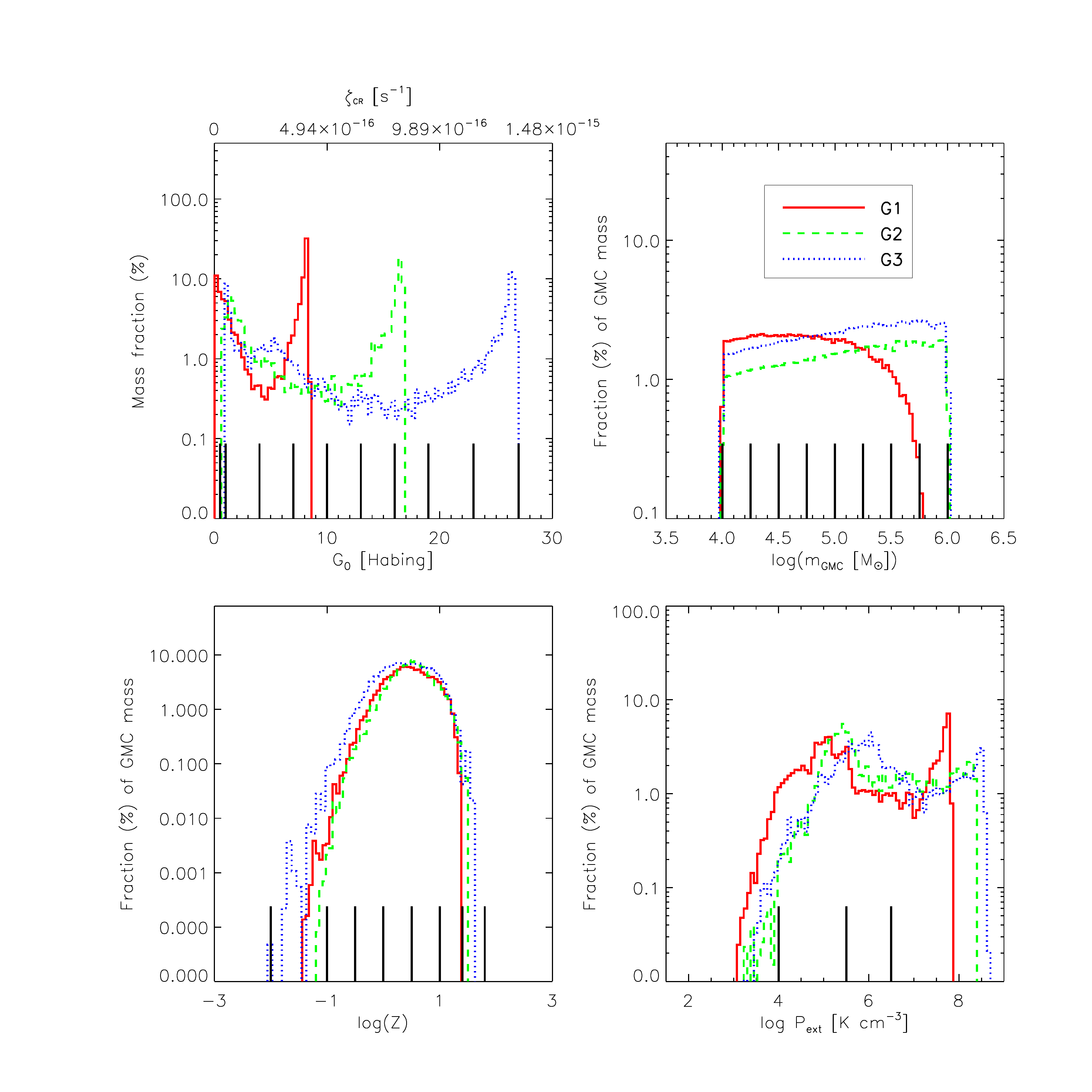}
\caption{Mass-weighted histograms of the basic parameters of the GMCs in G1 (red
solid), G2 (green dashed), and G3 (blue dotted). From top left and
clockwise: the local far-UV field (\g0) (and CR ionization rate since
$\cri\propto\g0$), GMC mass (\Mgmc), metallicity (\Z) and external pressure
(\Pe).  Black vertical lines indicate the \g0, \Mgmc, \Z, \Pe-values for which
$\Tk-\nh2$ curves were calculated (see Figure \ref{apD3}) -- a total of 630 GMCs
which make up our grid GMC models. Each GMC in the galaxies is assigned the
$\Tk-\nh2$ curve of the GMC model at the closest grid point.} 
\label{apD1}
\end{figure*}

\begin{figure*} 
\centering
\hspace{-1cm}
\includegraphics[width=2.2\columnwidth]{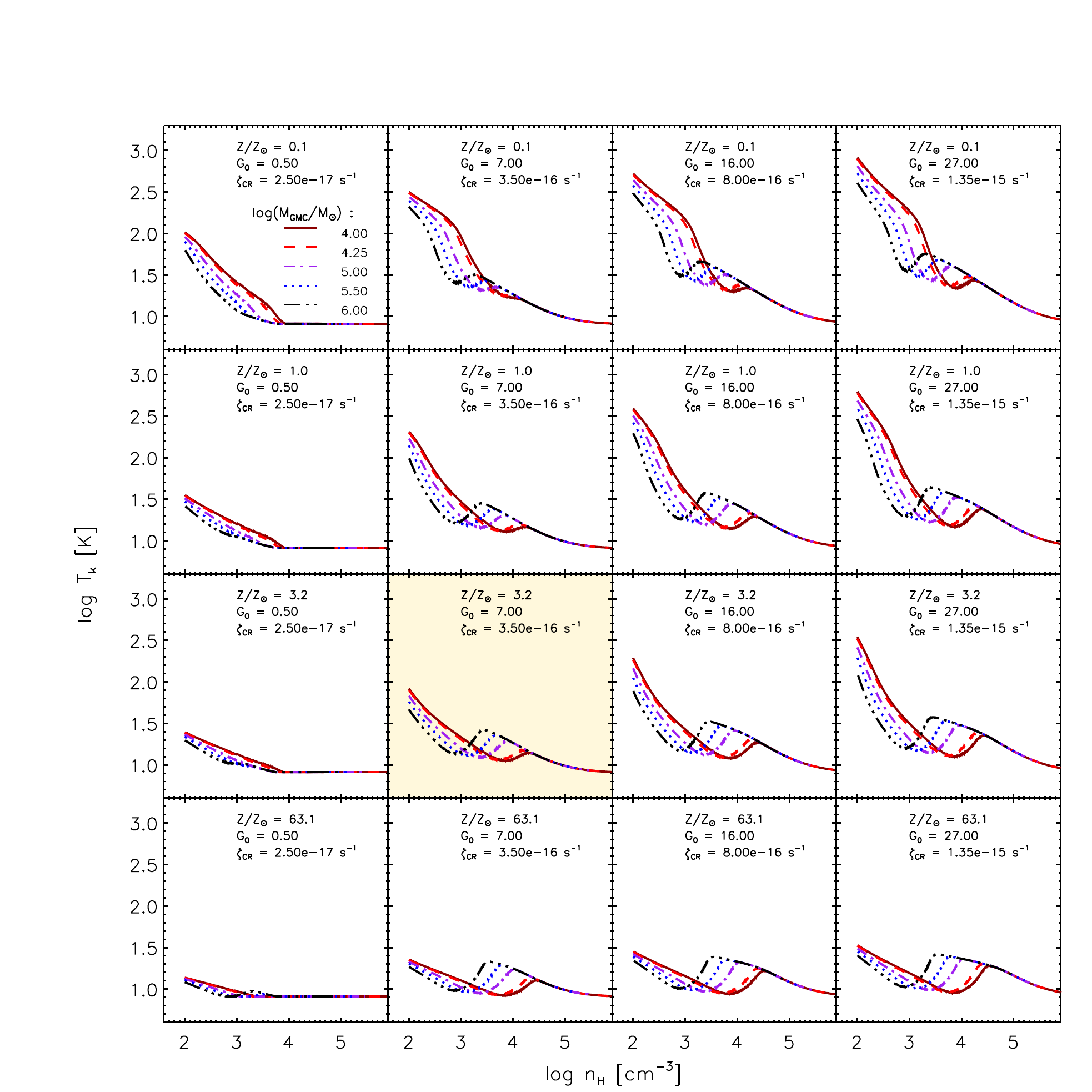} 
\caption{Kinetic temperature versus \h2 density curves for 80 out of the 630
grid model GMCs that span the full (\g0, $M_{\rm GMC}$, \Z) parameter space set
by and marked on top of the distributions in Figure \ref{apD1} (see also
Section \ref{gmcgrid}) for a pressure of $\Pe=10^4$\,K\,\cmpc.  The grid model
most often assigned to GMCs in G1 is indicated by the red dashed curve in the
highlighted panel and corresponds to $\g0=7.0$ ($\cri=3.5\e{-16}$\,\ps), $\log
\Mgmc/\msun =4.25$, and $\Z=3.2$.  In general, higher metallicity (from top to
bottom) leads to more cooling via emission lines of ions, atoms and molecules,
and hence lower temperatures.  On the other hand, higher UV and CR fields (from
left to right) cause more heating and therefore higher \Tk. The decreasing
trend of \Tk with higher values of \nh2 is mainly caused by the gradual
attenuation of the UV field as one moves into the cloud.  The `bump' at
$\nh2\sim10^3-10^4$\,\cmpc corresponds to the transition from CII line cooling
to the less efficient CO line cooling at higher densities.  At densities above
$\nh2=10^4$\,\cmpc, gas-dust interactions set in and eventually cools the gas
down to the CMB temperature in all GMC cores.}
\label{apD3}
\end{figure*}

Each SPH particle is divided into several GMCs as described in \S\,\ref{split},
and we derive the molecular gas density and temperature within each from three
basic parameters which are SFR density, GMC mass, \Mgmc, and metallicity,
$\Z/Z_{\odot}$.  Derived from these basic parameters are the far-UV and cosmic
ray field strengths, the \h2 gas mass fraction of each SPH particles, as well as
the GMC properties used to derive the CO excitation and emission; \h2 density
and temperature.  Histograms of the basic parameters are shown in
Figure\,\ref{apD1}, while properties derived thereof can be found in Figure\,\ref{apD3}.

\begin{figure*} 
\centering
\hspace{-1cm}
\includegraphics[width=2.2\columnwidth]{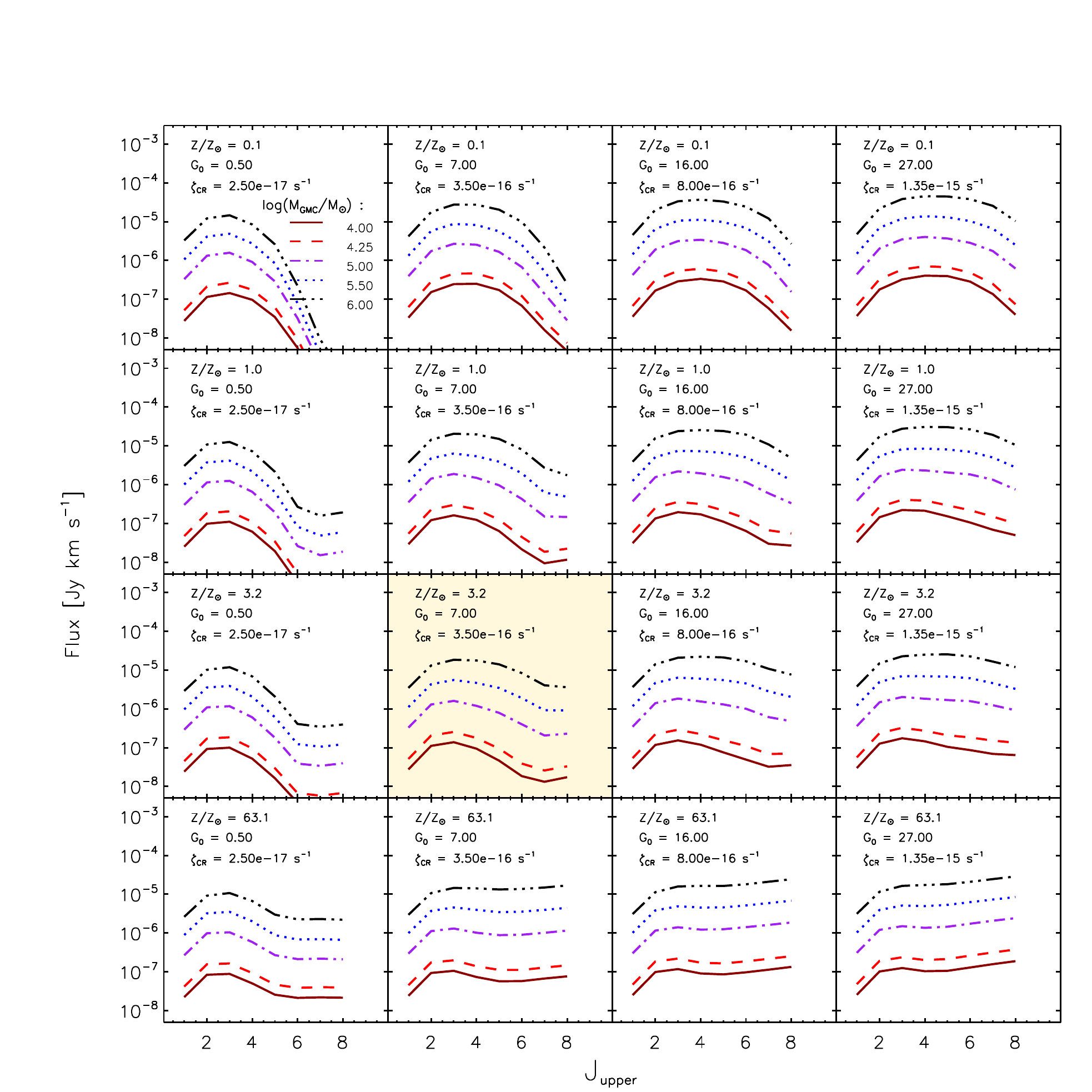}   
\caption{CO SLEDs obtained with \lime for the 80 model GMCs whose
$\Tk-\nh2$ curves are shown in Figure \ref{apD3}. The CO SLED used most often in
G1 is shown as the red dashed curve in the highlighted panel.  These CO
SLEDs were made for a fixed external pressure of $\Pe/k_{\rm B}=10^4\,{\rm
\cmpc\,K}$ and a default Plummer density profile.}
\label{apD4}
\end{figure*}

\begin{figure*} 
\centering
\hspace{-1cm}
\includegraphics[width=2.2\columnwidth]{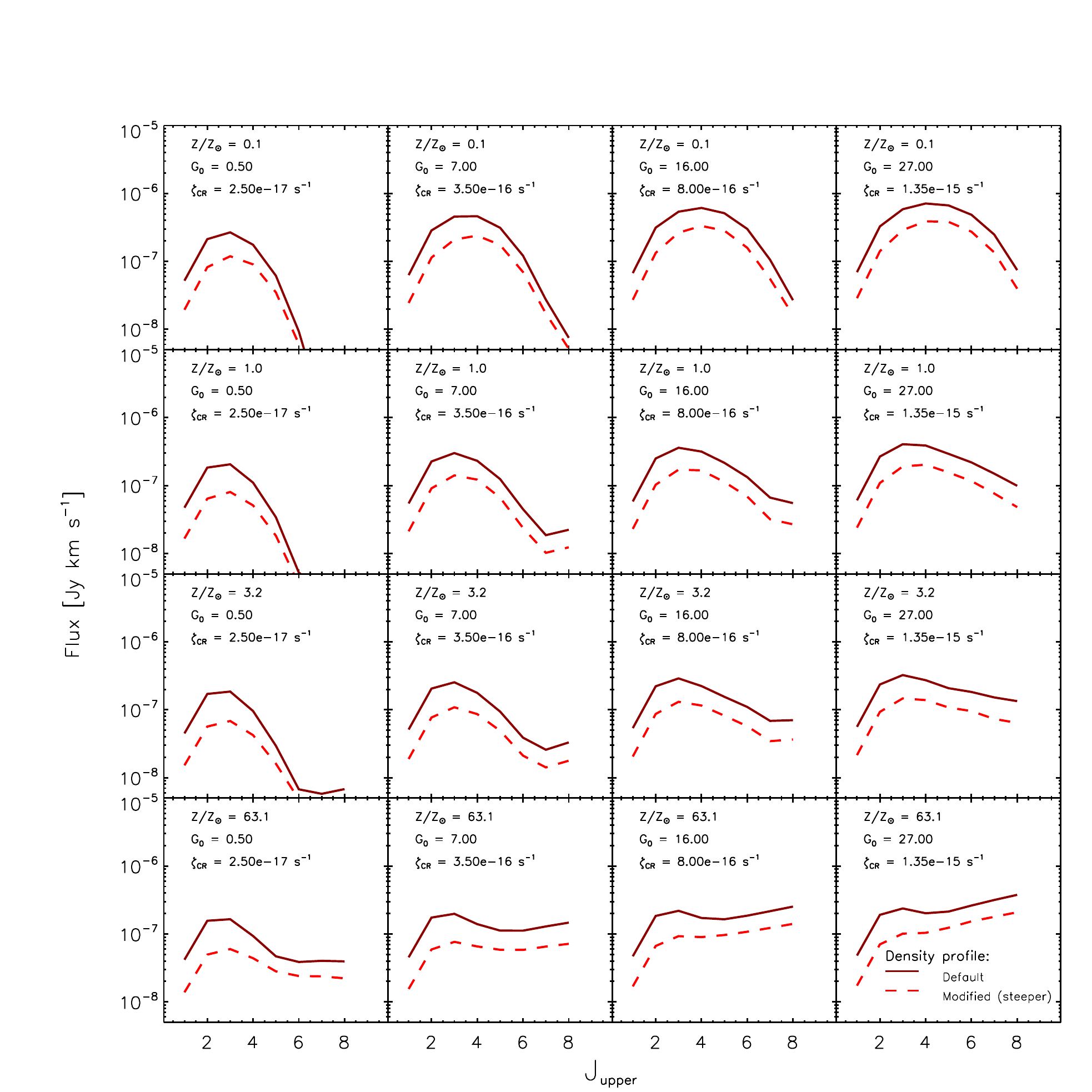}   
\caption{CO SLEDs obtained with \lime for the same [\g0,\Zn] values as 
shown in the panels of Figure \ref{apD4} for Plummer density profiles with power-law
index $-5/2$ (solid curve) and $-7/2$ (dashed curve).  In all panels, the
external pressure has been fixed to $\Pe/k_{\rm B}=10^4\,{\rm \cmpc\,K}$ and the GMC mass 
to $\Mgmc=10^{4.25}\,{\rm \msun}$.}
\label{apD5}
\end{figure*}
\bsp

\begin{figure*} 
\centering
\hspace{-1cm}
\includegraphics[width=2.2\columnwidth]{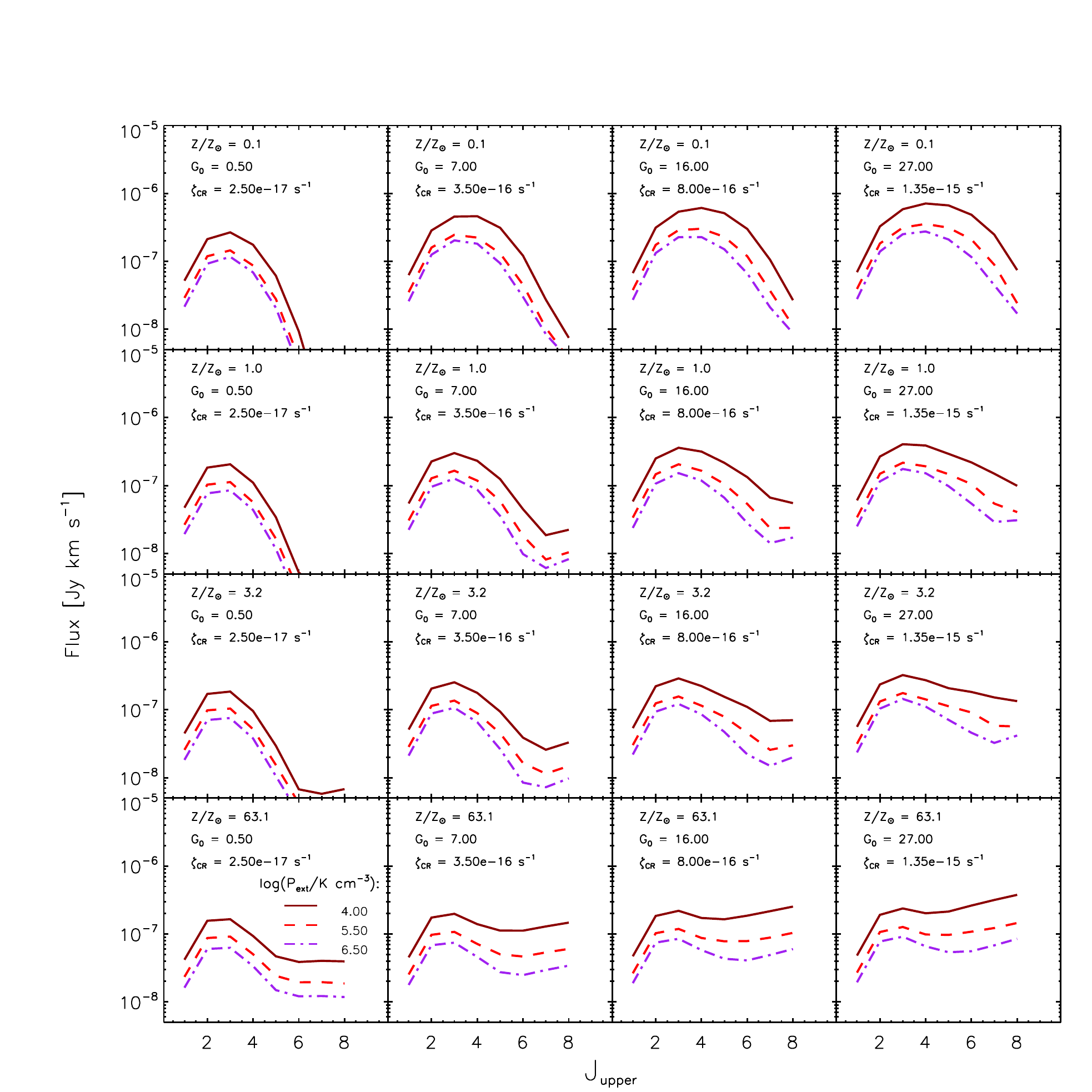}   
\caption{CO SLEDs obtained with \lime for the same [\g0,\Zn] values as
shown in the panels of Figure \ref{apD4} for $\Pe/k_{\rm B} = 10^4\,{\rm
cm^{-3}\,K}$ (solid curves), $10^{5.5}\,{\rm cm^{-3}\,K}$ (dashed curves), and
$10^{6.5}\,{\rm cm^{-3}\,K}$ (dot-dashed curves).  In all panels, the GMC mass
is fixed to $10^{4.25}\,\msun$.  Higher pressure environments are seen to lead
to a decrease in luminosity for all transitions.}
\label{apD6}
\end{figure*}

\clearpage
\clearpage

\section{Testing \sigame on MW-like galaxies}
\label{apMW}

As a benchmark test, \sigame was applied to galaxy simulations at $z=0$ with
properties similar to those of the MW. The reasoning here was that since much of
the sub-grid physics adopted by \sigame relies on empirical relations observed in
the MW (e.g., the GMC mass spectrum index), the method ought to come
close to re-producing the observed CO properties of spiral galaxies, such as our
MW, in the local universe.

The three galaxies (hereafter denoted MW1, MW2, and MW3 in order of increasing
SFR) were selected from a more recent version of the SPH simulation described
in Section\,\ref{cosmological_simulations}. The properties of MW1, MW2, and MW3
are listed in Table\,\ref{MWgal} and are seen to be within a factor $\sim 2-6$
of the stellar mass \citep[${\rm M}_{*,\rm{MW}}=6\e{10}$\,\msun;][]{mcmillan11}
and SFR \citep[$1.9\pm0.4$\,\sfru; ][]{chomiuk11} of the MW.

The steps outlined in Section \ref{subsection:methodology} (and further described in
Sections \ref{WCNM} to \ref{Tk_GMC}) were followed, but with two minor modifications: 1) the
CMB temperature at $z=0$ was set to 2.725\,K rather than 8.175\,K used at
$z=2$, and 2) the resolved FUV fields of our MW-like model galaxies span a
range below that seen in our $z=2$ model galaxies (cf. Figure\,\ref{apD1})
meaning that the \g0 parameter values for our GMC grid had to be adjusted
correspondingly. The following \g0 grid points were chosen: $[0.05,
0.1, 0.15, 0.1, 0.2, 0.4, 0.6, 0.8, 1.0, 1.2]$\,Habing.

\begin{figure*} 
\centering
\hspace{-1cm}
\includegraphics[width=2.2\columnwidth]{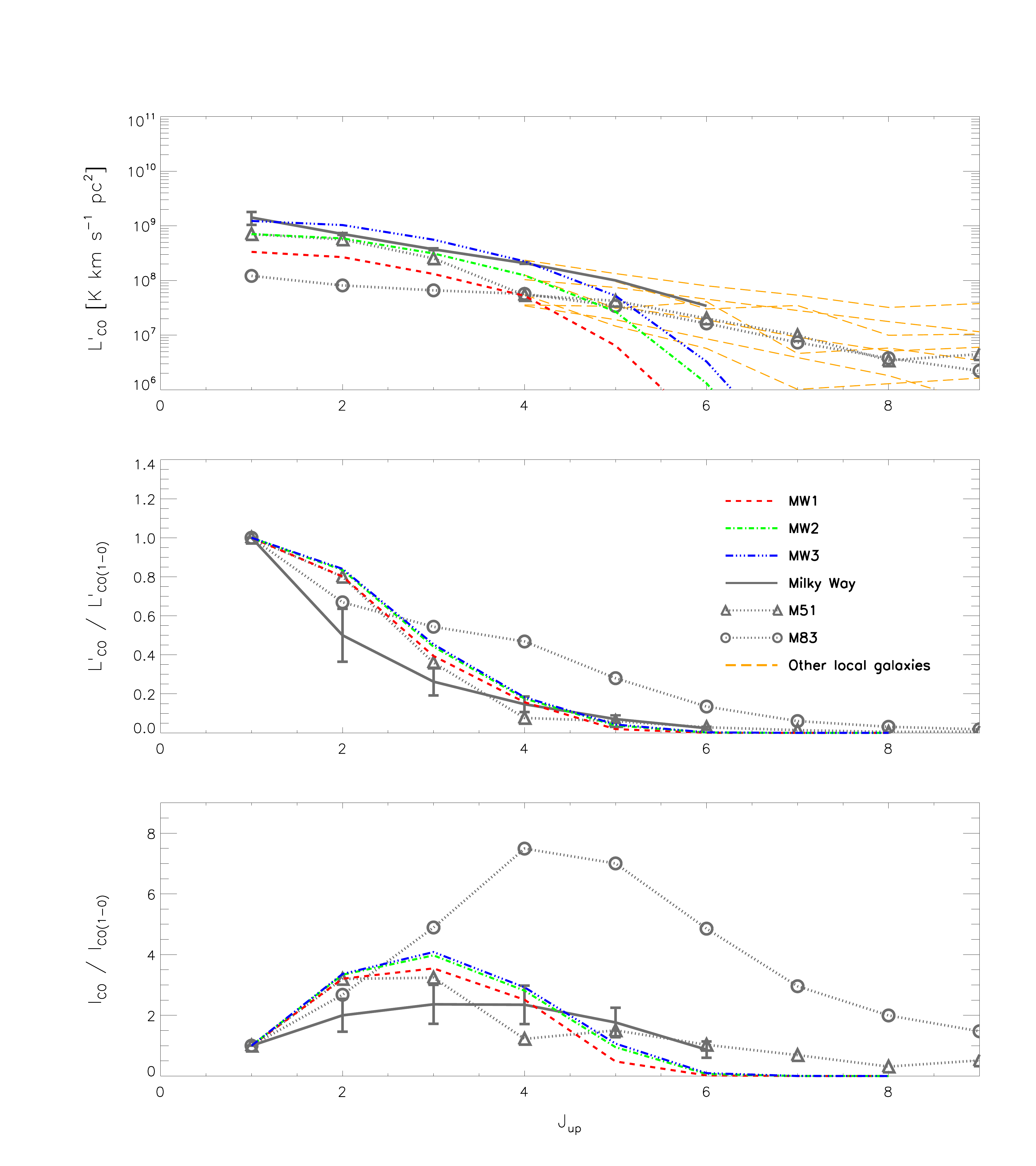}   
\caption{Global CO SLEDs of our three model galaxies MW1, MW2 and MW3 shown as
red (dashed), green (dash-dot) and blue (dash-dot-dot-dot) curves,
respectively, selected to represent the MW in terms of stellar mass and SFR.
The SLEDs are given as absolute line luminosities in units of
K\,\kms\,pc$^{-2}$ (top panel), as brightness temperature ratios normalised to
the CO(1$-$0) transition (middle panel), and as velocity-integrated intensity
ratios normalised to CO(1$-$0) (bottom panel).  The observed global CO SLED of
the MW is shown in grey \citep{fixsen99}, and global CO measurements of the
local star-forming galaxy M\,51 are displayed with triangles. 
The CO observations for M\,51 are from
\citet{vlahakis13} and \citet{hughes13}.  Also shown, with
long-dashed orange lines, are observed CO SLEDs of local galaxies with IR luminosities within $20\%$
of that of the MW \citep{kamenetzky15}.} 
\label{apD7}
\end{figure*}

In Figure\,\ref{apD7} we compare the resulting CO
SLEDs with that of the MW from \cite{fixsen99}, as well as with those of M\,51
($M_{\ast}=4.7\e{10}\,\msun;$ \cite{cooper12}, ${\rm SFR}=2.6$\,\sfru;
\citealt{schuster07}) from \cite{vlahakis13}, \cite{hughes13} and 
\cite{kamenetzky15}, M\,83 
($M_{\ast}=7.9\e{10}\,\msun$ and ${\rm SFR}=3.2$\,\sfru;
\citealt{jarrett13}) from \cite{wu15} 
and other local galaxies of IR luminosity within 20\% of that of 
the MW ($1.8\e{10}$\,\lsun; \citealt{wright91}) from \cite{kamenetzky15}. 
M\,51 is a nearby galaxy slightly smaller in size and stellar mass than the MW, 
whereas M\,83 is more massive and star-forming than the MW. 
The dispersion in observed CO SLEDs among the MW, M\,83, M\,51 
and other local galaxies, allows for 
a large range in normalization and shape of the CO SLED within 
which our model galaxies find themselves.

The CO line luminosities of the MW lie roughly in the range spanned by 
the CO SLEDs of MW2 and MW3 up to and including $J_{\rm up} = 4$, 
but significantly above that of MW1.
For the higher $J$ transitions the line luminosities of our simulations 
drop off more rapidly than the MW, 
meaning that our model galaxies have lower luminosities compared to
the MW (and other local galaxies of IR luminosities within 20\% from that
of the MW; long-dashed orange lines) at $J_{\rm up} > 4$.
This suggests that our simulations are missing a warm/dense component which is
required to significantly excite these high-$J$ lines. 
The agreement between the CO SLED of MW2 and that of M\,51 is good at 
low transitions ($J_{\rm up} < 6$), which is encouraging given that the two galaxies
have nearly identical SFR and $M_{\ast}$. 
At higher $J$ values, the CO SLED of M\,51 has an excess of CO emission when compared with MW2, 
corresponding to a warm gas phase not captured by our models. 
This warm phase is more pronounced in signature when looking at the CO SLED of M\,83, 
the line ratios of which are above our model galaxies for all $J_{\rm up} > 2$. 
Our models MW1, MW2 and MW3 only agree in CO luminosity with M\,51 at the 
$J_{\rm up} = 4$ and $5$ transitions. 

As for brightness temperature ratios (see middle panel) of our simulated MW-like 
galaxies only agree with the MW within the observational errors at CO(4$-$3), 
and other wise display a CO SLED shape noticably different from the MW. 
Where our model galaxies peak at the CO(3$-$2) transition in 
velocity-integrated intensity (see bottom panel), similar to M\,51, 
the MW and M\,83 peak at CO(4$-$3).
However, nowhere do the simulated line ratios differ by more than a factor of 2. 

The molecular gas masses of MW1, MW2 and MW3 are $4.6$, $5.4$ and 
$8.5\e{9}\,\msun$, respectively, significantly above $(1.0\pm0.3)\e{9}\,\msun$ 
for the MW \citep{heyer15}. 
The corresponding global ($R<20\,$kpc) \aCO factors are 
$13.6$, $7.7$ and $7.0\,\msun\,\rm{pc}^{-2}$\,(K\,km\,s$^{-1}$)$^{-1}$. 
For the inner disk ($R<10\,\rm{kpc}$) of the MW, $\alpha_{\rm CO,MW} 	
\simeq 4.3\pm 0.1\,{\rm \msun\,pc^{-2}\,(K\,km\,s^{-1})^{-1}}$ 
\citep[e.g.,][]{strong96,dame01,pineda08,bolatto13}.
However, typically the measurements of \aCO in the MW do not probe the central 
($<1\,$kpc) region. 
Excluding the central $R<1\,$kpc region and going out to $R=10\,$kpc, 
the \aCO factors of MW1, MW2 and MW3 are 
$11.6$, $6.9$ and $6.2\,\msun\,\rm{pc}^{-2}$\,(K\,km\,s$^{-1}$)$^{-1}$, 
i.e. factors of $2.7$, $1.6$ and $1.4$ above $\alpha_{\rm CO,MW}$, respectively.

\begin{center}
\begin{table}
\centering
\caption{Properties of the three $z=0$ MW-like galaxies (MW1, MW2, and MW3)
used to benchmark \sigame.} 
\begin{tabular}{p{0.6cm}p{1.5cm}p{1.4cm}p{1.4cm}p{0.4cm}p{0.4cm}p{0.4cm}p{0.4cm}}
\hline
\hline
 	 			&	SFR 		& 	$M_{\ast}$	 		& $M_{\rm SPH}$ 		 &	$f_{\rm SPH}$	&	$\Z$	& 	$R_{\rm cut}$  \\ 
			 	&	[\sfru]		& 	[$10^{10}$\,\msun]	& [$10^{10}$\,\msun]	 &					&		& 	[kpc]  \\ 
\hline
MW1				&	2.2			&	$0.97$				&	$1.17$				 &	55\%			&	2.07	&	$20$		\\ 
MW2				& 	4.1			&	$3.12$				&	$1.27$				 &	29\%			&	3.31	&	$20$		\\ 
MW3 			& 	6.2			&	$3.83$				&	$1.69$				 &	31\%			&	3.90	&	$20$		\\ 
\hline
\end{tabular}
\\
	{\bf Notes.} All quantities are determined within a radius $R_{\rm cut}$,
	which is the radius where the cumulative radial stellar mass function of
	each galaxy becomes flat. The gas mass ($M_{\rm SPH}$) is the total SPH gas
	mass within $R_{\rm cut}$.  The metallicity ($\Z=Z/Z_{\odot}$) is the mean
	of all SPH gas particles within $R_{\rm cut}$.
\label{MWgal}
\end{table}
\end{center}

\end{document}